\documentclass[twocolumn]{aastex6}
\pdfoutput=1 
\usepackage{amsmath,amstext}
\usepackage[T1]{fontenc}
\usepackage{apjfonts} 
\usepackage[figure,figure*]{hypcap}
\usepackage{graphicx}
\usepackage{affils}
\usepackage{hyperref}
\usepackage{microtype}
\usepackage{caption}
\usepackage{subcaption}
\usepackage{xspace}

\def\Mej{$M_{\rm ej}$}
\def\P{$P$}
\def\B{$B_{\perp}$}
\def\vp{$v_{\rm phot}$}
\def\vej{$v_{\rm ej}$}
\def\k{$\kappa$}
\def\kg{$\kappa_\gamma$}
\def\Ek{$E_{\rm K}$}
\def\Mns{$M_{\rm NS}$}
\def\Tf{$T_{\rm f}$}
\def\Av{$A_V$}

\def\Ni{$^{56}$Ni}

\def\M{M$_\odot$}
\def\Z{Z$_\odot$}
\def\kms{km\,s$^{-1}$}
\def\cmsqperg{cm$^{2}$\,g$^{-1}$}
\def\ergpers{erg\,s$^{-1}$}

\def\mosfit{\texttt{MOSFiT}\xspace}

\shorttitle{MCMC modelling of SLSNe}
\shortauthors{M.~Nicholl et al.}
\begin{document}

\title{The magnetar model for Type I superluminous supernovae I: Bayesian analysis of the full multicolour light curve sample with MOSFiT}

\DeclareAffil{cfa}{Harvard-Smithsonian Center for Astrophysics, 60 Garden Street, Cambridge, Massachusetts 02138, USA; \href{mailto:matt.nicholl@cfa.harvard.edu}{matt.nicholl@cfa.harvard.edu}}

\affilauthorlist{Matt Nicholl\affils{cfa},
James Guillochon\affils{cfa} and
Edo Berger\affils{cfa}
}

\begin{abstract}

We use the new Modular Open Source Fitter for Transients (\mosfit) to model 38 hydrogen-poor superluminous supernovae (SLSNe). We fit their multicolour light curves with a magnetar spin-down model and present the posterior distributions of magnetar and ejecta parameters. The colour evolution of all SLSNe can be well matched with a simple absorbed blackbody. We find the following medians (1$\sigma$ ranges) for the key parameters: spin period 2.4\,ms (1.2--4\,ms); magnetic field $0.8\times 10^{14}$\,G (0.2--1.8\,$\times 10^{14}$\,G); ejecta mass 4.8\,\M\ (2.2--12.9\,\M); kinetic energy $3.9\times 10^{51}$\,erg (1.9--9.8\,$\times 10^{51}$\,erg). This significantly narrows the parameter space compared to our uninformed priors, showing that although the magnetar model is flexible, the parameter space relevant to SLSNe is actually well constrained by existing data. The requirement that the instantaneous engine power is $\sim 10^{44}$\,erg at the light curve peak necessitates either a large rotational energy ($P<2$\,ms), or more commonly that the spin-down and diffusion timescales be well-matched. We find no evidence for separate populations of fast- and slow-declining SLSNe, which instead form a continuum both in light curve widths and inferred parameters. Variations in the spectra are well explained through differences in spin-down power and photospheric radii at maximum-light. We find no correlations between any model parameters and the properties of SLSN host galaxies. Comparing our posteriors to stellar evolution models, we show that SLSNe require rapidly rotating (fastest 10\%) massive stars ($\gtrsim 20$\,\M), and that this is consistent with the observed SLSN rate. High mass, low metallicity, and likely binary interaction all serve to maintain rapid rotation essential for magnetar formation. By reproducing the full set of SLSN light curves, our posteriors can be used to inform photometric searches for SLSNe in future survey data.

\end{abstract}


\section{Introduction}

Superluminous supernovae (SLSNe) are a class of stellar explosions originally defined by absolute magnitudes of $M<-21$ at the peak of their light curves \citep{gal2012}. This corresponds to a luminosity of $\sim10^{44}$\,erg\,s$^{-1}$, and over rest-frame durations of several months they radiate a total of $\sim10^{51}$\,erg --- about 100 times more energy than any normal supernova (SN). Since their relatively recent discovery \citep{qui2011,chom2011}, the mechanism by which SLSNe produce this copious UV-optical emission has been one of the most hotly-debated topics in time-domain astronomy.

SLSNe come in at least two spectroscopic flavours. Type I SLSNe do not show hydrogen lines \citep[e.g.][]{pas2010,ins2013}, but instead have very hot spectra with \ion{O}{2} absorption at peak. Indeed, with the discovery of several Type I SLSNe with $M\gtrsim-21$ \citep{ins2013,lun2014,chen2016b}, these events are now generally defined by their unique spectra rather than by a magnitude cut. Type II SLSNe do exhibit hydrogen in their spectra, usually in the form of relatively low-velocity emission lines with broad bases \citep[e.g.][]{smi2007b}, and appear to be an extension of the lower-luminosity Type IIn SN population. A few Type II SLSNe have had Balmer lines with weak or no narrow emission component \citep{gez2009,mil2009,ins2016}, but it is not yet clear how these are related to the other classes. Generally, the hydrogen rich SLSNe II are thought to be powered by the interaction of fast SN ejecta with a dense shell or wind surrounding the progenitor \citep{che2011,ben2014}.

The interpretation of Type I SLSNe (hereafter, SLSNe) has not yet reached a consensus. However, in the last few years, a number of clues have emerged that these events are most likely powered by an internal heat source generically termed a `central engine'. Spectroscopic models of H-poor SN ejecta strongly illuminated from within have been successful in reproducing the early spectra of SLSNe \citep{des2012,how2013,maz2016}. A consistency in the strengths of UV absorption lines between SLSN spectra with very different continuum temperatures can be well explained by a central energy source, but is hard to reconcile with `top-lighting' of the ejecta by circumstellar interaction \citep{nic2017}. Moreover, radio \citep{nic2016b} and X-ray \citep{margutti2017} observations of SLSNe favour low-density environments similar to Type Ic SNe, rather than the dense mass-loss required by interaction-powered models.

It has recently been shown by \citet{nic2016c} and \citet{jer2016b} that nebular-phase 
spectra of SLSNe show similar properties to the hyper-energetic SNe that accompany long gamma-ray bursts (GRBs), indicating that a similar engine to that operating in GRB-SNe may also apply to SLSNe. A narrow \ion{O}{1} recombination line at this phase could indicate a high-density region in the inner ejecta resulting from a central overpressure. This connection with long GRBs is reinforced by the preference of both classes for metal-poor dwarf galaxies \citep{chen2013,lun2014,per2016,chen2016}---though exactly how similar these host populations are is still debated \citep{lel2015,ang2016,schu2016}---and the discovery of a borderline-superluminous SN associated with an ultra-long GRB \citep{gre2015}. Additionally, polarimetry of one nearby object has shown a dominant axis and an increase in polarisation with time --- properties that are consistent with GRB-SNe and with engine-powered models \citep{ins2016,lel2017}.

Specifically, the best candidate for a central engine is the spin-down power from a millisecond pulsar with a magnetic field of $\sim10^{13}-10^{15}$\,G (a magnetar) formed in the stellar core-collapse \citep{kas2010,woo2010,met2015}. It has been known for some time that this model can reproduce the rather diverse bolometric light curves of SLSNe \citep{ins2013,cha2013,nic2014,nic2015b}, including those with slow declines that have also been suggested to be nickel-powered explosions \citep{gal2009,nic2013,lun2016}. However, while many authors have now presented magnetar model fits to individual SLSNe or small population samples, there exist many systematic differences between the codes, input parameters and bolometric corrections used in the literature. This has so far precluded much systematic analysis of the model parameter space occupied by observed SLSNe. Looking for correlations between parameters, and comparing observed parameter ranges to those suggested by theory, is an essential test not only of which models are correct, but also of what kinds of stars actually lead to these explosions. For example, if multiple stellar evolutionary channels can lead to engine formation, this could be borne out through multimodality in the distributions of one or more parameters. This can only be tested by studying a maximal sample in a homogeneous fashion.

As supernova science moves into the era of `big data' (and even rare subclasses like SLSNe approach a statistically-meaningful sample), such population studies will become increasingly important. One promising catalyst to facilitate this is the \textit{Open Supernova Catalog}\footnote{\url{https://sne.space/}} \citep[\textit{OSC};][]{gui2017}, which aims to collect all supernova data and metadata in an accessible format. To encourage the application of theoretical modeling to this dataset, we have developed a code to fit supernova data that interacts directly with the \texttt{Astrocats}\footnote{\url{https://github.com/astrocatalogs/astrocats}} platform used by the \textit{OSC} (and related catalogs such as the \textit{Open TDE Catalog}\footnote{\url{https://tde.space/}}). This code, the Modular Open Source Fitter for Transients, or \mosfit\footnote{\url{https://github.com/guillochon/MOSFiT}}, will be described in detail by \citet{gui2017}.

Here we present the first fits to data, using \mosfit to model the full published sample of SLSNe, including all available photometry to model the \emph{multicolour} light curve evolution (as opposed to only the bolometric properties as in most previous studies). We find that the UV-optical-infrared data of all SLSNe can be well fit with the magnetar model, while this approach removes systematic uncertainties associated with constructing bolometric light curves from observations with widely variable time and wavelength coverage. 
We significantly constrain the free parameters of the model, and show that although magnetar-powered light curves are indeed diverse, in fact only a relatively small region of parameter space actually corresponds to observed SLSNe.

Our SLSN sample is described in section \ref{sec:sample}. In section \ref{sec:code}, we detail our implementation of the magnetar model in \mosfit, and explain the caveats as well as the advantages compared to existing bolometric light curve fits. We present the fits and posteriors for all objects in section \ref{sec:fits}. We analyse the derived parameters in section \ref{sec:analysis}. We search for correlations between properties of SLSNe as well as their host galaxies, relate explosion properties to observables, and investigate the existence of any possible sub-groups within the sample. Comparing our derived parameters to published stellar evolution models, we discuss progenitor scenarios in section \ref{sec:progenitors}. We conclude in section \ref{sec:conc} and briefly discuss the path forward in understanding SLSNe.

\section{SLSN sample}
\label{sec:sample}

Previous sample studies of the light curves of SLSNe using magnetar models have been carried out by \citet[22 events]{nic2015b}, \citet[15 events]{pra2016} \citet[31 events]{yu2017} and \citet[19 events]{liu2017}. Here we expand the sample size to 38 SLSNe, all observed in at least two filters, by including more recently published events and high-redshift SLSNe. Our sample encompasses all SLSNe fulfilling 3 simple criteria: 
\begin{enumerate}
\item Spectroscopic classification as a Type I SLSN;
\item Published light curves; 
\item At least some data close to maximum light. For 32 of 38 SLSNe, there is sufficient data on both the rise and decline of the light curves, in a either a single band or two very similar bands (e.g.~$r$ and $R$) to clearly identify a maximum. For the remaining 5 events, the date of the peak is less certain, but we believe the light curves begin within $\lesssim 10$\,d of peak because the observed spectra show typical SLSN maximum light features (\ion{O}{2} lines on a blue continuum).\footnote{This criterion excludes LSQ14an, which is well observed at late times \citep{ins2017} but has no data around maximum light to constrain our model fits.}
\end{enumerate}
These objects are listed in Table \ref{tab:sample}.
As part of this study, we have ensured that all data used in our fits are publicly available\footnote{\href{https://sne.space/?event=PTF10hgi\%2CGaia16apd\%2CPTF12dam\%2CSN2015bn\%2CSN2007bi\%2CSN2011ke\%2CSSS120810\%2CSN2012il\%2CPTF11rks\%2CSN2010gx\%2CSN2011kf\%2CLSQ14mo\%2CLSQ12dlf\%2CPTF09cnd\%2CSN2013dg\%2CSN2005ap\%2CiPTF13ehe\%2CiPTF15esb\%2CiPTF16bad\%2CLSQ14bdq\%2CPTF09cwl\%2CSN2006oz\%2CiPTF13dcc\%2CPTF09atu\%2CPS1-14bj\%2CPS1-11ap\%2CDES14X3taz\%2CPS1-10bzj\%2CDES13S2cmm\%2CiPTF13ajg\%2CPS1-10awh\%2CPS1-10ky\%2CPS1-10ahf\%2CSCP-06F6\%2CPS1-10pm\%2CSNLS-07D2bv\%2CPS1-11bam\%2CSNLS-06D4eu&instruments=g\%2Cr\%2CB\%2CV\%2Cz&redshift=\%3C2&visible=name\%2Cdiscoverdate\%2Cmaxappmag\%2Cmaxabsmag\%2Chost\%2Cra\%2Cdec\%2Cinstruments\%2Credshift\%2Cclaimedtype\%2Cphotolink\%2Cspectralink\%2Cradiolink}{This hyperlink displays our sample on the \textit{OSC}}} from the \textit{OSC}.
This should facilitate future statistical studies of the SLSN population.

We include both `fast' and `slow' evolving SLSNe. It was originally thought that the SLSNe with the longest light curve timescales, which decline at close to one magnitude per 100 days, were powered by radioactive $^{56}$Ni / $^{56}$Co decay \citep{gal2009,gal2012}, and were the observational manifestation of the long-predicted pair-instability SNe from Population III stars \citep{bar1967,rak1967}. This picture has since changed significantly due to the modest ejecta masses inferred from slow-evolving SLSN rise times (\citealt{nic2013,nic2015b}, but see \citealt{koz2017}) and their blue spectral energy distributions \citep{des2012,nic2013,lun2016} that argue against extreme abundances of iron-group elements. Nebular-phase modelling \citep{jer2016,jer2016b} and late-time decline rates \citep{nic2016c,ins2017} also disfavour a pair-instability origin for these events.

Spectroscopic similarity between fast and slow SLSNe suggests that these events are likely variations on a theme rather than entirely separate classes \citep{nic2013}. However, the previous SLSN sample study by \citet{nic2015b} did show hints of a gap in the light curve timescale distribution between typical fast and slow events, but the distinction was not statistically significant. Moreover, \citet{kan2016b} recently showed that Gaia16apd (SN2016eay) was a SLSN with an intermediate timescale. Finally, host galaxy properties of fast and slow SLSNe are indistinguishable on average \citep{lel2015,schu2016}. In this study, we fit all SLSNe with the same magnetar model, and will test whether the inferred parameters of fast and slow events indicate separate populations or not, or reflect interesting correlations between the engine and explosion properties.

It has recently been suggested that many SLSNe show a brief initial peak in the light curve immediately after explosion and prior to the much slower rise to the true superluminous maximum \citep{lel2012,nic2015b,nic2016a,smi2016}. This has been variously interpreted as shock cooling in low-density extended material around the progenitor \citep{piro2015}, a second shock breakout driven by the hydrodynamic impact of a central engine \citep{kas2015}, or a sign of an off-axis jet driving a mildly relativistic wind \citep{mar2017}. All of these explanations are compatible with magnetar-powered explosions, but the simple light curve model we use here can only capture the primary peak.

We mark events with a clear such `bump' in Table \ref{tab:sample} \citep[though many others may have a bump that went undetected due to survey limitations;][]{nic2016a}. We exclude data during the bump phase from our fits. The exception to this is iPTF13dcc \citep{vre2016}, which showed either a plateau or an unusually bright bump around the light curve peak. The data around this phase are noisy and highly oversampled, so we rebin to a daily cadence before fitting. With only 6 of 38 events showing a high-significance bump detection this should have little bearing on the overall statistics of our modeling. This fact will be demonstrated in detail in section \ref{sec:analysis}. Modeling of the bump itself will become more feasible as more events are discovered with well-sampled, multicolour data during this phase.

\begin{table}
\caption{List of SLSNe in our sample}
\label{tab:sample}
\begin{center}
\begin{tabular}{ccc}
\hline
SLSN & Redshift & Reference\\
\hline
PTF10hgi	&	0.0987	&	\citet{ins2013}\\
Gaia16apd$\dagger$	&	0.102	&	\citet{yan2016},\\
			&			&	\citet{nic2017}\\
			&			&	\citet{kan2016b}\\
PTF12dam$\dagger$	&	0.1073	&	\citet{nic2013}\\
			&			&	\citet{chen2015}\\
			&			&	\citet{vre2016}\\
SN2015bn$\dagger$	&	0.1136	&	\citet{nic2016b,nic2016c},\\
SN2007bi$\dagger$	&	0.1279	&	\citet{gal2009}\\
SN2011ke	&	0.1428	&	\citet{ins2013}\\
SSS120810	&	0.156	&	\citet{nic2014}\\
SN2012il	&	0.175	&	\citet{ins2013}\\
PTF11rks	&	0.1924	&	\citet{ins2013}\\
iPTF15esb	&	0.224	&	\citet{yan2017}\\
SN2010gx	&	0.2297	&	\citet{pas2010},\\
			&			&	\citet{qui2011}\\
SN2011kf	&	0.245	&	\citet{ins2013}\\
iPTF16bad	&	0.2467	&	\citet{yan2017}\\
LSQ14mo		&	0.253	&	\citet{chen2016b}\\
LSQ12dlf	&	0.255	&	\citet{nic2014}\\
PTF09cnd	&	0.2584	&	\citet{qui2011}\\
SN2013dg	&	0.265	&	\citet{nic2014}\\
SN2005ap	&	0.2832	&	\citet{qui2007}\\
iPTF13ehe	&	0.3434	&	\citet{yan2015}\\
LSQ14bdq*$\dagger$	&	0.345	&	\citet{nic2015a}\\
PTF09cwl	&	0.3499	&	\citet{qui2011}\\
SN2006oz*	&	0.376	&	\citet{lel2012}\\
iPTF13dcc*$\dagger$	&	0.431	&	\citet{vre2016}\\
PTF09atu	&	0.5015	&	\citet{qui2011}\\
PS1-14bj$\dagger$	&	0.5215	&	\citet{lun2016}\\
PS1-11ap$\dagger$	&	0.524	&	\citet{mcc2014}\\
DES14X3taz*	&	0.608	&	\citet{smi2016}\\
PS1-10bzj	&	0.650	&	\citet{lun2013}\\
DES13S2cmm	&	0.663	&	\citet{pap2015}\\
iPTF13ajg	&	0.740	&	\citet{vre2014}\\
PS1-10awh	&	0.908	&	\citet{chom2011}\\
PS1-10ky	&	0.956	&	\citet{chom2011}\\
PS1-10ahf$\dagger$	&	1.1	&	\citet{mcc2015}\\
SCP-06F6	&	1.189	&	\citet{bar2009}\\
PS1-10pm*	&	1.206	&	\citet{mcc2015}\\
SNLS-07D2bv	&	1.50	&	\citet{how2013}\\
PS1-11bam	&	1.565	&	\citet{ber2012}\\
SNLS-06D4eu*	&	1.588	&	\citet{how2013}\\
\hline
\end{tabular}
\end{center}

*SLSN showed strong evidence for an early-time light curve `bump' \citep{nic2016a}\\
$\dagger$ At least one spectrum at $t\gtrsim 200$\,d after explosion
\end{table}

\section{Description of our model}
\label{sec:code}

\subsection{Motivation}

The goal of our study is to provide a set of magnetar-powered model light curve fits for the entire existing sample of Type I SLSNe, and to use these fits to extract fundamental properties of the explosions and engines. While for many events there exist published fits with similar models, we seek to improve on the literature in several important ways:
\begin{itemize}
\item Increased sample size. So far only \citet{pra2016}, \citet{yu2017} and \citet{liu2017} have applied model fits to a large ($>10$ events) sample of SLSNe.
\item Homogeneity. We fit all events with the same code and assumptions, which is essential for direct comparisons between SLSNe in the sample. This also avoids systematic differences in the bolometric corrections between events (see next section), which have been problematic for the previous sample studies.
\item Determining the solutions of highest likelihood using Bayesian analysis, in order to determine realistic error bars for derived quantities. Almost all analytic models of SLSNe have been fit using $\chi^2$ minimization, with the recent exception of \citet{liu2017}.
\item Using all available information, including priors on the velocity from spectra, and fitting the observed colour evolution. Other than \citet{pra2016}, all magnetar models to date have been fit only to bolometric light curves, though \citet{liu2017} also fit to temperature and velocity measurements.
\item Marginalizing over the `nuisance' parameters---opacity, neutron star mass, high-energy leakage coefficient, host galaxy extinction---that are generally set to fixed values in other studies. This allows a better determination of the true parameter space for the more fundamental parameters---ejecta mass, energy, magnetic field and spin period.
\item Physical constraints to force consistency between model predictions and observed properties not captured directly by the light curve, i.e.~photospheric velocity and ejecta optical depth.
\end{itemize}

\subsection{Multicolour vs bolometric light curves}

In the existing literature, it is standard to estimate bolometric luminosities of SNe from observations, and to fit model bolometric light curves. However, there are only a few SLSNe for which the full UV-optical-NIR light curves are available with adequate sampling to construct reliable bolometric light curves, and even in this case it is necessary to make some assumptions to account for missing flux \citep[see e.g.][]{lym2014,bro2016,lusk2017}. For objects with more typical datasets comprising only a few filters generally at observer-frame optical wavelengths, these corrections are significant. Moreover, the uncertainty in the bolometric light curve is difficult to quantify when derived from only a small number of filters.

The alternative approach, which we employ here, is to fit the multicolour light curves directly. This introduces some additional complexity to the problem---and additional free parameters---as one must then include a model for the spectral energy distribution (SED). However, taking this approach has two important advantages. First, it becomes possible to model events for which limited data make it extremely challenging to derive a bolometric light curve from observations. Second, retaining colour information can be very helpful in constraining models. A subtle but related point, rarely discussed in the literature, is that when one assumes an SED in order to convert filtered observations to a bolometric light curve for model fitting, in principle this SED should have to be consistent with the SED implied by the output model. This is sometimes done in an approximate sense (i.e.~by comparing the temperature evolution in the model to the blackbody used in deriving the luminosity), but consistency is not strictly enforced in general. The remainder of this section describes the implementation of the multicolour magnetar model in \mosfit.

\subsection{Overview of MOSFiT}\label{sec:mosfit}

\mosfit is a \texttt{Python}-based modular code to provide flexible fits to astrophysical light curves. The structure, usage and philosophy will be described in detail by \citet{gui2017}, but we summarise a few key points here.

The inputs to any implementation of \mosfit are model and parameter files, and a series of \texttt{Python} modules containing the physics. These files specify the list of modules and variables needed in the problem, which are then chained together in order to produce the model light curves. These are fit to observed data (which can be downloaded automatically from the \textit{OSC}) using a Markov Chain Monte Carlo (MCMC) fitter based around the popular \texttt{emcee} package \citep{for2013}, which uses an augmented version of the affine-invariant ensemble sampling method of \citet{goo2010}. The objective function we use includes a modeled white noise error term $\sigma$, with the log likelihood being
\begin{equation}
\ln {\cal L} = -\frac{n}{2}\ln\left[2\pi\sigma^2\right]
   - \frac{1}{2}\sum_{i=1}^{n} \left[\frac{\left(O_i-M_i\right)^2}{\sigma_i^2 + \sigma^2} - \ln\left[2\pi\sigma_i^2\right]\right],
\end{equation}
where $O_i$, $\sigma_i$, and $M_i$ are the $i$th of $n$ observed magnitudes, errors, and model magnitudes, respectively. This error model is more commonly known as ``maximum likelihood analysis'', and is a subset of Gaussian process models with no explicitly-modeled covariances (other Gaussian process error models do account for covariance). In the appendix, we show that including covariances within the more general Gaussian process framework does not have a significant effect on our results.

For each light curve fit, the code was run in parallel using 8 nodes for a duration of 48 hours on Harvard University's Odyssey computer cluster. This typically equated to $\sim 30,000-60,000$ iterations of the MCMC algorithm. The first 10,000 iterations are used to burn in the ensemble, during which minimization is employed periodically as the ensemble evolves to the global minimum; the remainder of the runtime is used to ensure convergence about that minimum. Convergence was measured by calculating the Gelman-Rubin statistic, or Potential Scale Reduction Factor \citep{gel1992}, which estimates the extent to which the full parameter space has been explored. \citet{bro1998} suggest that PSRF\,$<1.2$ should indicate reliable convergence; we terminate our simulations when we reach PSRF\,$<1.1$. Additionally, the simulations in this paper have been repeated numerous times to ensure reliable convergence to the same solution.

We describe below the \mosfit chain used in our fits. All of the \texttt{Python} modules described below are included in the distribution of \mosfit and can be imported from \texttt{mosfit.modules}; the full list of modules used in our fits is distributed along with the model light curves to the \textit{OSC}. While much of the physics below has been detailed in the existing literature, our intention here is to be as explicit as possible about the modules we choose. Because \mosfit is open and modular, one can then trivially implement an alternative scheme for any particular stage of the model to see how sensitive the fits and derived parameters are to each assumption.

\subsection{Modules: engine}
\label{sec:eng}

The basic form of the magnetar engine model has been described numerous times in the literature \citep{ost1971,kas2010,cha2012,ins2013}\footnote{The analytic model by \citet{ins2013} is available from \href{https://star.pst.qub.ac.uk/webdav/public/ajerkstrand/Codes/Genericarnett/}{https://star.pst.qub.ac.uk/webdav/public/ajerkstrand/Codes/Genericarnett/}}. We implement this in \mosfit through the module \texttt{engines.magnetar}. The typical assumption is that the magnetar energy input follows the functional form appropriate for magnetic dipole radiation:
\begin{equation}
F_{\rm mag}(t) = \frac{E_{\rm mag}}{t_{\rm mag}} \frac{1}{(1+t/t_{\rm mag})^2}.
\label{eq:fmag}
\end{equation}
In this case the rotational energy of the magnetar \citep[for plausible neutron star equations of state; see][]{lat2005} is
\begin{equation}
E_{\rm mag} = \frac{1}{2} I \omega^2 = 2.6 \times 10^{52} \left(\frac{M_{\rm NS}}{1.4 {\rm M}_\odot}\right)^{3/2}  \left(\frac{P}{1 {\rm ms}}\right)^{-2} {\rm erg},
\label{eq:emag}
\end{equation}
and it spins down on a timescale 
\begin{equation}
t_{\rm mag} \simeq \frac{P}{2\dot{P}} = 1.3 \times 10^5 \left(\frac{M_{\rm NS}}{1.4 {\rm M}_\odot}\right)^{3/2}  \left(\frac{P}{1 {\rm ms}}\right)^{2} \left(\frac{B_\perp}{10^{14}{\rm G}}\right)^{-2} {\rm s},
\label{eq:tmag}
\end{equation}
where $I$ is the neutron star moment of inertia and $\omega$ its angular frequency for spin period \P\ and mass \Mns, and \B\ is the component of the magnetic field perpendicular to the spin axis. The free parameters in this module are \P, \B, and \Mns. All priors on these and other parameters are given in Table \ref{tab:priors}.

\begin{table}
\caption{Free parameters and priors used in the model}
\label{tab:priors}
\begin{center}
\begin{tabular}{cccccc}
\hline
Parameter & Prior & Min & Max & Mean & Std. Dev.\\
\hline
$P$ / ms & Flat & 0.7 & 20 &  & \\
\B / $10^{14}$G & Log-flat & 0.01 & 10 &  & \\
\Mej / \M & Log-flat & 0.1 & 100 &  & \\
\vp / $10^{4}$\kms & Gaussian & 0.1 & 3.0 & 1.47* & 4.3\\
\k / g\,cm$^{-2}$ & Flat & 0.05 & 0.2 &  & \\
\kg / g\,cm$^{-2}$ & Log-flat & 0.01 & 100 &  & \\
\Mns / \M & Flat & 1.4 & 2.2 &  & \\
\Tf / $10^{3}$K & Gaussian & 3.0 & 10.0 & 6.0 & 1.0\\
\Av / mag & Flat & 0 & 0.5 &  & \\
Explosion time / d & Flat & $-$100 & 0 &  & \\
Variance & Log-flat & $10^{-3}$ & 100 &  & \\
\hline
\end{tabular}
\end{center}
*Mean spectroscopic absorption velocity at 15 days after maximum light from \citet{Liu&Modjaz16}. If a measurement at this time existed in their sample for an individual SLSN, we used that value instead.
\end{table}

The timescale above is derived by setting $dE_{\rm mag}/dt$ equal to the Larmor formula for a magnetic dipole in vacuum. Our definitions here match the original definitions used by \citet{ost1971}. However, the literature contains a number of definitions for the spin-down time, which lead to systematic offsets between the derived magnetic fields (spin period and neutron star mass are independently constrained by $E_{\rm mag}$). For a fixed angle $\theta$, the dipole field is $B = B_\perp / \sin \theta$. Thus our derived field \B\ gives a strict lower limit on the dipole moment of the magnetar. 

In contrast, the model of \citet{kas2010} assumes an angle of 45$^\circ$, while \citet{met2015} use a force-free (rather than vacuum) definition of the spin-down time. Equating our equation \ref{eq:tmag} with equation 2 of \citeauthor{kas2010} yields a conversion $B_{\rm KB10}/B_\perp = 2.5$, while comparison to equation 3 of \citeauthor{met2015} gives $B_{\rm M15}/B_\perp = 0.48$. Therefore systematic differences in magnetic field caused by discrepant definitions of $t_{\rm mag}$ are not more than a factor $\sim2$. Here we carry out a uniform investigation using a single definition.

\subsection{Modules: diffusion}

The spin-down luminosity output from the magnetar engine model is then fed into a module to simulate diffusion of this energy through the ejecta (\texttt{transforms.diffusion}). This takes the form of the common analytic solution derived by \citet{arn1982}, giving an output luminosity
\begin{equation}
L_{\rm out}(t) = e^{-\left( t/t_{\rm diff} \right)^2} (1-e^{-At^{-2}}) \int_0^t 2 \, F_{\rm in}(t') \frac{t'}{t_{\rm diff}} e^{\left( t'/t_{\rm diff} \right)^2} \frac{dt'}{t_{\rm diff}},
\end{equation}
where in this case $F_{\rm in}= F_{\rm mag}$ (equation \ref{eq:fmag}). The diffusion time is given by
\begin{equation}
t_{\rm diff} =\left(\frac{2 \kappa M_{\rm ej}}{\beta c v_{\rm ej}}\right)^{1/2}
\label{eq:tdiff}
\end{equation}
and the leakage parameter \citep{wang2015} is
\begin{equation}
A = \frac{3 \kappa_\gamma M_{\rm ej}}{4 \pi v_{\rm ej}^2},
\end{equation}
with ejecta mass \Mej, velocity \vej, optical opacity \k\ and opacity to high-energy photons \kg (all free parameters in the model). The leakage term, $1-e^{-At^{-2}}$, controls the fraction of the magnetar input energy that is thermalised in the ejecta (and thus observable as UV-optical-NIR luminosity)---this declines with time as the density in the ejecta decreases. 

The main weakness in this approach is the assumption of a grey and constant opacity; however, an accurate treatment of time-variable opacity requires a computation of the ionization state of the ejecta with detailed radiation transport that would vastly increase the complexity of the model. We assume the opacity is dominated by electron scattering, as is common in analytic modelling of SNe. The value of this opacity is $\kappa_{\rm es} = 0.2(\bar{Z}/\bar{A})(x_e/\bar{Z})$\,\cmsqperg, where $\bar{Z}$ and $\bar{A}$ are the mean nuclear charge and mass, and $x_e$ is the ionization fraction of the ejecta. For hydrogen-free material, $\bar{Z}/\bar{A} \simeq0.5$, setting $\kappa_{\rm es} = 0.2$\,\cmsqperg\ as the upper limit on electron-scattering opacity for fully ionized ejecta \citep[see also][for a detailed discussion]{ins2013}.

This method is virtually ubiquitous in analytic modeling of SLSNe, but there is a decision to be made in terms of the free parameters to use: equation \ref{eq:tdiff} can be formulated using any two of \Mej, \vej\ and \Ek. The advantage to using \vej\ rather than \Ek\ is that the velocities have a useful prior from absorption line widths in the spectra. However, it is important to distinguish between the characteristic velocity of the ejecta \citep{arn1982} and the velocity at the photosphere, \vp, only the latter of which is measurable. Thes are not necessarily the same, but for the purposes of this work, we assume that \vej\,$\simeq$\,\vp, as is standard in the literature.

We use the results of \citet{Liu&Modjaz16}, who used a template-fitting method to provide reliable velocity estimates and errors for a large literature sample of SLSNe. They measured the \ion{Fe}{2}\,$\lambda$5169 line width, which is thought to be a reasonable tracer of the photosphere. In SLSNe this line is contaminated by \ion{Fe}{3} at early times, as also noted by \citeauthor{Liu&Modjaz16}. To avoid contamination, we take the values at 15 days after maximum light. When measurements are provided for individual SLSNe, we use those values; otherwise we use their average value of 14700$\pm$4300\,\kms (see Table \ref{tab:priors}).

The engine input and diffusion are sufficient to calculate the bolometric luminosity of the model. The following sections describe the conversion between luminosity and broadband magnitudes for comparison to multi-colour data, which is one of the key differences between this study and previous analytic models of SLSNe.

\subsection{Modules: photosphere}

The first step in deriving the SED is to determine the temperature and radius of the photosphere. We use the module \texttt{photospheres.temperature\_floor}. This model simply assumes that the photospheric radius expands at a constant velocity, \vp, with a temperature derived from the model luminosity and the Stefan-Boltzmann law, until the ejecta cool to some critical (constant) temperature at which the photosphere then recedes into the ejecta (for example due to recombination, or fragmentation of a dense shell inflated by the central magnetar). The temperature and radius are therefore given by:
\begin{eqnarray}
T_{\rm phot}(t) = 
\left\{
\begin{array}{lr}
\left(\frac{L(t)}{4 \pi \sigma v_{\rm phot}^2 t^2}\right)^{\frac{1}{4}},\ &
\quad \left(\frac{L(t)}{4 \pi \sigma v_{\rm phot}^2 t^2}\right)^{\frac{1}{4}} > T_{\rm f} \\
T_{\rm f},&
\left(\frac{L(t)}{4 \pi \sigma v_{\rm phot}^2 t^2}\right)^{\frac{1}{4}} \le T_{\rm f} \\
\end{array}
\right.
\end{eqnarray}
\begin{eqnarray}
R_{\rm phot}(t) = 
\left\{
\begin{array}{lr}
v_{\rm phot} t, &
\quad \left(\frac{L(t)}{4 \pi \sigma v_{\rm phot}^2 t^2}\right)^{\frac{1}{4}} > T_{\rm f} \\
\left(\frac{L(t)}{4 \pi \sigma T_{\rm f}^4}\right)^{\frac{1}{2}},&
\left(\frac{L(t)}{4 \pi \sigma v_{\rm phot}^2 t^2}\right)^{\frac{1}{4}} \le T_{\rm f} \\
\end{array}
\right.
\end{eqnarray}

This formulation is motivated by observations of SLSNe, which universally show a temperature that declines from maximum light before flattening at $\sim 4000-7000$ after $\sim 50$ days from maximum light \citep[e.g.][]{ins2013}. We recently showed that this prescription gives a reasonable match to the temperature and radius evolution in SLSNe \citep{nic2017}. The photosphere module introduces an additional free parameter: the final plateau temperature, \Tf. We note that while this parameter may appear slightly ad-hoc, it has little bearing on the posteriors of the important physical parameters in the fits, which are primarily determined by the light curve shape closer to maximum light rather than the late-time constant-colour phase. This parameter simply allows us to extend our fits to later times, where other photospheric models based on determining the optical depth break down \citep{ins2013}. This is useful as the late-time decline rate is important in determining \B. Our prior on \Tf\ is a Gaussian centered at 6000\,K. This choice improves the speed of convergence to the final fit, but the solution itself is largely insensitive to the choice of Gaussian or flat prior. A more detailed treatment of the nebular phase is beyond the scope of this study. However, we note that the typical temperature of $\sim 6000$\,K could be motivated physically as the approximate recombination temperature of \ion{O}{2}.

\subsection{Modules: spectral energy distribution}

We assume in our model that the SED is a modified blackbody. This is a reasonable choice for SLSNe. It has been shown that SLSN spectra are relatively smooth compared to other SN types (i.e.~the equivalent widths of absorption/emission features are generally much lower for SLSNe; e.g.~\citealt{yan2016}), and that blackbody curves can well reproduce the optical and NIR broadband colours of SLSNe throughout their evolution \citep[e.g.][]{nic2016b,nic2017}. However, the UV part of the observed SED is subject to significant absorption \citep{chom2011}. 

The blackbody SED is calculated according to the Planck formula using the temperature and radius of the photosphere, via the module \texttt{seds.blackbody\_cutoff}. This module applies linear flux suppression below a specified `cutoff' wavelength, in order to match the UV deficit described above. The chosen functional form, $F_{\lambda < \lambda,{\rm cut}} = F_\lambda (\lambda/\lambda_{\rm cut})$, gives 100\% transmission at the cutoff wavelength and 0\% transmission at 0\,\AA. The absorbed blackbody SED is renormalised such that the integrated flux is equal to the luminosity of the model. For SLSNe, the cutoff is observed to be approximately 3000\,\AA\ \citep{chom2011,nic2017}. This is qualitatively similar to the empirical cutoff employed by \citet{pra2016}.

\begin{figure}
\centering
\includegraphics[width=\columnwidth]{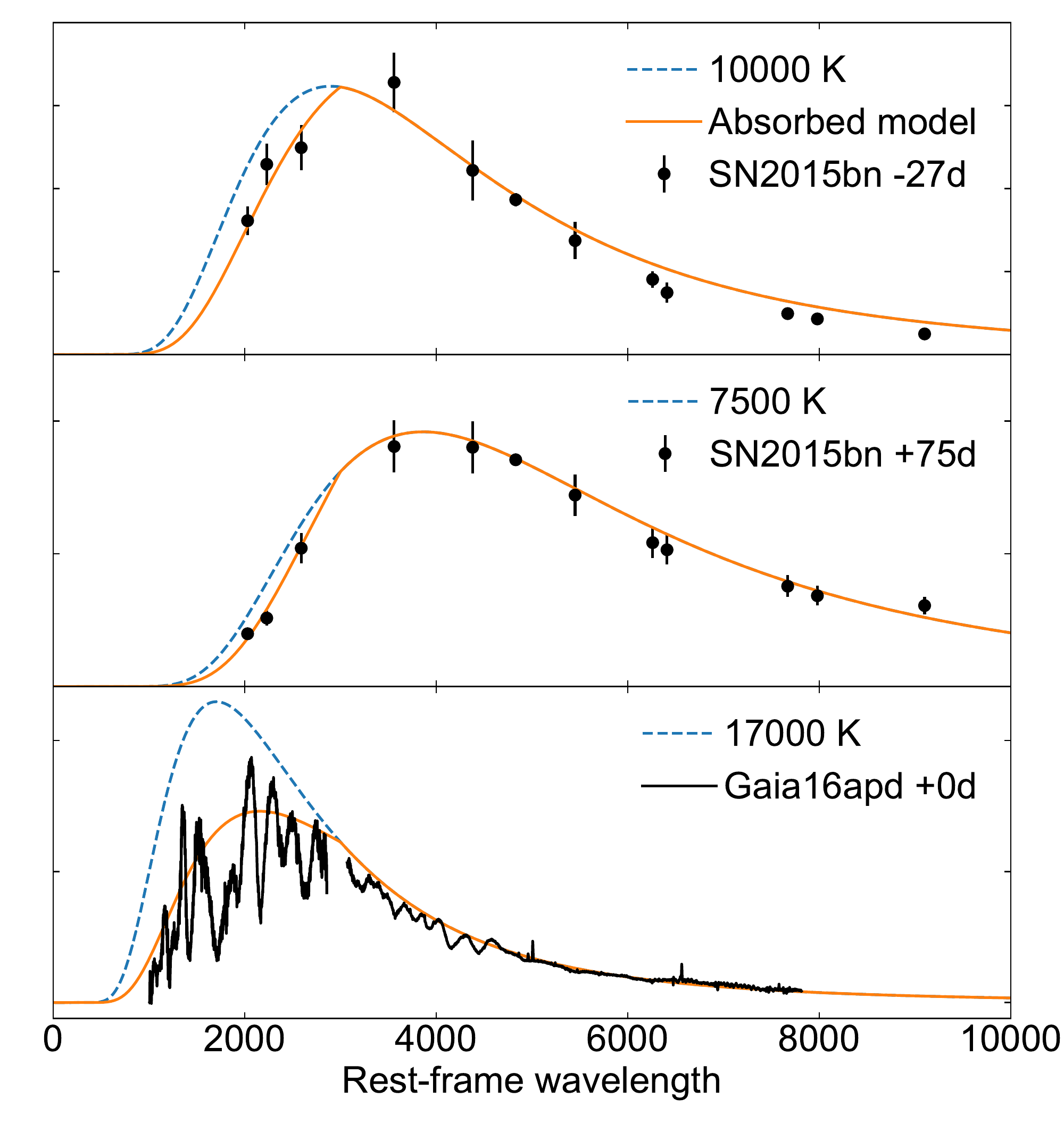}
\caption{The SED we use in our model. The underlying distribution is a blackbody, but between 0-3000\,\AA\ we linearly vary the transmission from 0-100\%. Dashed lines show blackbody SEDs for nominal temperatures, whereas solid curves show the SED after applying the UV absorption. Overlaid are photometry and spectra of well-observed SLSNe. Our simple model reproduces the SED at all epochs with \textit{Swift} photometry for SN\,2015bn \citep{nic2016b}, and matches the broad structure of the Gaia16apd UV spectrum \citep{yan2016}. Note that for temperatures below $\sim 8000$\,K the correction to a blackbody is quite minor.}
\label{fig:sed}
\end{figure}

\begin{figure}
\centering
\includegraphics[width=\columnwidth]{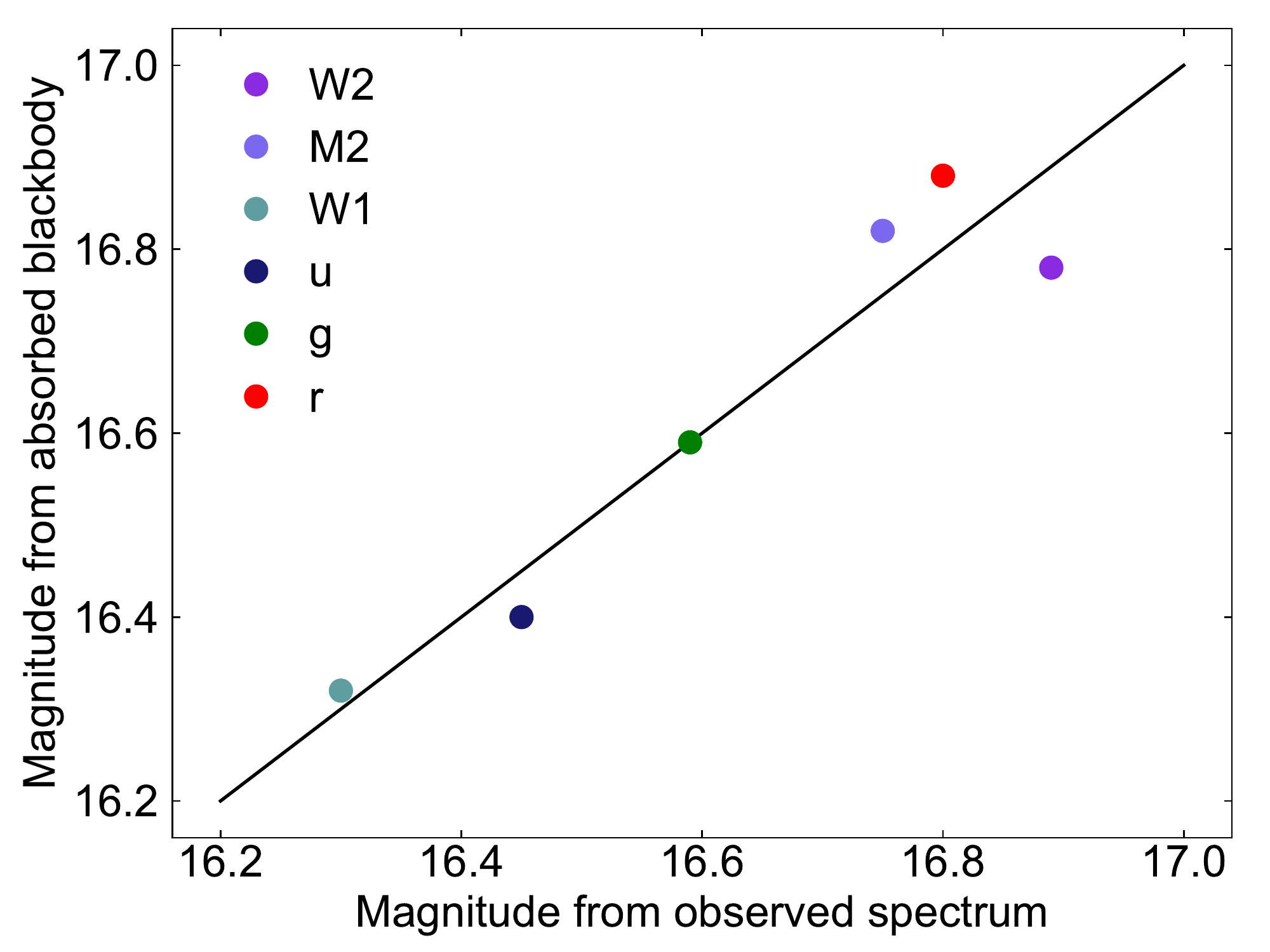}
\caption{Synthetic magnitudes from the observed spectrum of Gaia16apd and the absorbed blackbody model (Figure \ref{fig:sed}. The solid line shows where the two are equal. Averaged over each broad filter, the differences are all less than 0.1\,mag.}
\label{fig:mags}
\end{figure}

We further tested our prescription against SN\,2015bn, which has UV data from \textit{Swift} spanning 150 days, and Gaia16apd, which has a spectrum reaching rest-frame 1000\,\AA, and found that this could well reproduce the observed UV deficit at all times at the relevant wavelengths $1000-3000$\,\AA. The good agreement between our simple model and the data is shown in Figure \ref{fig:sed}. We show this even more explicitly in Figure \ref{fig:mags}. Here we calculate synthetic magnitudes using the tool \texttt{SMS} \citep{ins2016} on both the observed spectrum of Gaia16apd and our absorbed blackbody approximation. Across the full UV-optical range, the differences are less than 0.1 magnitudes in all bands. This confirms that our computationally-efficient approximation can accurately reproduce the broadband magnitudes of SLSNe.

Most of the SLSNe in the sample have the bulk of their data in the $r$ band and redwards. In these cases, the unabsorbed part of the blackbody can be reasonably well constrained. However, for SLSNe at $z\gtrsim1$ (8/38 events), observer-frame $r$ includes rest-frame flux below 3000\,\AA, such that the model magnitudes could be very sensitive to the UV absorption model. We tested our high-redshift SLSNe extensively and found no systematic differences in the fit quality or derived parameters compared to the lower-redshift events. Finally, we note that including time-series spectral templates, based on observed SNe, is a long-term goal in \texttt{MOSFiT}, but this is beyond the scope of our study here.

\subsection{Modules: physical constraints}
\label{sec:const}

Not all combinations of parameters yield realistic solutions. We add two additional constraints through the module \texttt{constraints.slsn\_constraints}. The constraints class in \mosfit is used to implement complex priors that depend on non-linear combinations of multiple input parameters.

The first requirement is that energy is conserved. Assuming the analytic density profile from equation 12 of \citet{mar2017}, and that the characteristic velocity can be well approximated by \vp, the kinetic energy is given by \Ek\,$=\frac{1}{2}$\,\Mej\,\vp$^2$ (a similar result is obtained if we instead assume that the ejecta mass is entirely concentrated in a thin shell). For a homologous density profile, the energy would instead be given by \Ek\,$=\frac{3}{10}$\,\Mej\,\vp$^2$. 

The total energy available is $E_{\rm mag} - E_{\rm rad} + E_\nu$, where $E_{\rm mag}$ is from equation \ref{eq:emag}, $E_{\rm rad} = \int L(t) dt$ and $E_\nu \approx 10^{51}$\,erg is the energy of a canonical core-collapse explosion, thought to be provided by neutrinos from the proto-NS \citep[for a recent review see][]{jan2017}. Thus models are strongly disfavoured if they violate $E_{\rm K} > E_{\rm mag} - E_{\rm rad} + E_\nu$. \citet{ins2013} used a fixed $E_{\rm K} = 10^{51}\,{\rm erg} + \frac{1}{2}(E_{\rm mag} - E_{\rm rad})$, based on carefully calibrating their light curves to hydrodynamic simulations by \citet{kas2010}. By calculating \Ek\ from \Mej\ and \vej, we introduce an additional free parameter, but this allows us to use the prior information about \vej\ to ensure consistency between light curves and spectra.

The second requirement is that the ejecta do not become optically thin too quickly, as this could contradict spectroscopic observations. For a constant opacity, the optical depth in the ejecta reaches $\tau = 1$ at a time
\begin{equation}
t_{\rm neb} = (3 \kappa M_{\rm ej} / 4 \pi v_{\rm ej}^2)^{1/2}.
\label{eq:tneb}
\end{equation}
As no SLSN has exhibited a spectrum with a strong nebular component earlier than 100 days after explosion, we introduce a prior to modify the likelihood score of any fit with $t_{\rm neb} < 100$\,d. The scaling is chosen conservatively to incur only a mild penalty for a violation of a few days, but an increasing penalty for large violations. Most SLSNe do not have spectroscopy beyond $\sim100$\,d, but for some nearby and/or slowly-evolving SLSNe, spectroscopy at this time shows nebular features gradually developing between $200-400$\,d after explosion \citep{gal2009,nic2013,nic2016b,nic2017,jer2016b,lun2016,kan2016b,ins2017} --- if an event has such a late-time spectrum, we instead set the constraint to 200\,d (these are marked in Table \ref{tab:sample}). In practice, the nebular-time constraint only affects the best fit for a few objects.

\subsection{Modules: extinction}

Before comparison to data, both host galaxy and Milky Way extinction are applied to the model light curves. Host extinction is applied in the rest frame, while Milky Way extinction is applied in the observer frame. For the Milky Way, \Av\ is taken from the dust maps of \citet{schlaf2011}, via the \textit{OSC}, assuming the usual total-to-selective extinction ratio $R_V=3.1$. \mosfit uses the reddening curve from \citet{odo1994} implemented in the \texttt{extinction} package\footnote{http://extinction.readthedocs.io}, which is a slightly modified version of that from \citet{car1989}.

In general the host galaxy extinction is not known (but can be estimated from galaxy spectra using e.g.~the Balmer decrement). SLSN host galaxy studies indicate an extinction \Av\,$<0.5$\,mag \citep{lun2014}; we therefore leave \Av\ free to vary, with a flat prior between 0 and 0.5 magnitudes. The ratio $R_V$ is also uncertain. Given the dwarf nature of SLSN hosts, an LMC- or SMC-like extinction curve may be more applicable. \citet{pei1992} found $R_V=3.16$ for the LMC and $R_V=2.93$ for the SMC. Testing with $R_V$ as a free parameter, we found that it was usually poorly constrained by the SLSN light curves, and had little effect on the other parameters. We therefore fix its value at 3.1 for simplicity.

\subsection{Modules: other modules}

The remaining modules in the chain are much more general and account for conversion of the SED to broadband photometry (\texttt{observables.photometry}) and calculating the likelihood score for each iteration of the fitting process (\texttt{objectives.likelihood}, see \ref{sec:mosfit}). These generic modules will be described in detail by \citet{gui2017}, but we summarize a few relevant points here.

The conversion from the SED to magnitudes is carried out by redshifting the SED to the observer frame, and diluting the flux per unit wavelength by a corresponding factor $1+z$, then convolving with a filter function for each observed band. The zeropoint used to normalize the flux depends on the photometric system used (AB or Vega). Data are published in a variety of systems, but in general $ugriz$ data are in the AB system and $UBVRIJHK$ are in the Vega system. \textit{Swift} UVOT data often appear in both systems (though Vega is the default). For any non-trivial combination of filter and magnitude system, we have tagged the \textit{OSC} data with the system used, so that \mosfit knows the correct zeropoint. The distance modulus is determined using the \textit{OSC} redshift and a standard Planck cosmology \citep{planck2016}.

\subsection{Summary}

In total, our model has 11 free parameters, of which 9 are physical. These are summarized in Table \ref{tab:priors}. The most important physical parameters that we wish to constrain are \Mej, \P\ and \B. A somewhat unique parameter is \vp, which has a fairly tight prior from spectroscopy \citep{Liu&Modjaz16}. However, we assume a constant average velocity, whereas in reality the photospheric velocity decreases with time --- we therefore expect our typical posteriors for \vp\ to be somewhat lower than those of \citeauthor{Liu&Modjaz16}. To derive realistic posteriors for these parameters, we marginalize over a number of poorly-constrained parameters: \k, \kg, \Mns.  Two additional parameters are necessary to match the colour evolution (but not the luminosity): \Tf\ and \Av. The late-time temperature, in particular, has little effect on the important parameters. We also fit for the explosion epoch (given here relative to the first datapoint). Finally, an additional white-noise variance term $\sigma$ is employed in calculating the model's likelihood score.

As shown in Table \ref{tab:priors}, different functional forms are used for the priors on different parameters. If a parameter is already reasonably well constrained from other observations (\vp, \Tf) we use a Gaussian prior. If a parameter can span a range covering several orders of magnitude, we use a prior that is flat in logarithmic space; otherwise we employ a prior that is flat in linear space. For some of the most important parameters (\Mej, \B) we tested both logarithmic and linear priors. The results were largely consistent between the two cases. In the case of linear priors, the posteriors looked closer to log-normal than to normal, indicating that the logarithmic prior is likely more appropriate.

\begin{figure*}
\centering
\includegraphics[width=5.6cm]{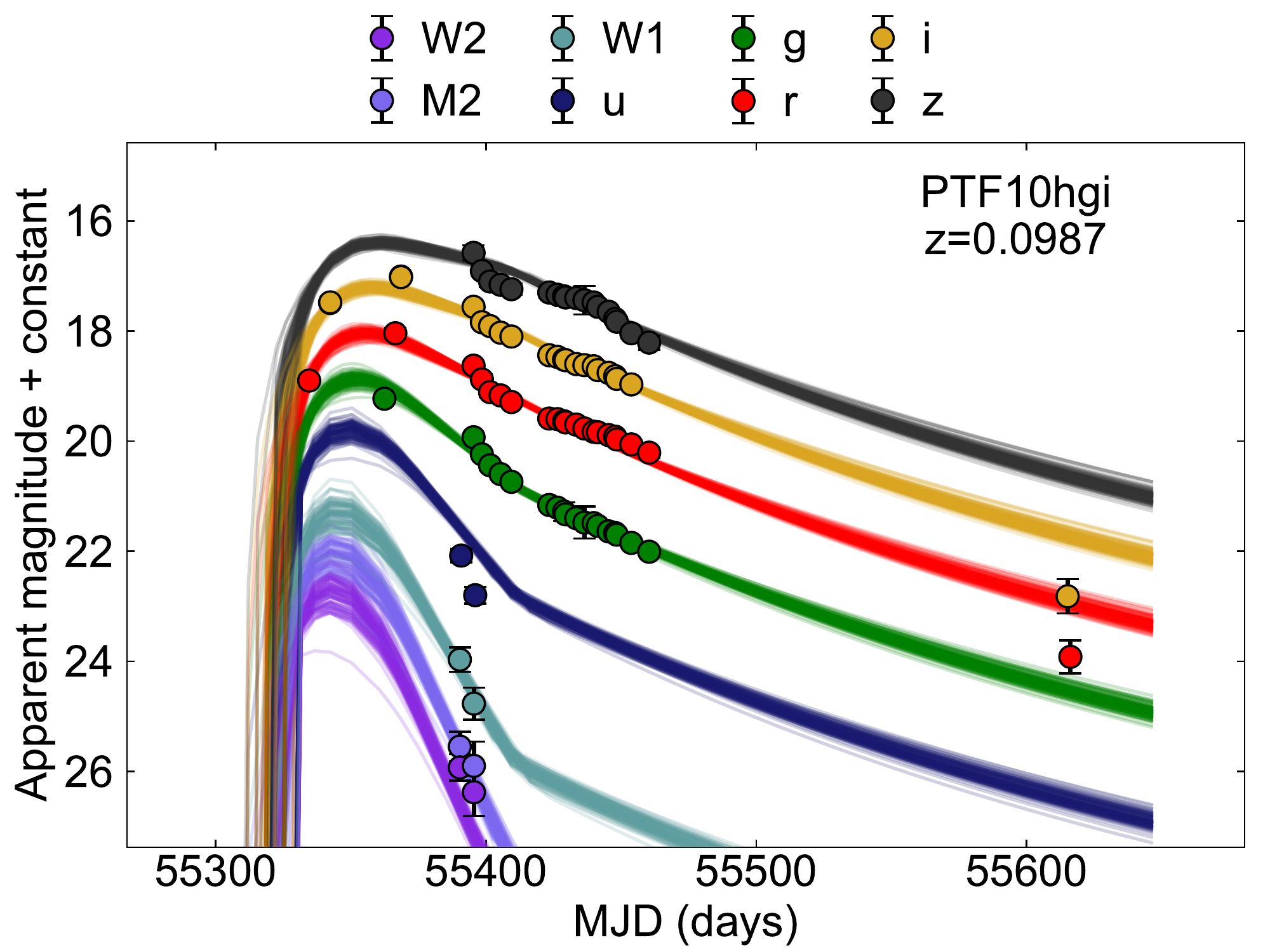}
\includegraphics[width=5.6cm]{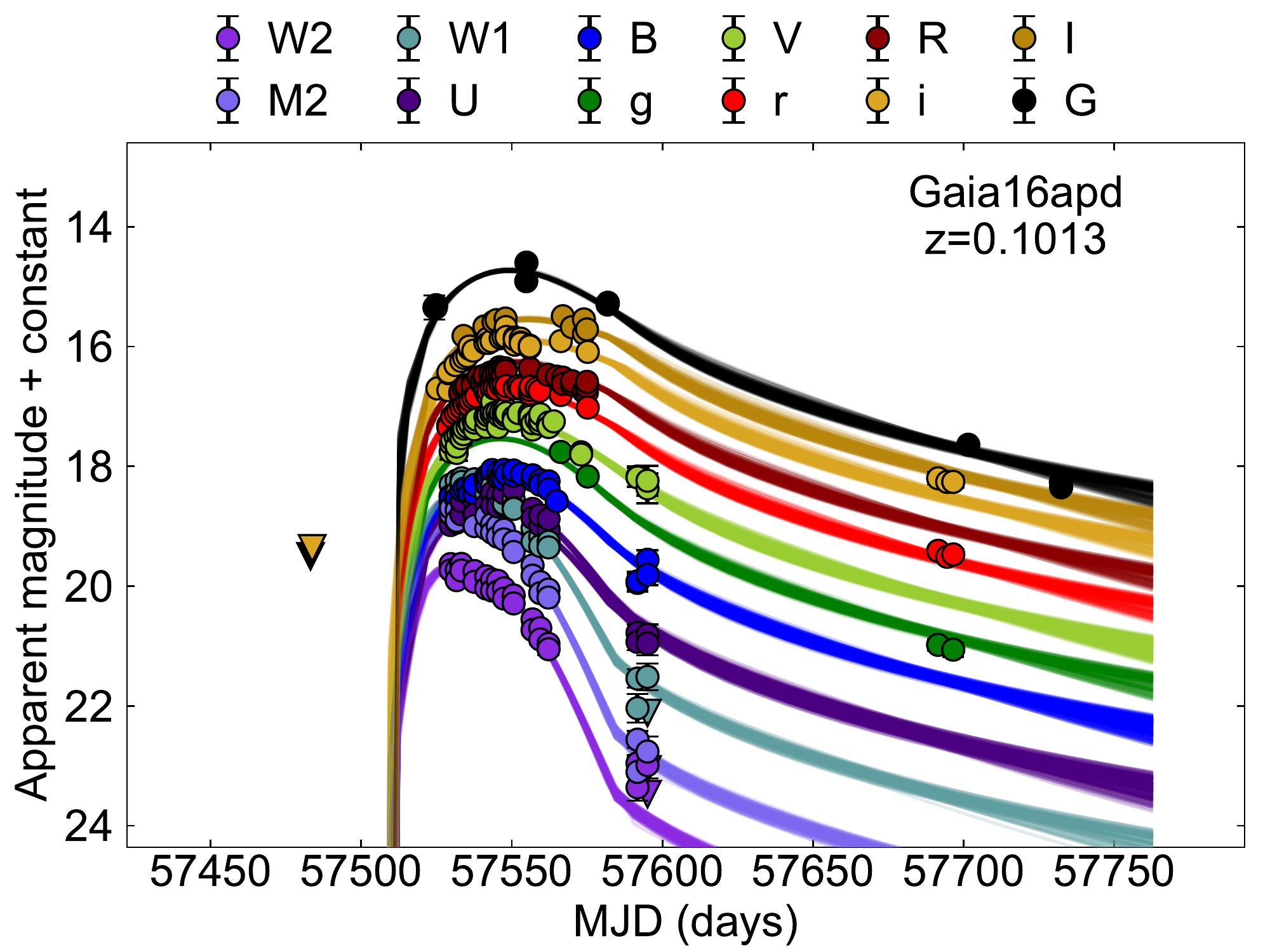}
\includegraphics[width=5.6cm]{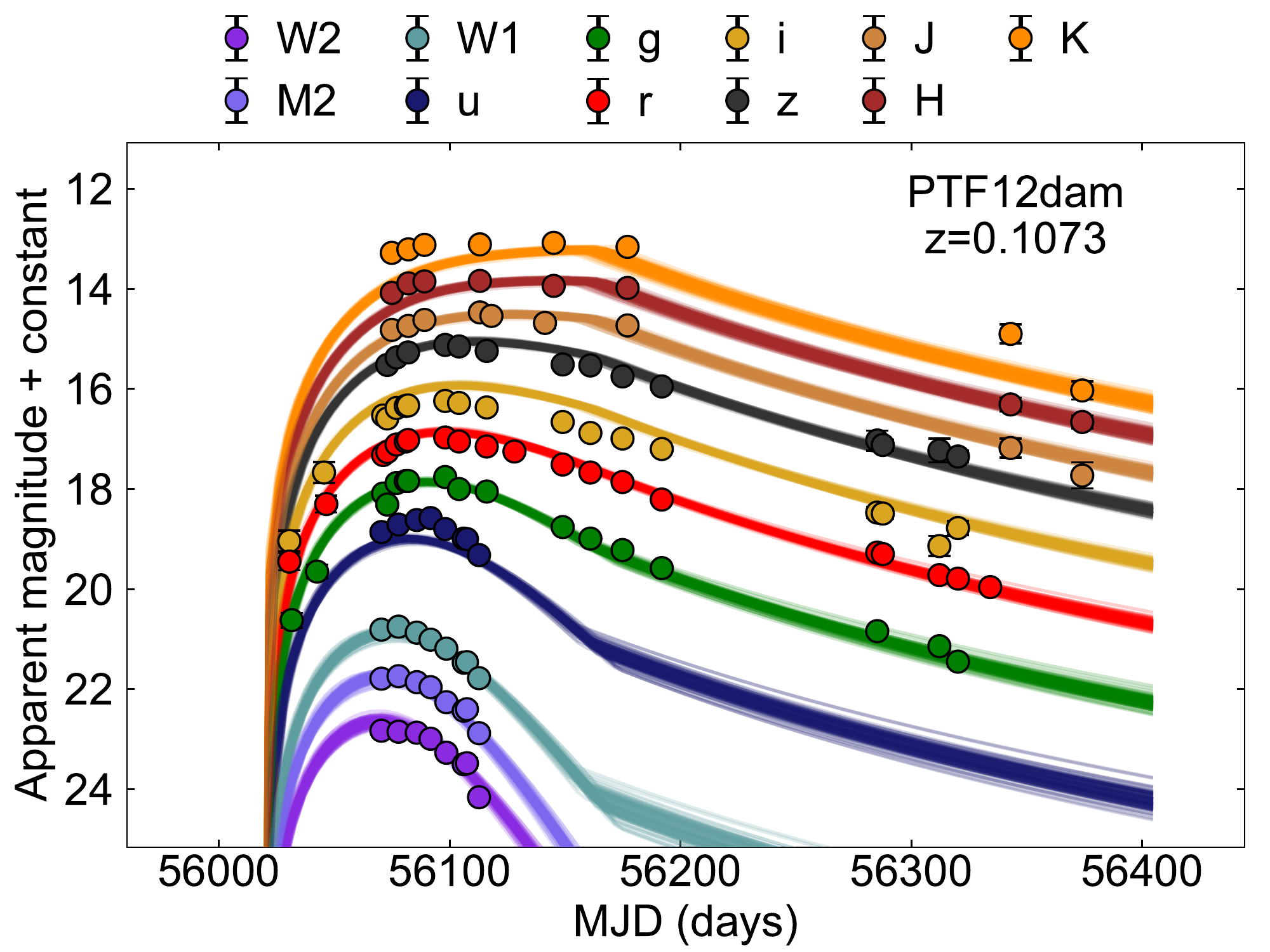}
\includegraphics[width=5.6cm]{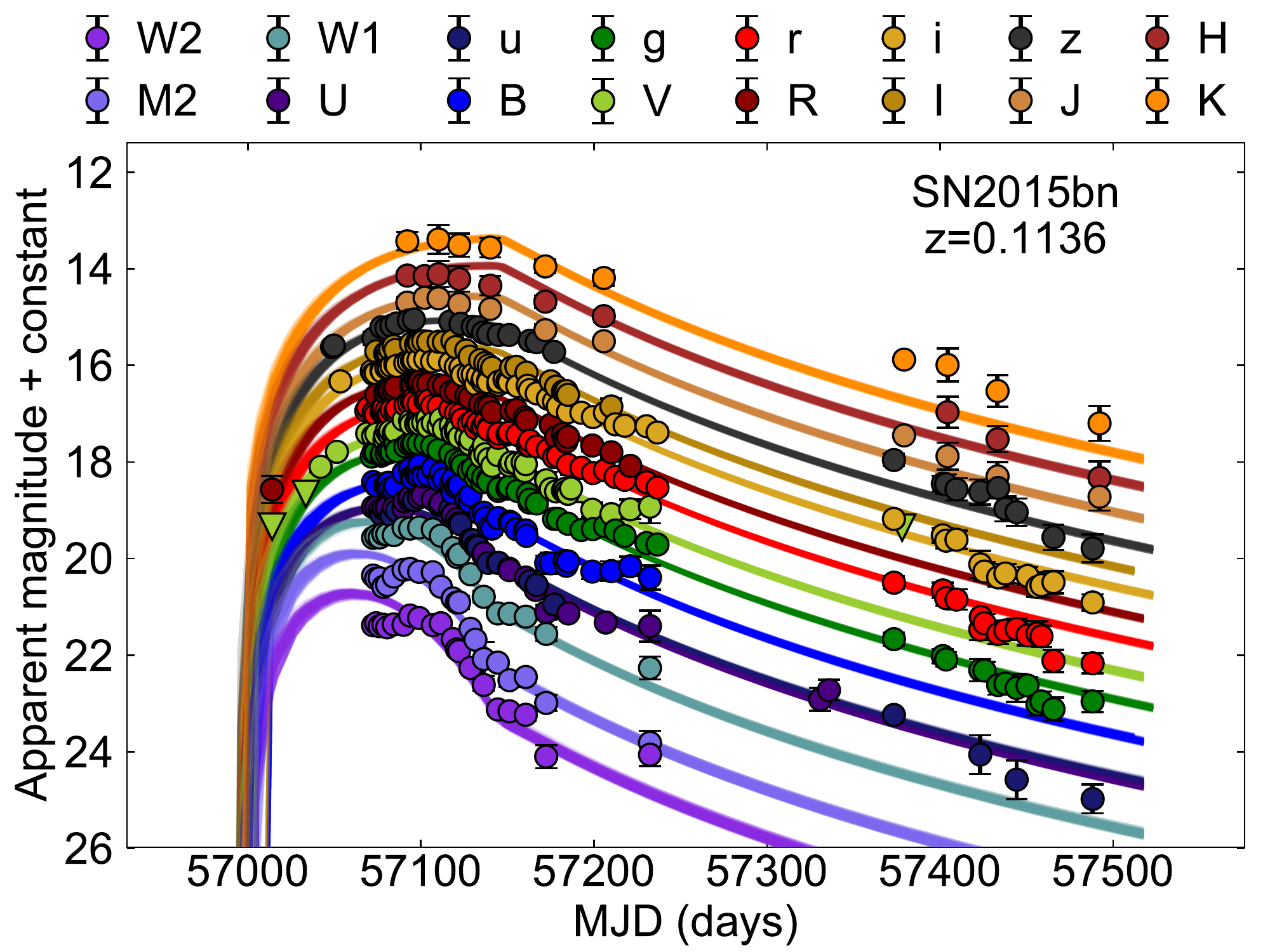}
\includegraphics[width=5.6cm]{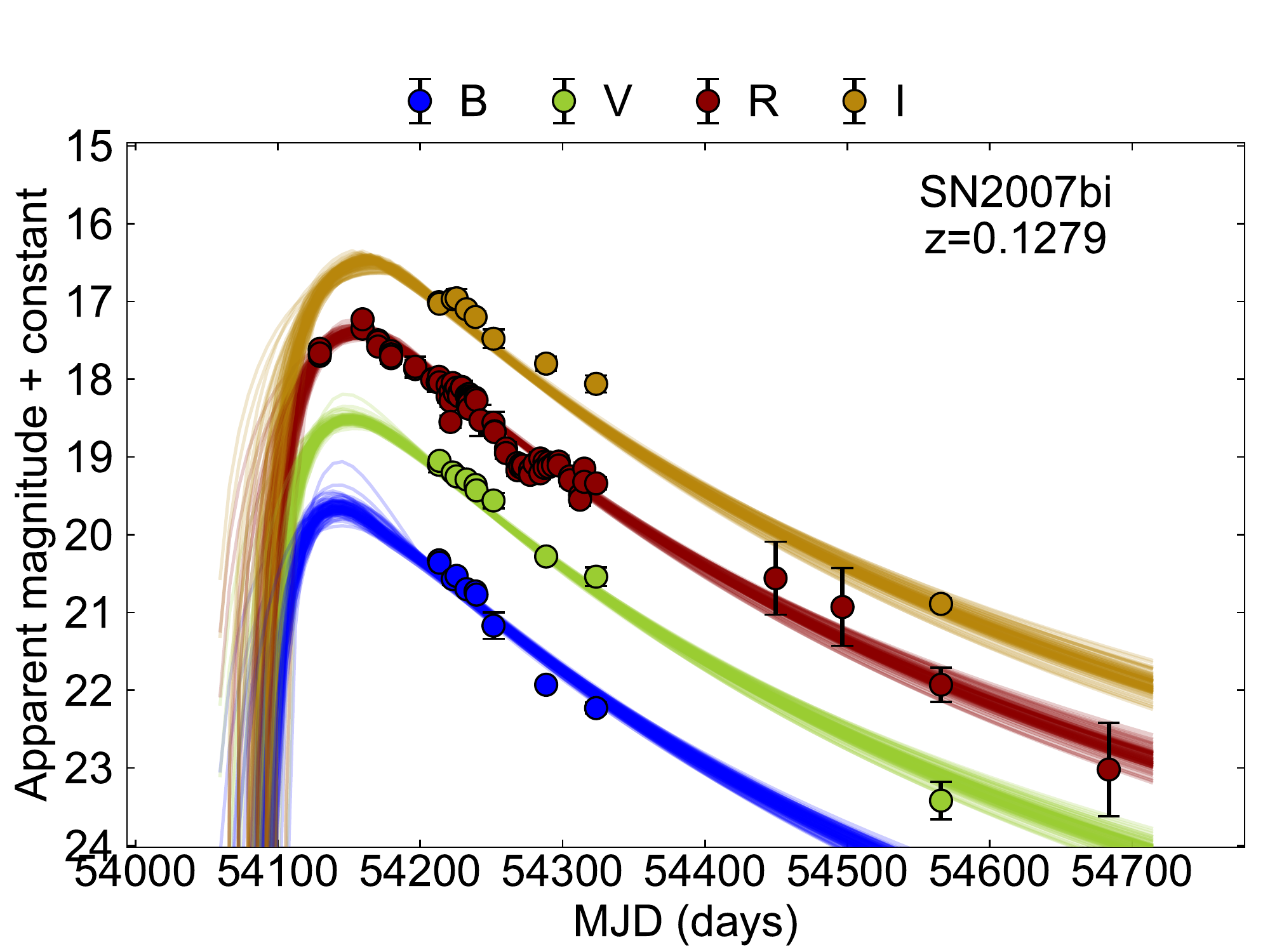}
\includegraphics[width=5.6cm]{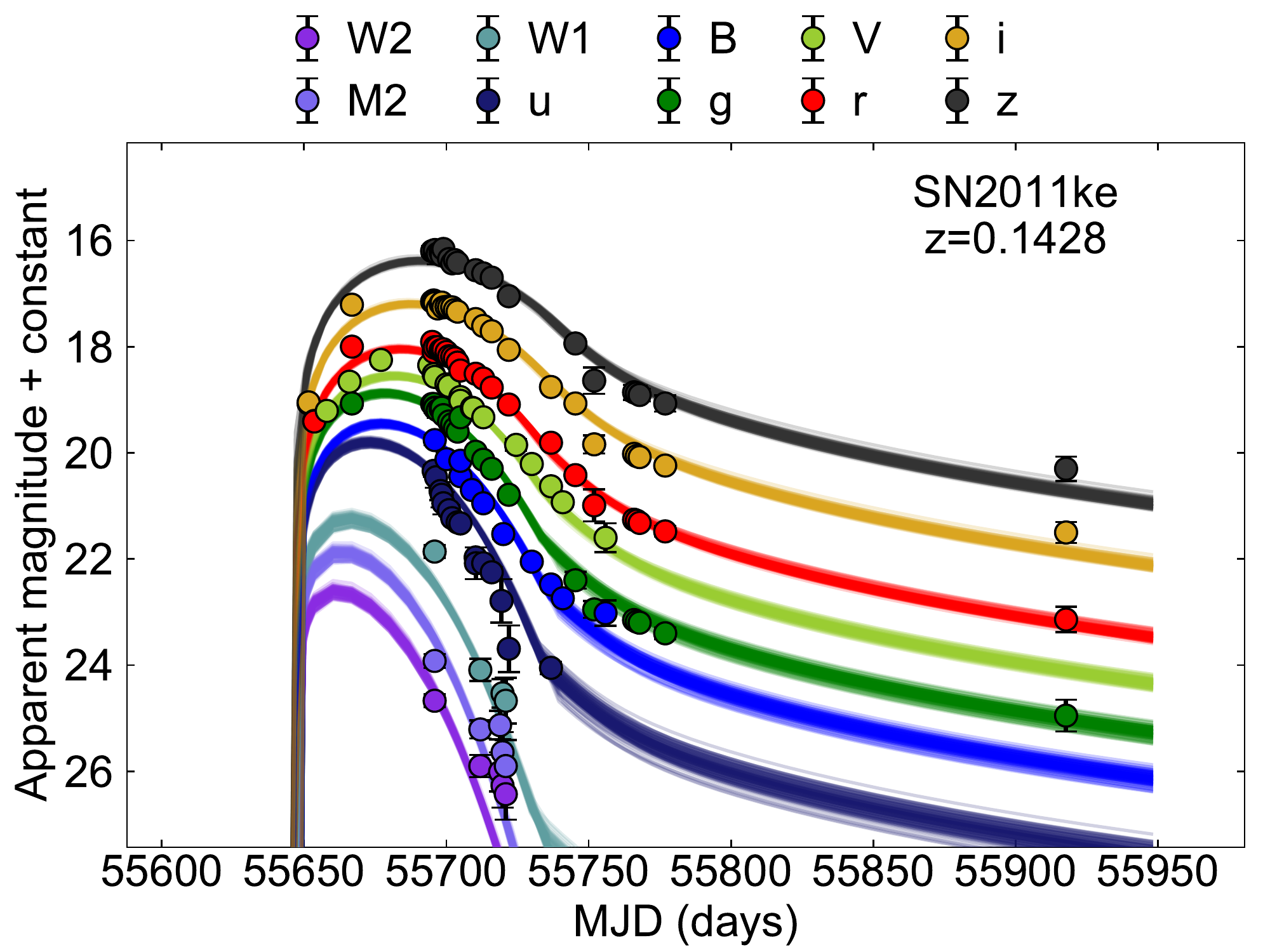}
\includegraphics[width=5.6cm]{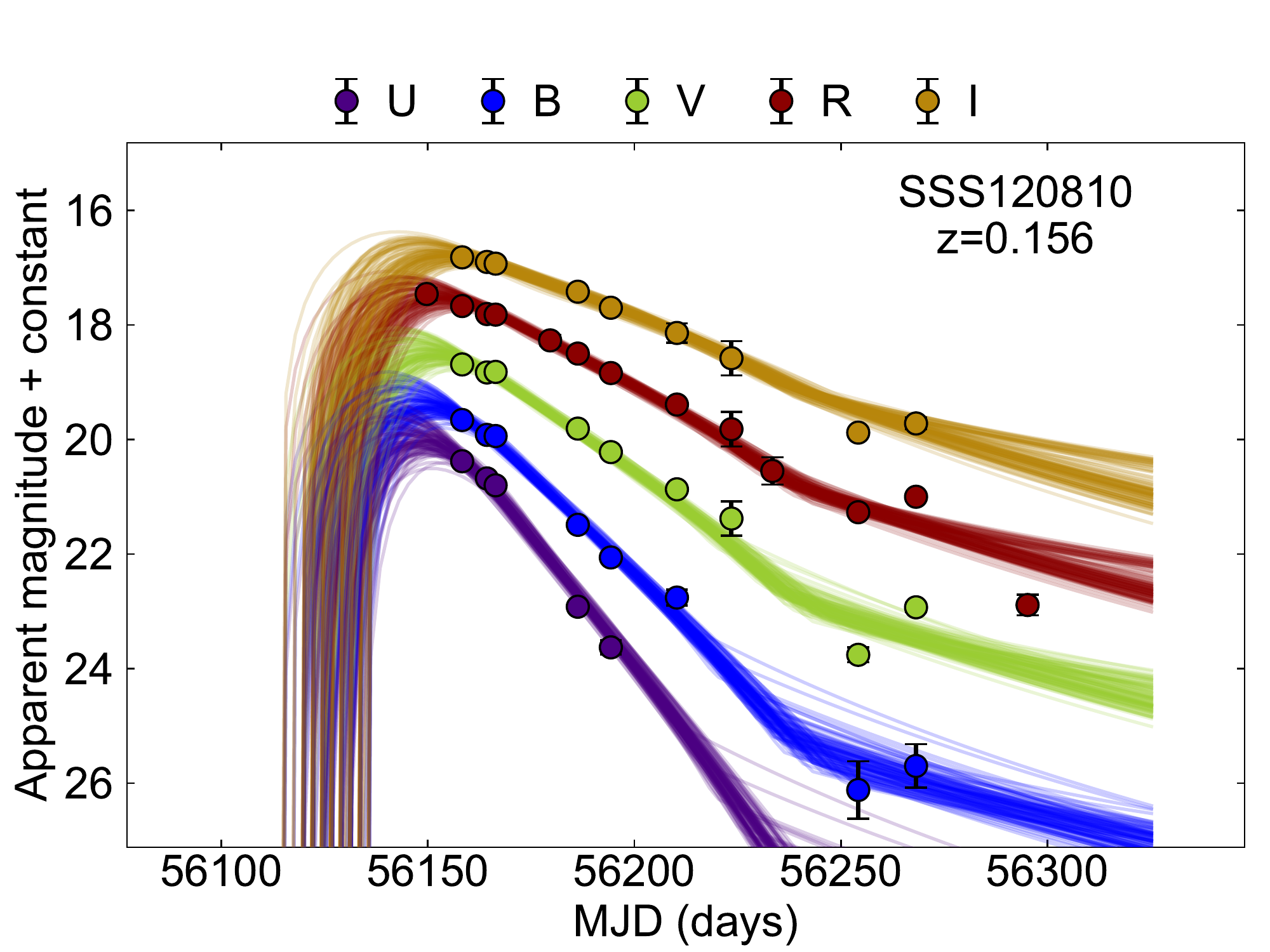}
\includegraphics[width=5.6cm]{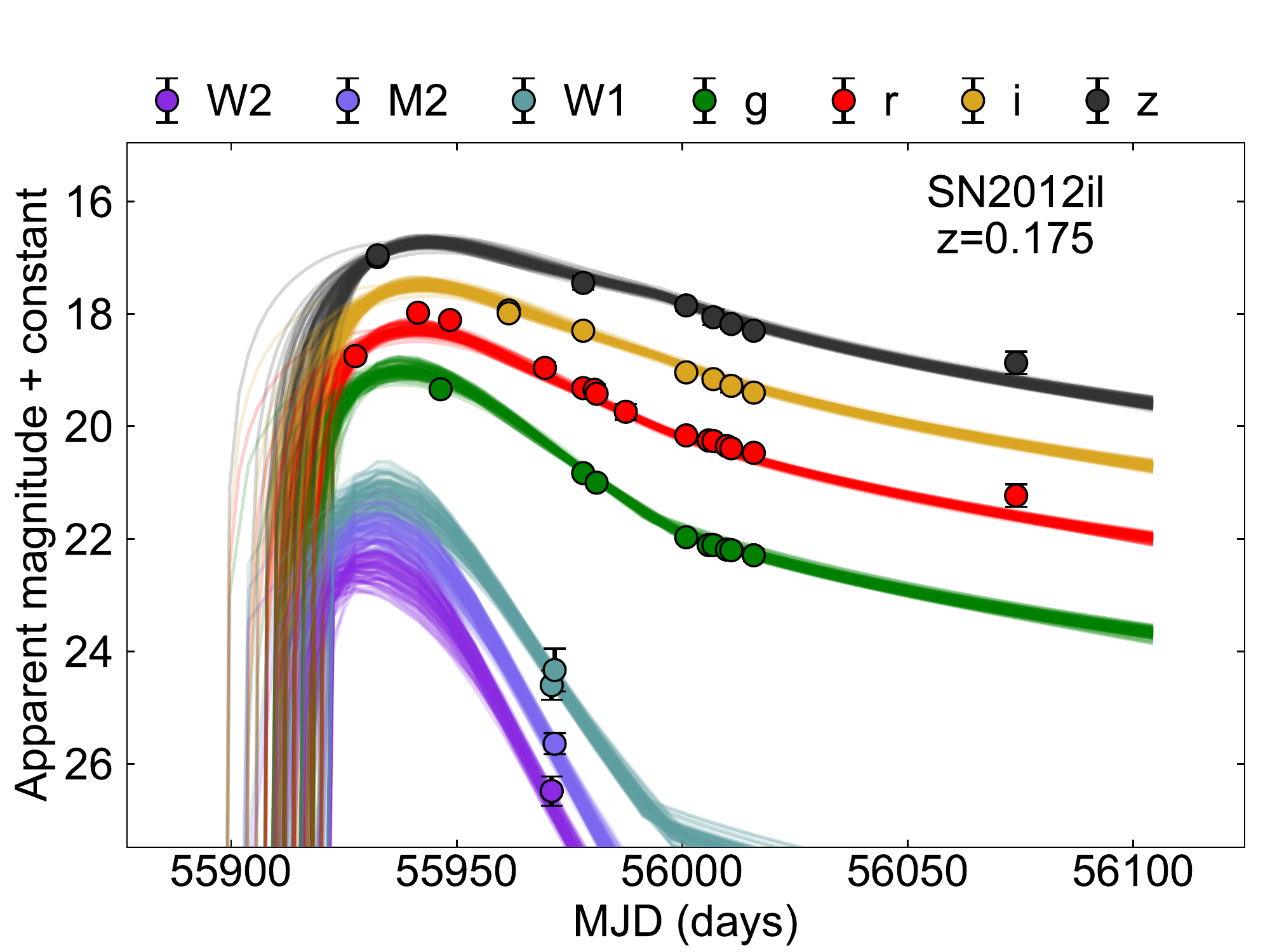}
\includegraphics[width=5.6cm]{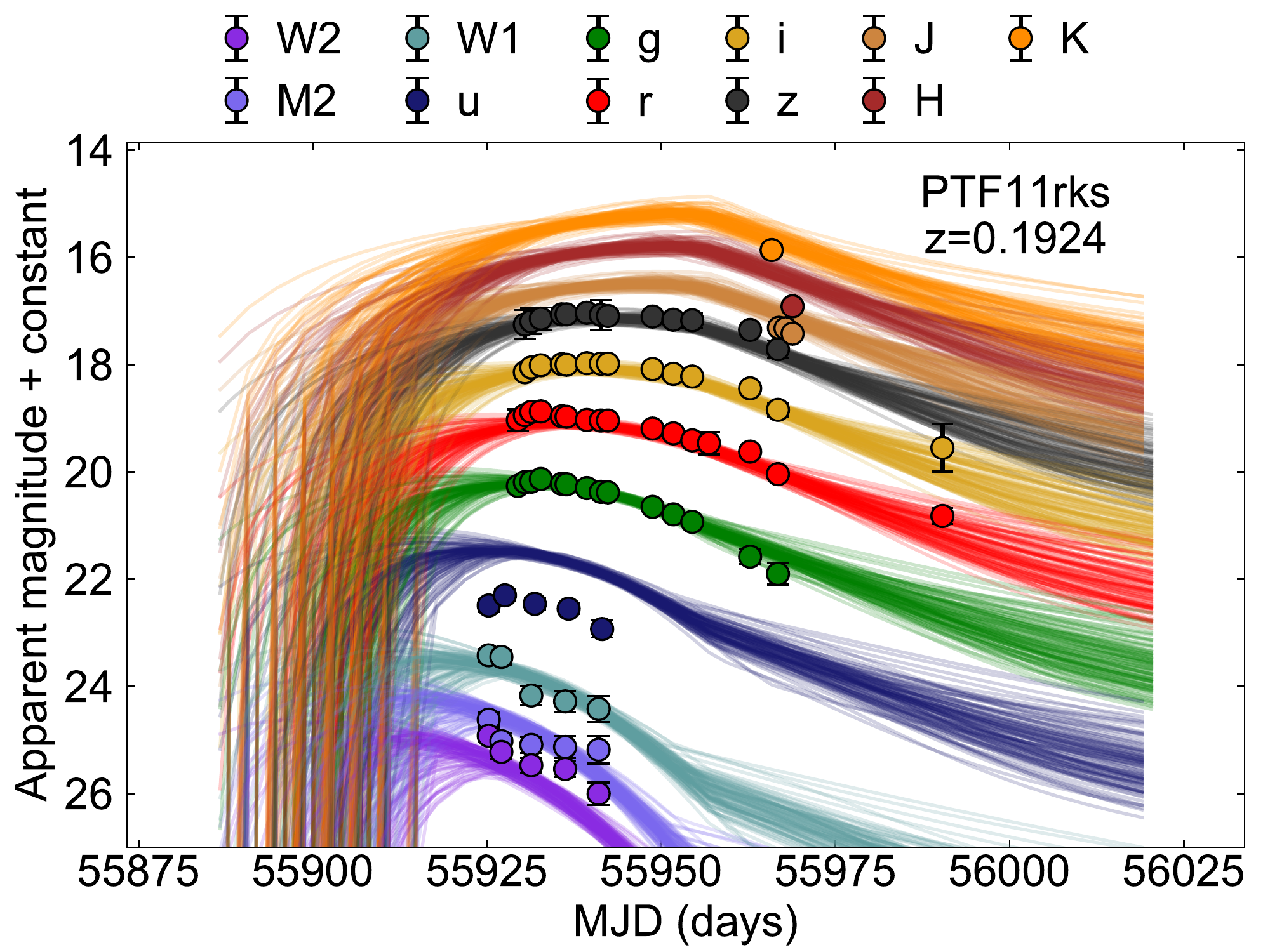}
\includegraphics[width=5.6cm]{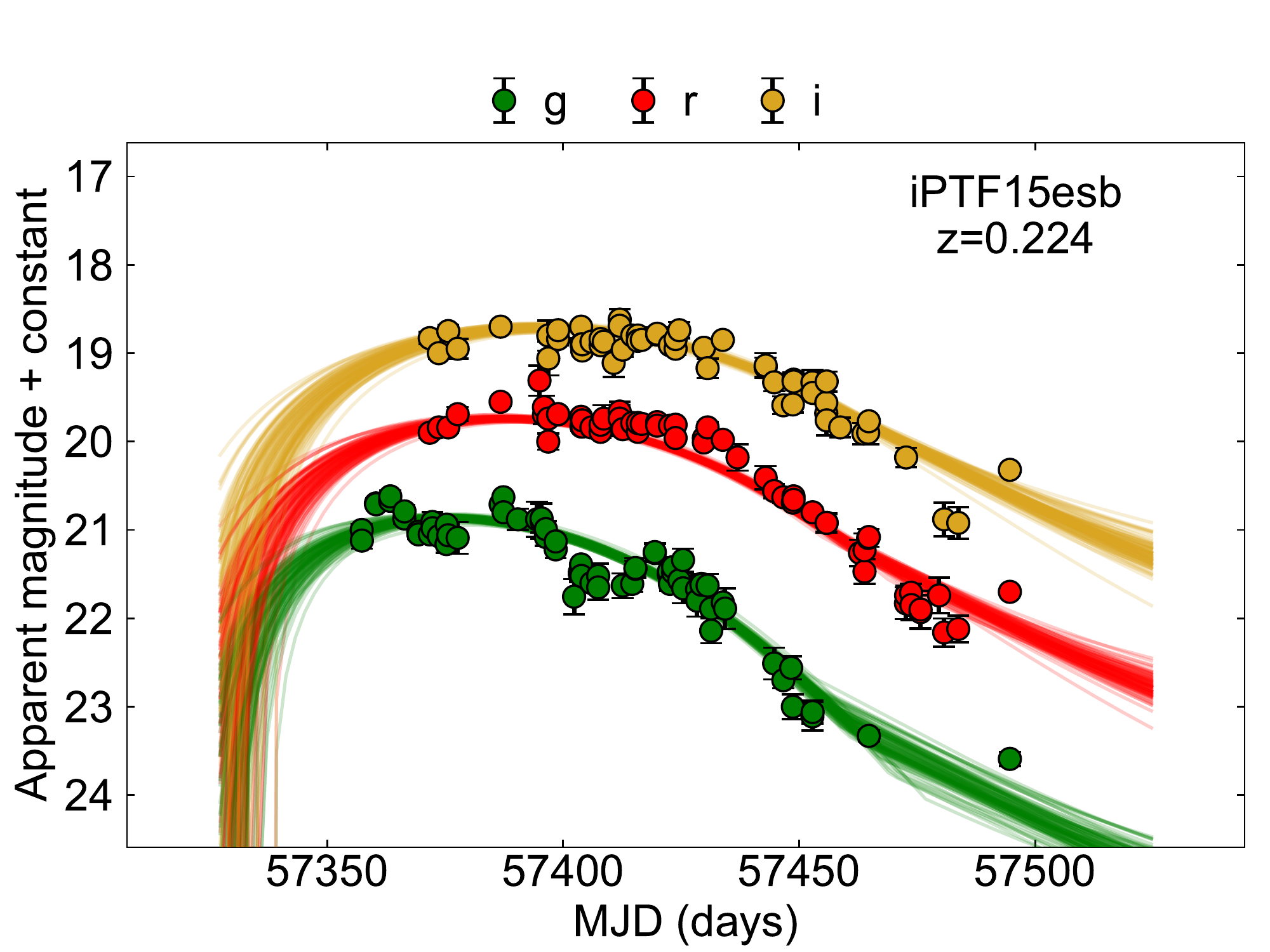}
\includegraphics[width=5.6cm]{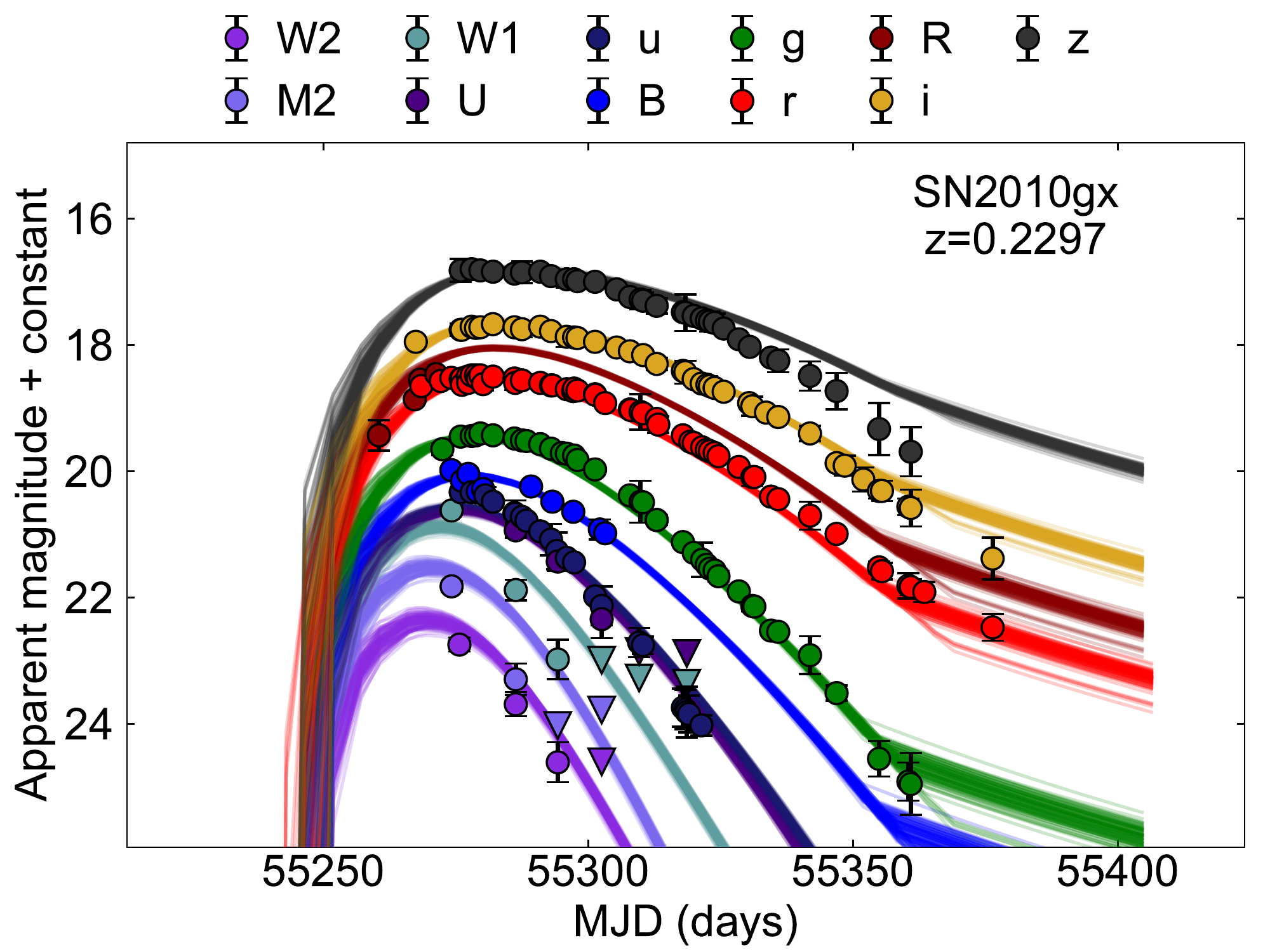}
\includegraphics[width=5.6cm]{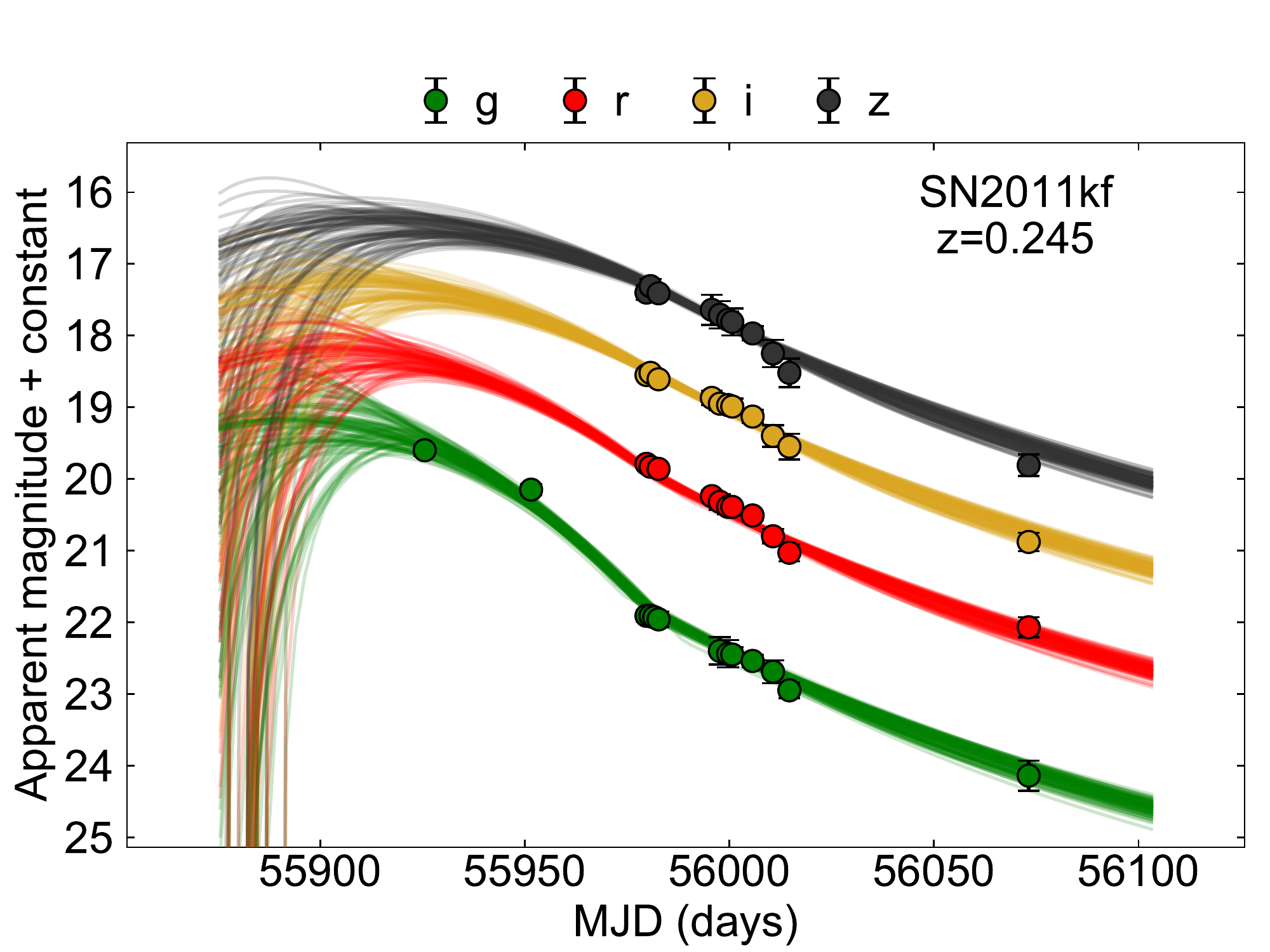}
\includegraphics[width=5.6cm]{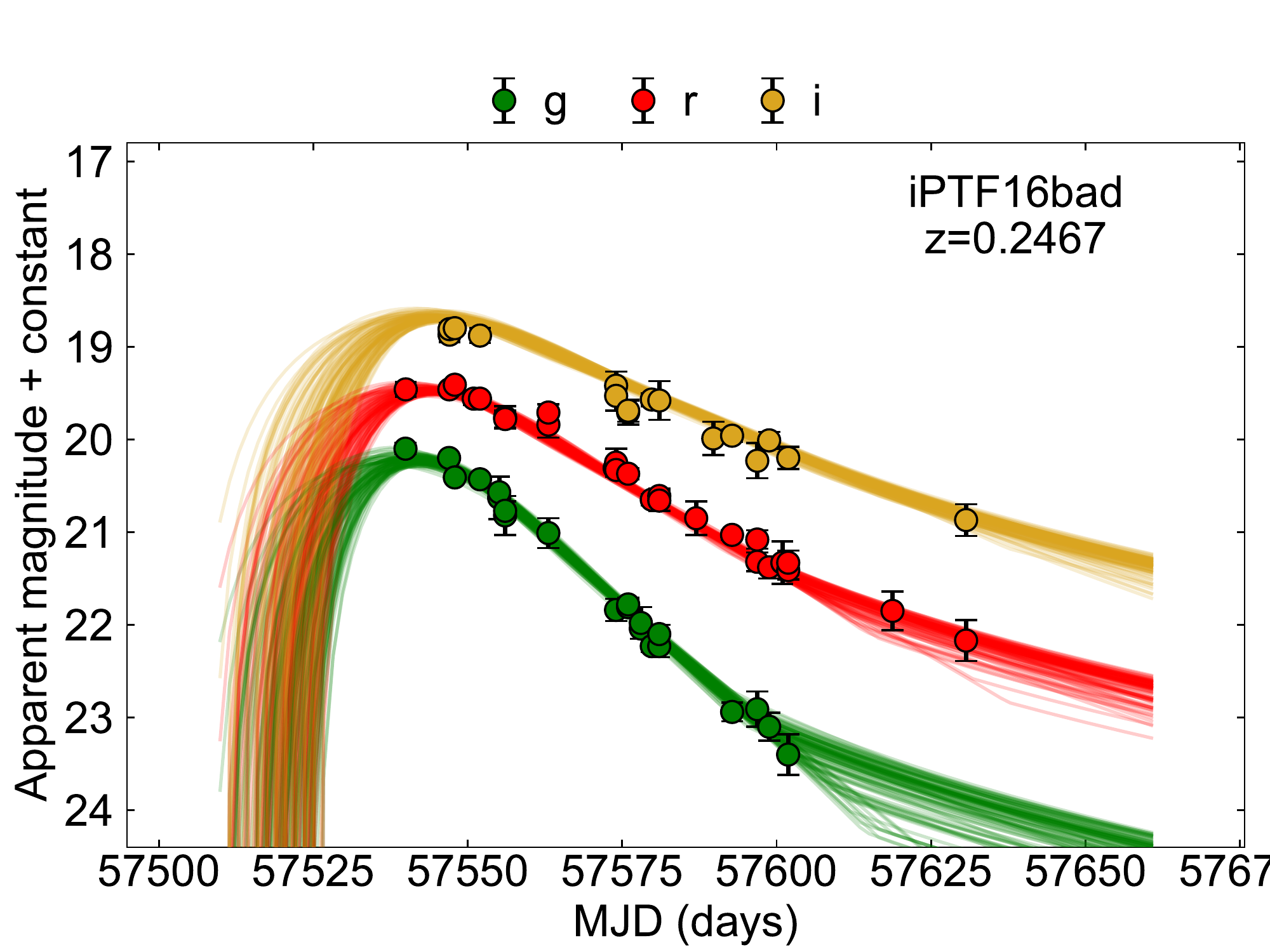}
\includegraphics[width=5.6cm]{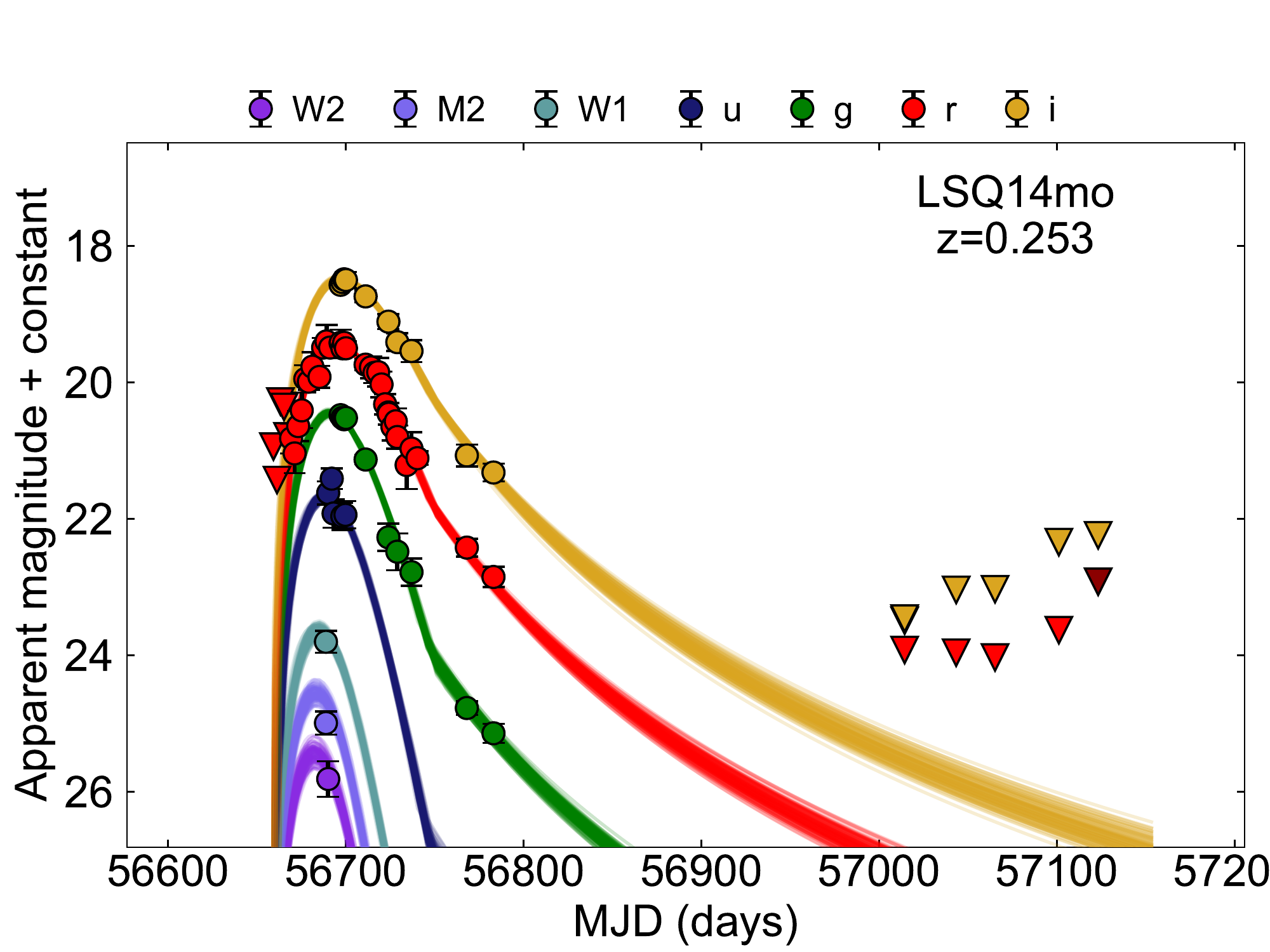}
\includegraphics[width=5.6cm]{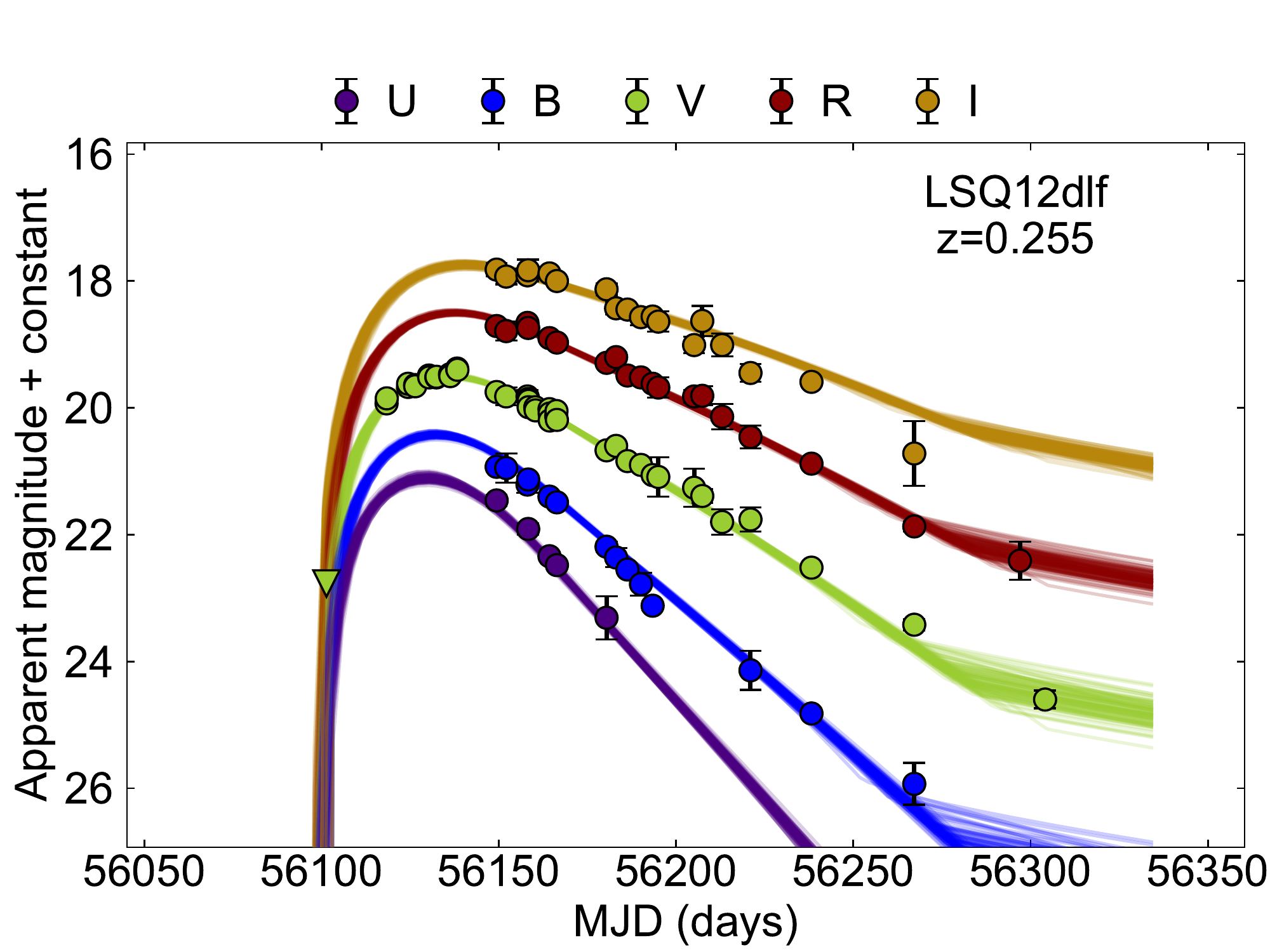}
\caption{Magnetar model fits to the full SLSN sample using \mosfit. Band offsets for display are:
$uvw2+4;\,
uvm2+3.5;\,
uvw1+3;\,
U+3;\,
u+2;\,
B+1.5;\,
g+1;\,
V+0.6;\,
r+0;\,
R-0.3;\,
i-1;\,
I-1;\,
z-2;\,
y-2.5;\,
J-2;\,
H-2.5;\,
K-3$.
}
\label{fig:all_lcs}
\end{figure*}
\begin{figure*}
\ContinuedFloat
\centering
\includegraphics[width=5.6cm]{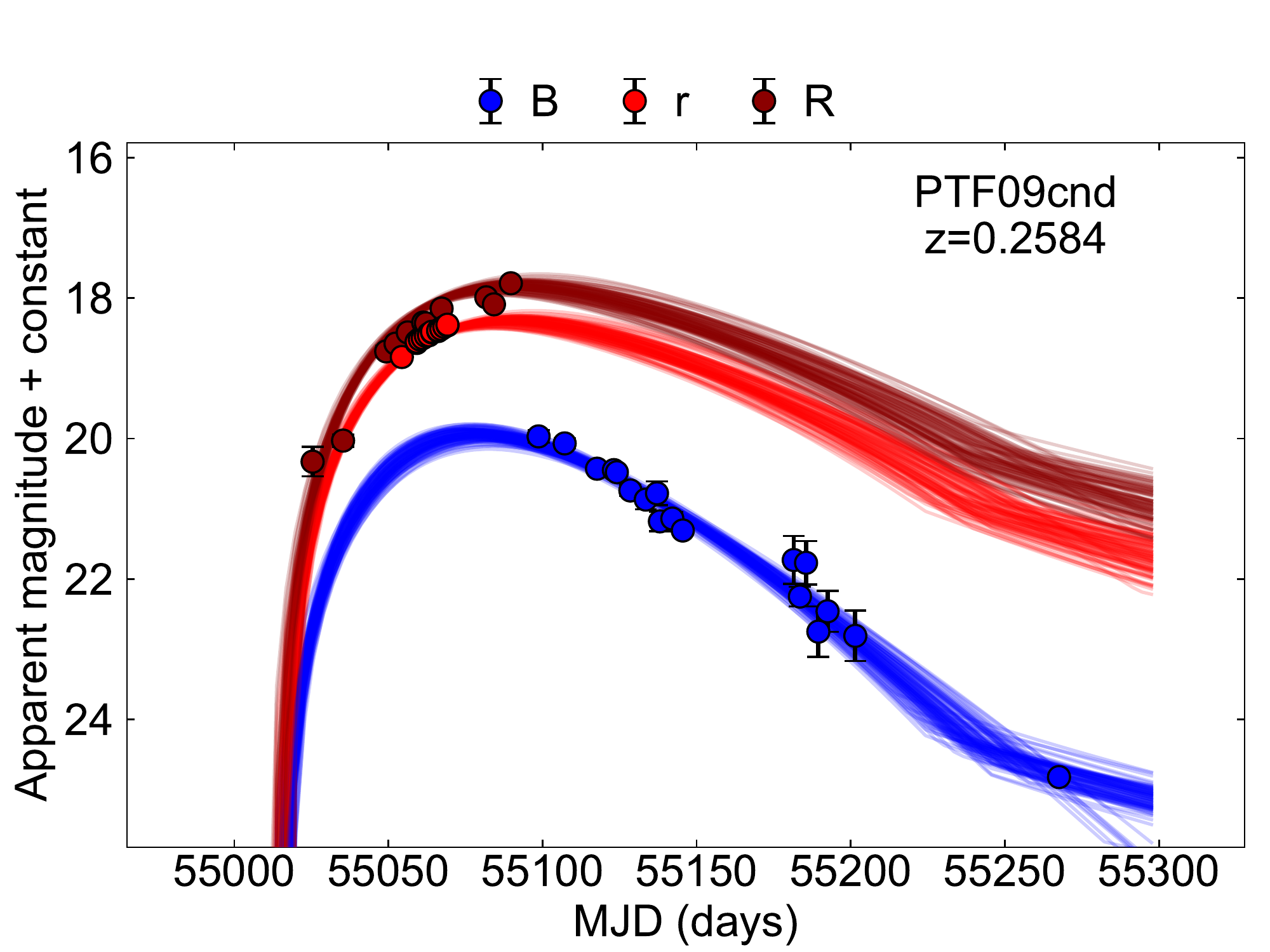}
\includegraphics[width=5.6cm]{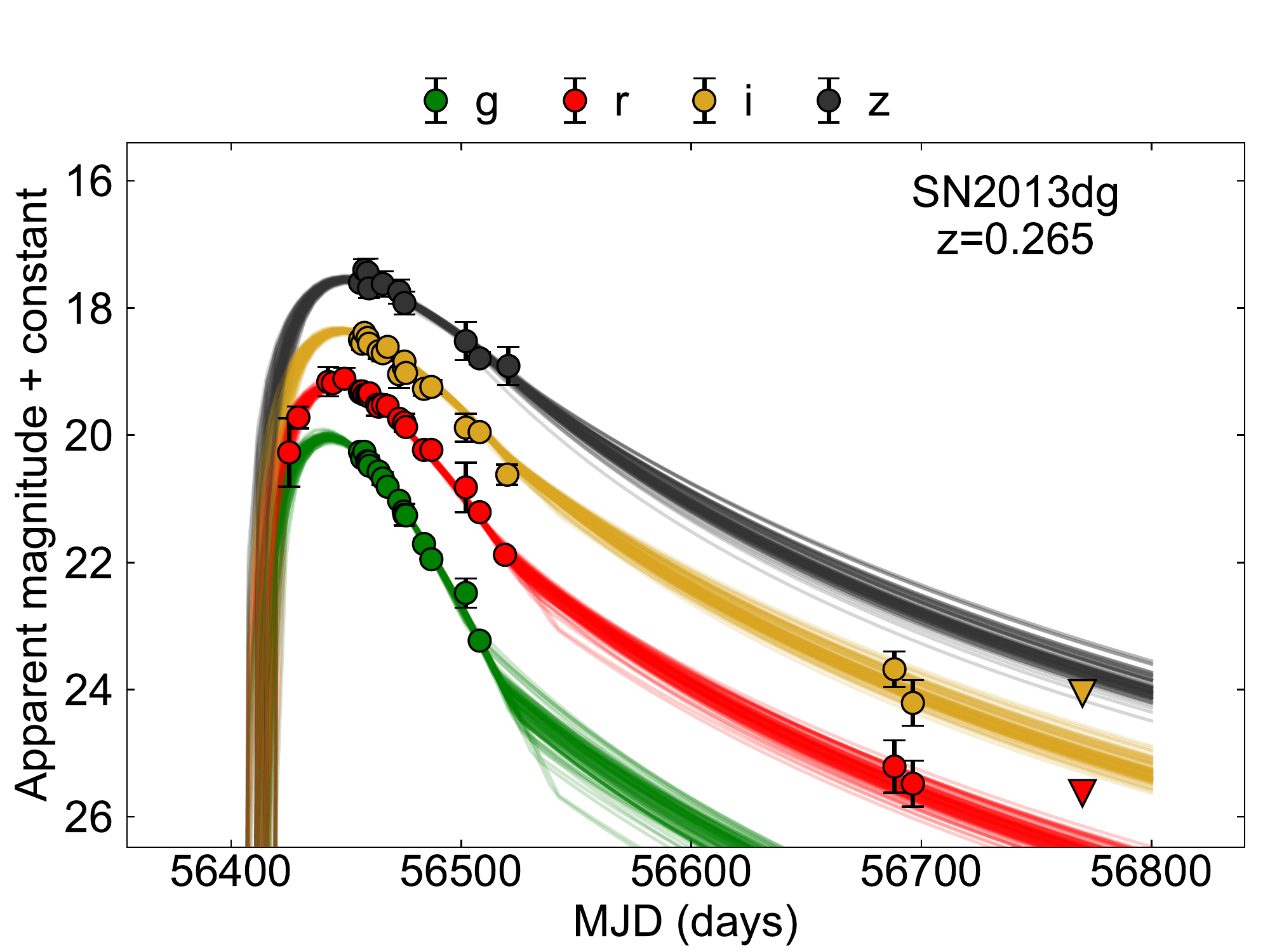}
\includegraphics[width=5.6cm]{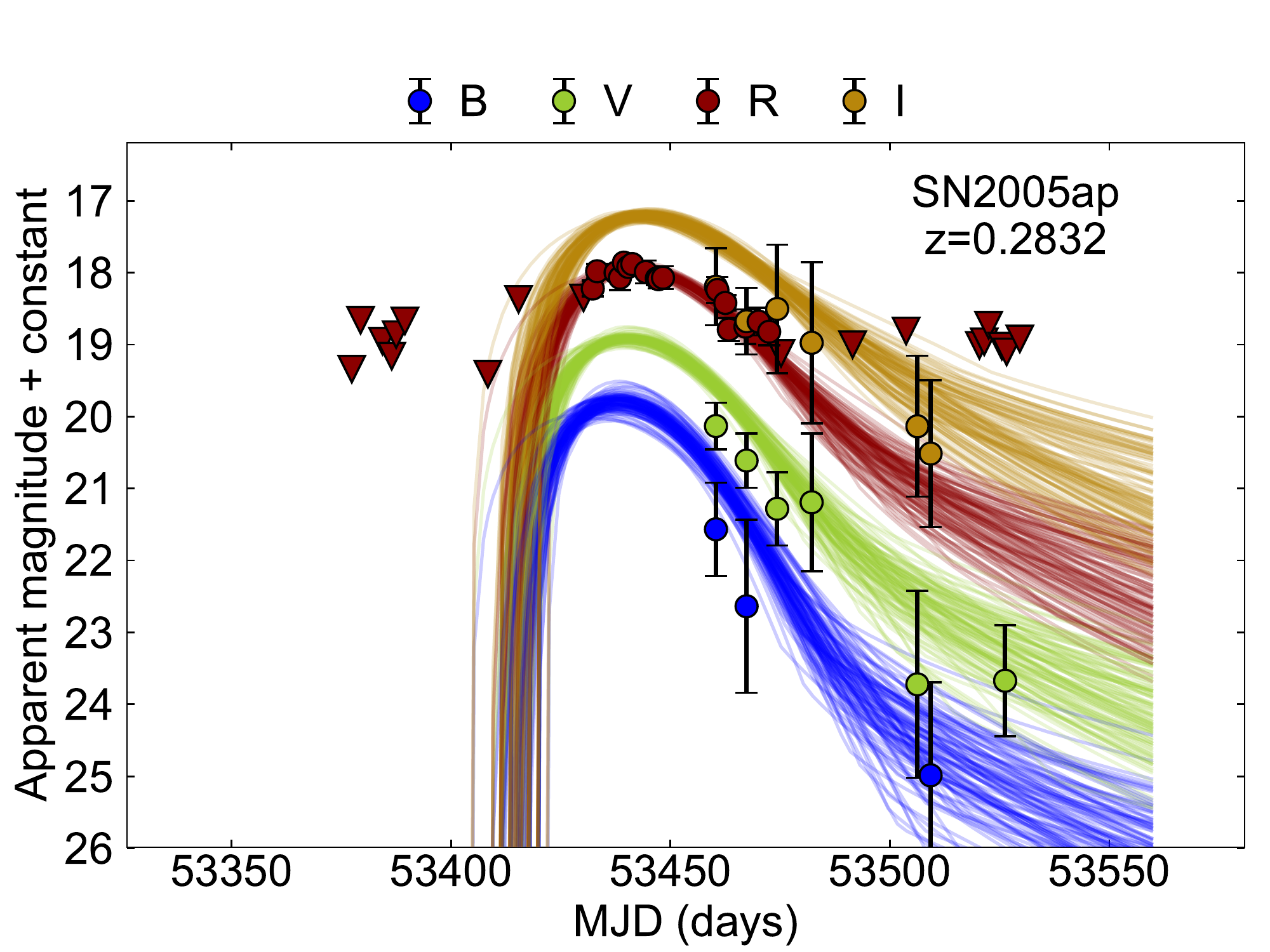}
\includegraphics[width=5.6cm]{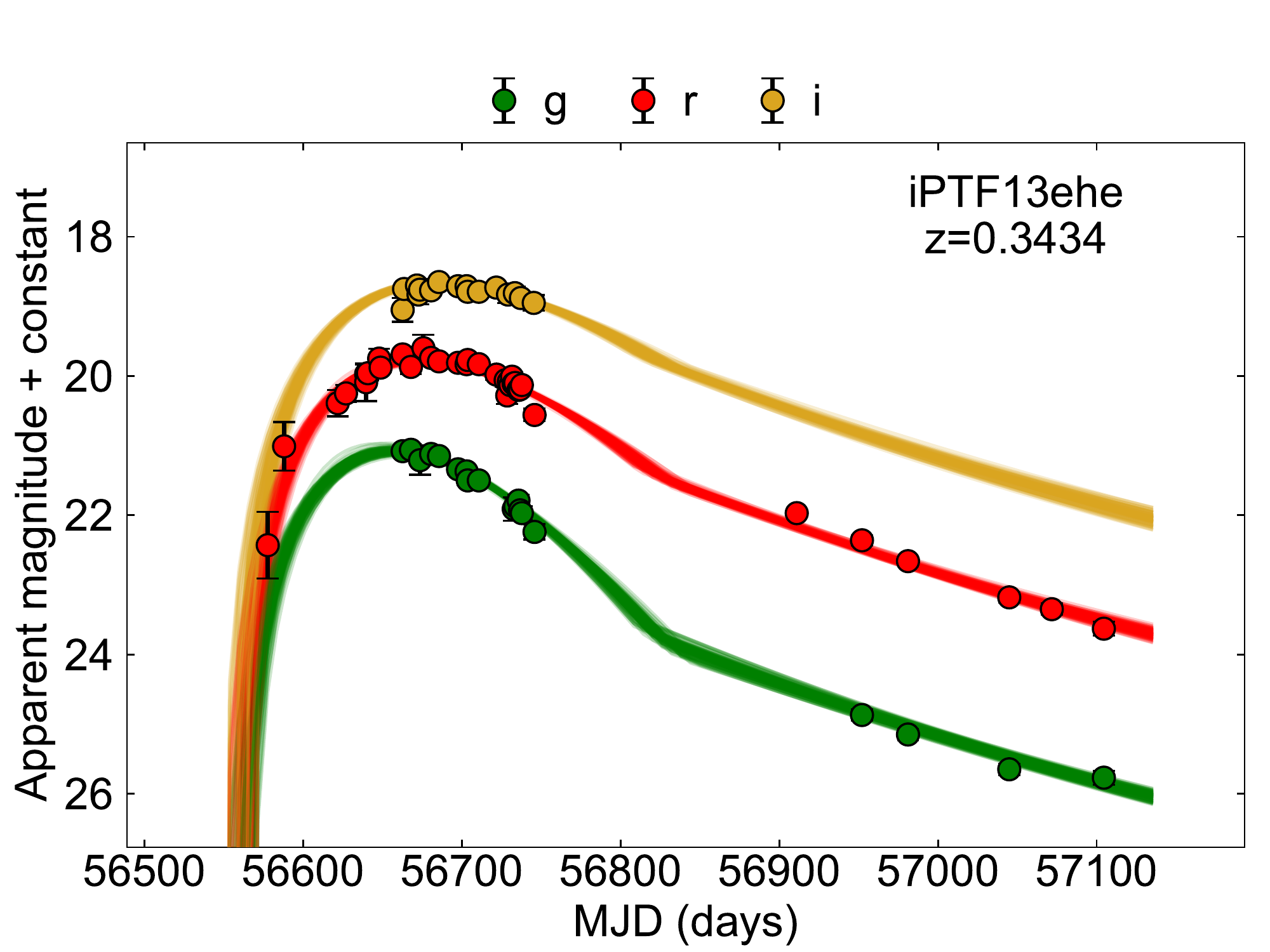}
\includegraphics[width=5.6cm]{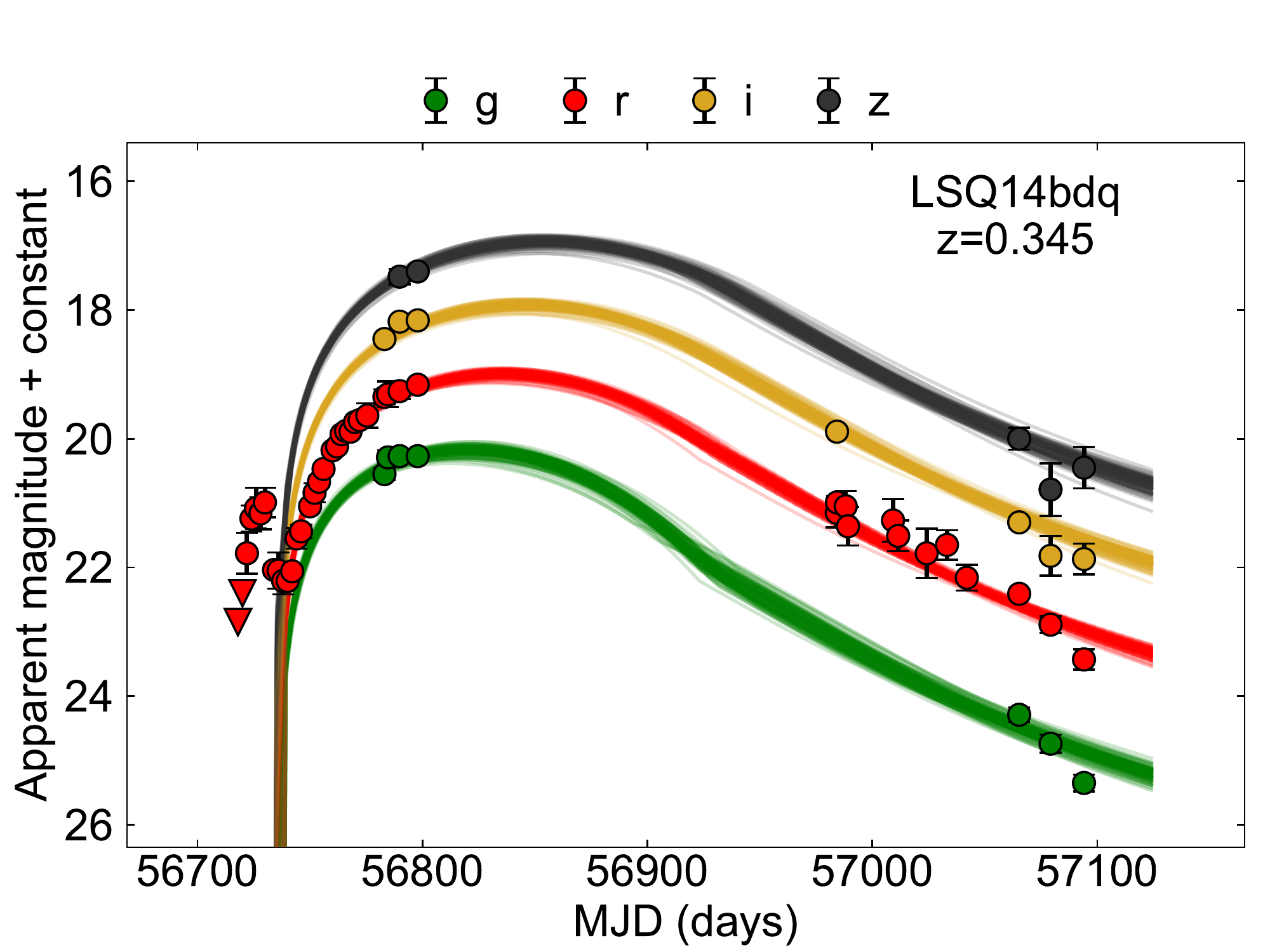}
\includegraphics[width=5.6cm]{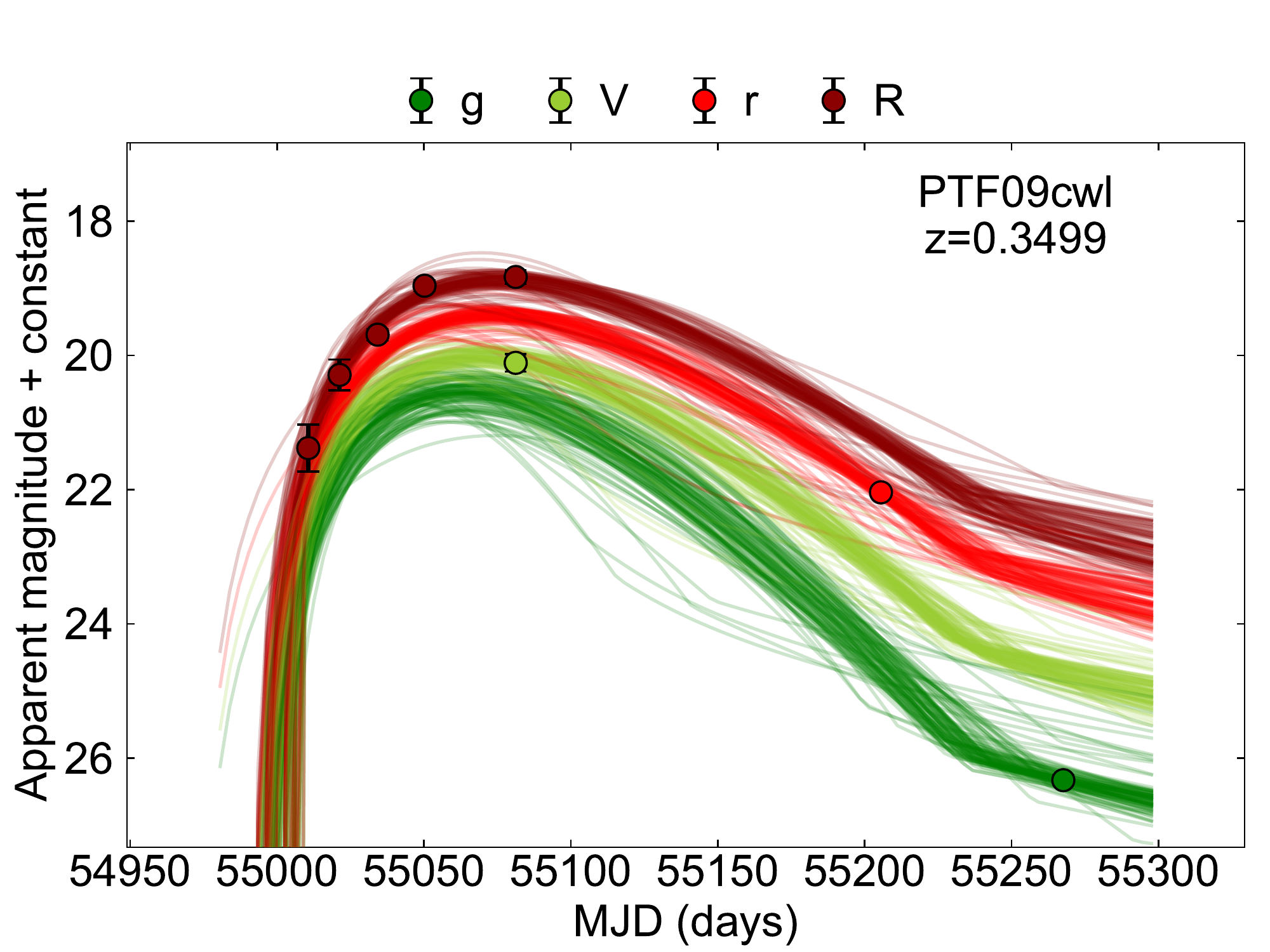}
\includegraphics[width=5.6cm]{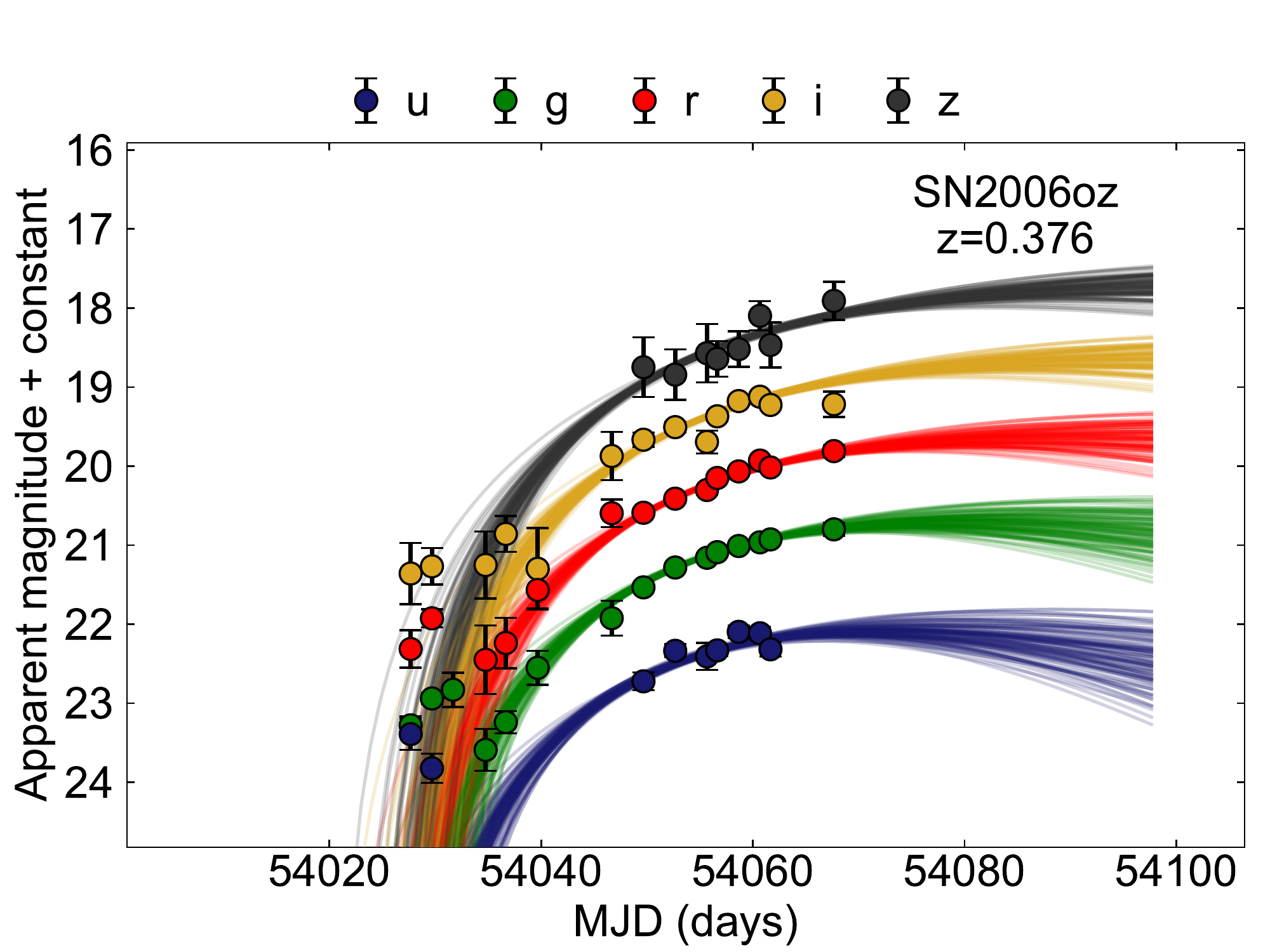}
\includegraphics[width=5.6cm]{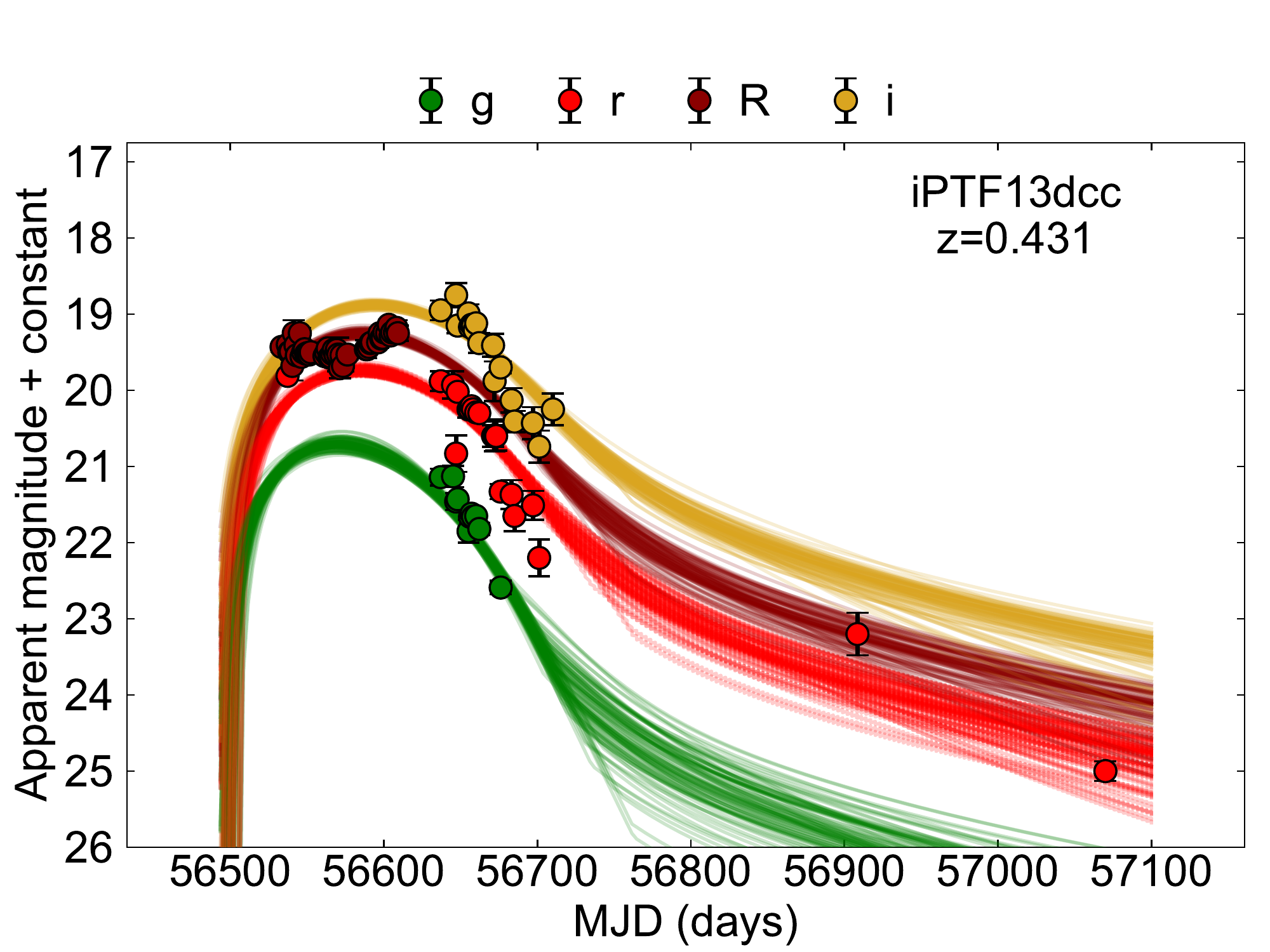}
\includegraphics[width=5.6cm]{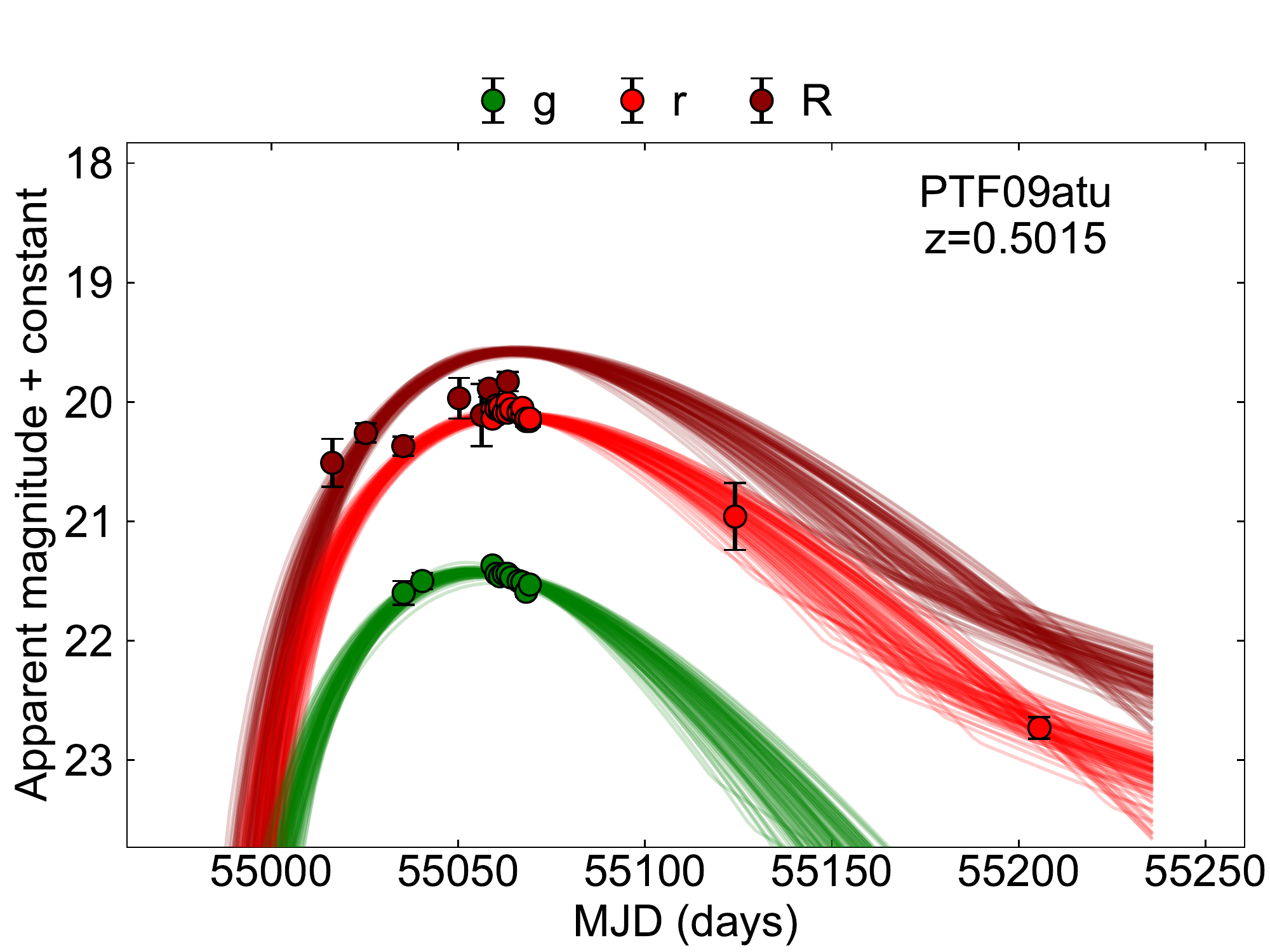}
\includegraphics[width=5.6cm]{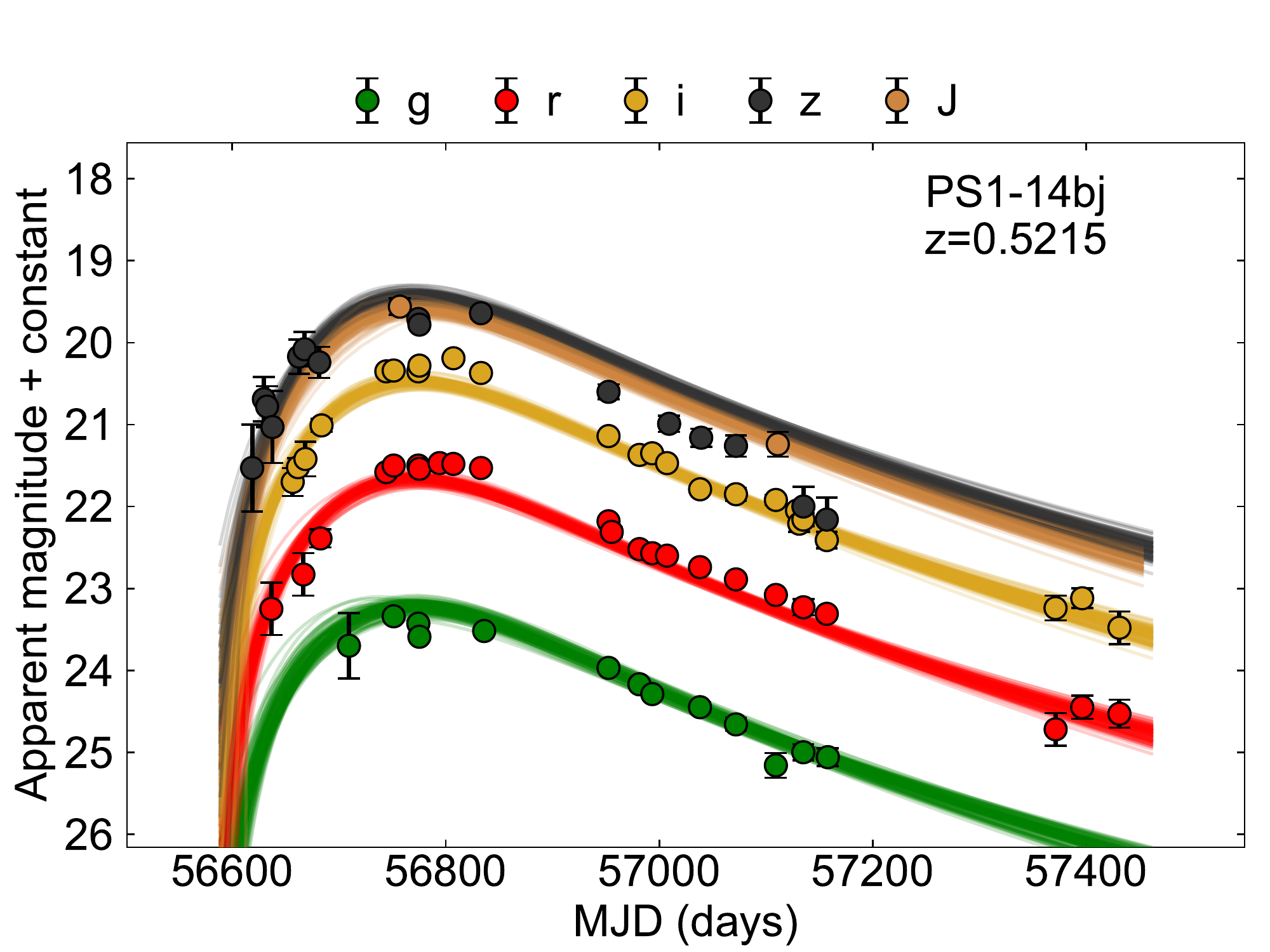}
\includegraphics[width=5.6cm]{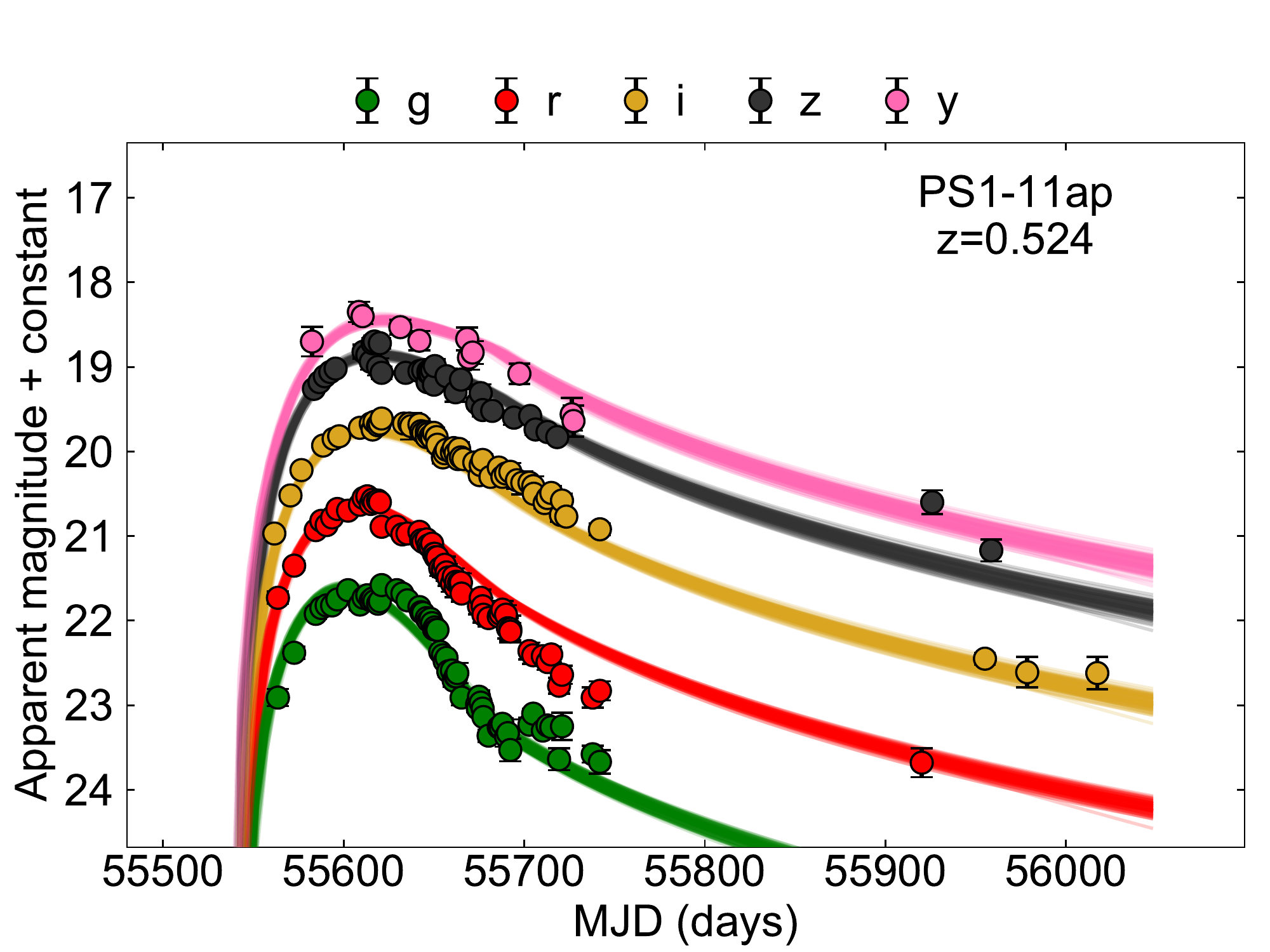}
\includegraphics[width=5.6cm]{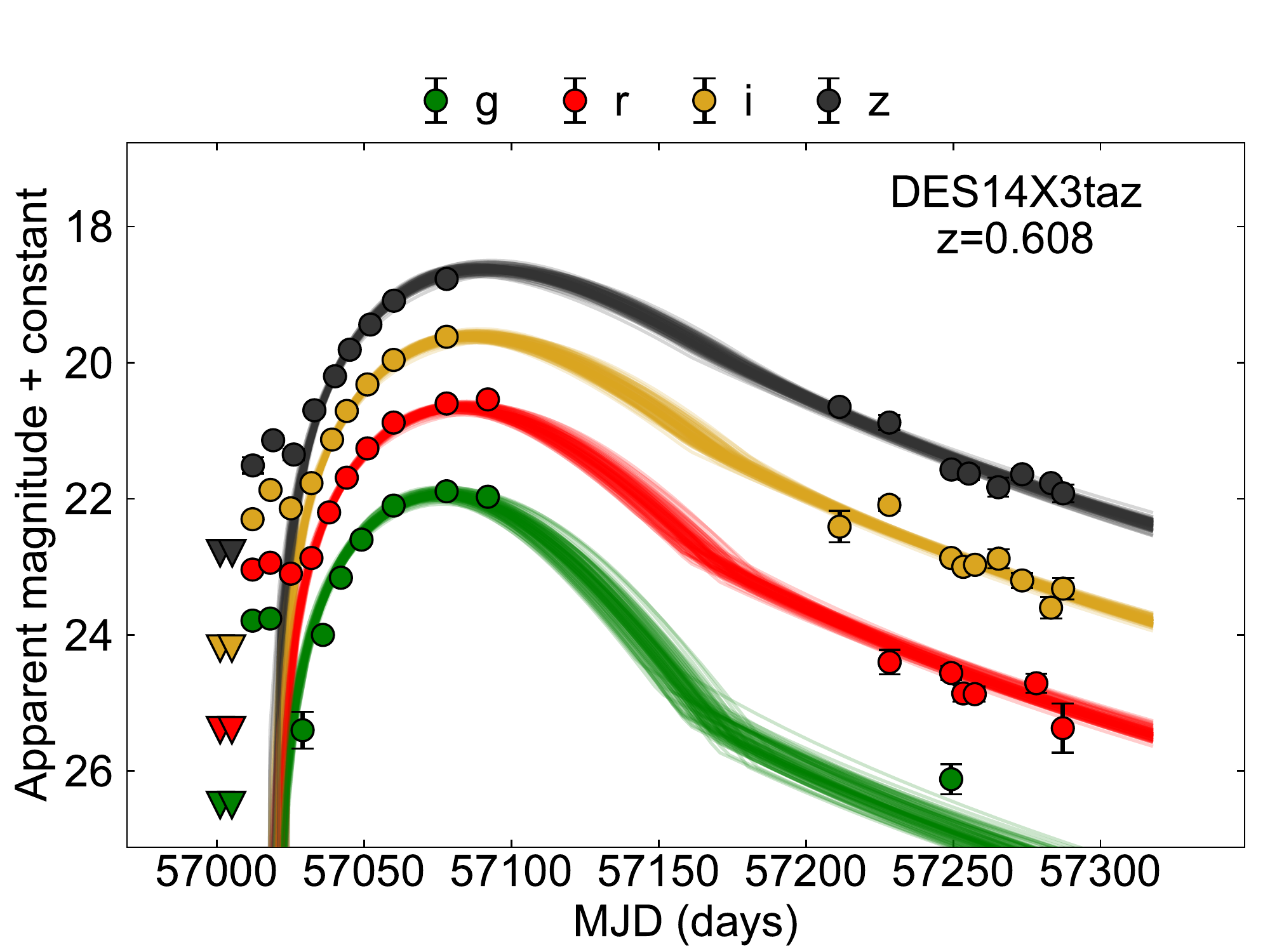}
\includegraphics[width=5.6cm]{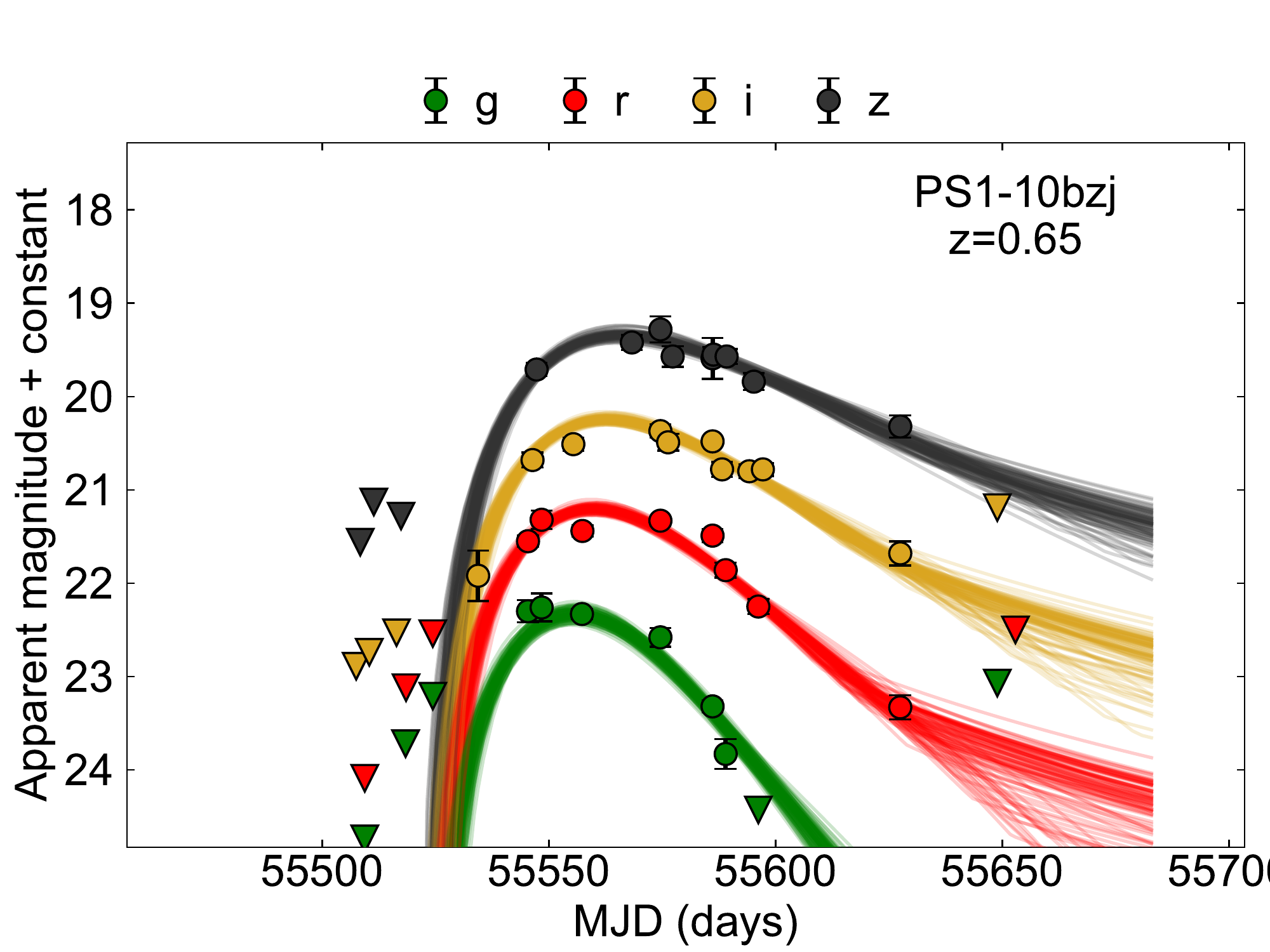}
\includegraphics[width=5.6cm]{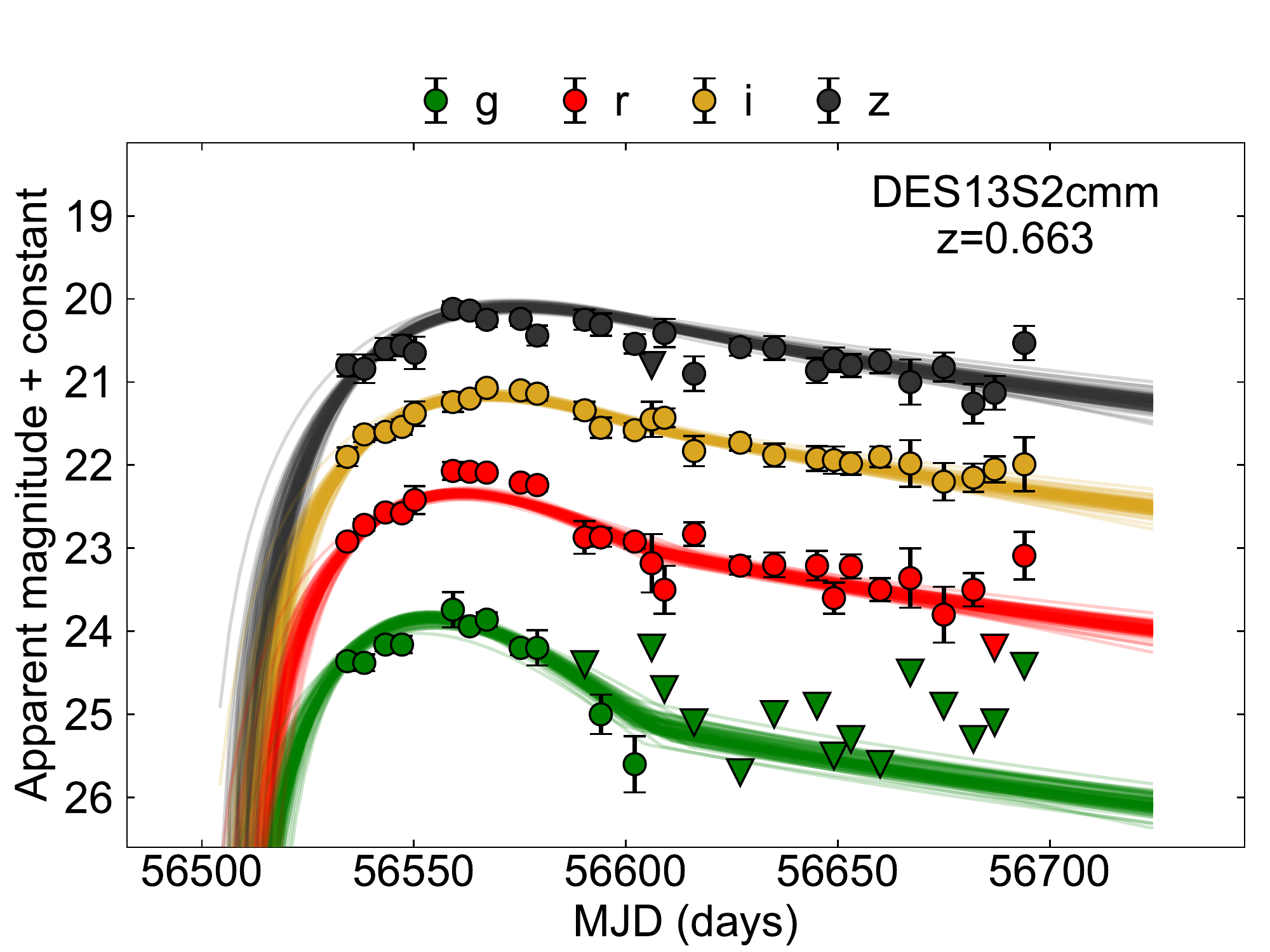}
\includegraphics[width=5.6cm]{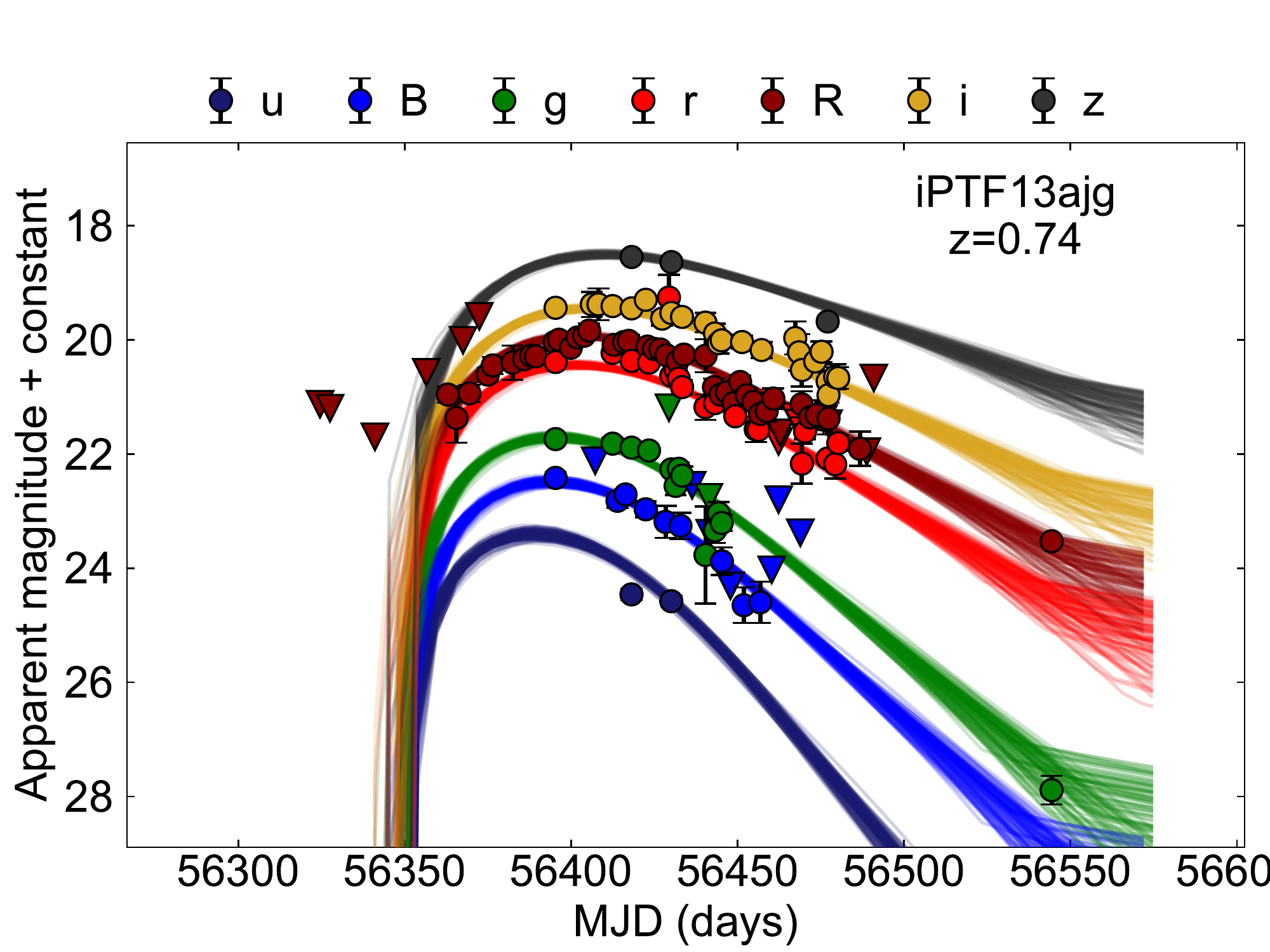}
\caption{Model fits---continued}
\end{figure*}
\begin{figure*}
\ContinuedFloat
\centering
\includegraphics[width=5.6cm]{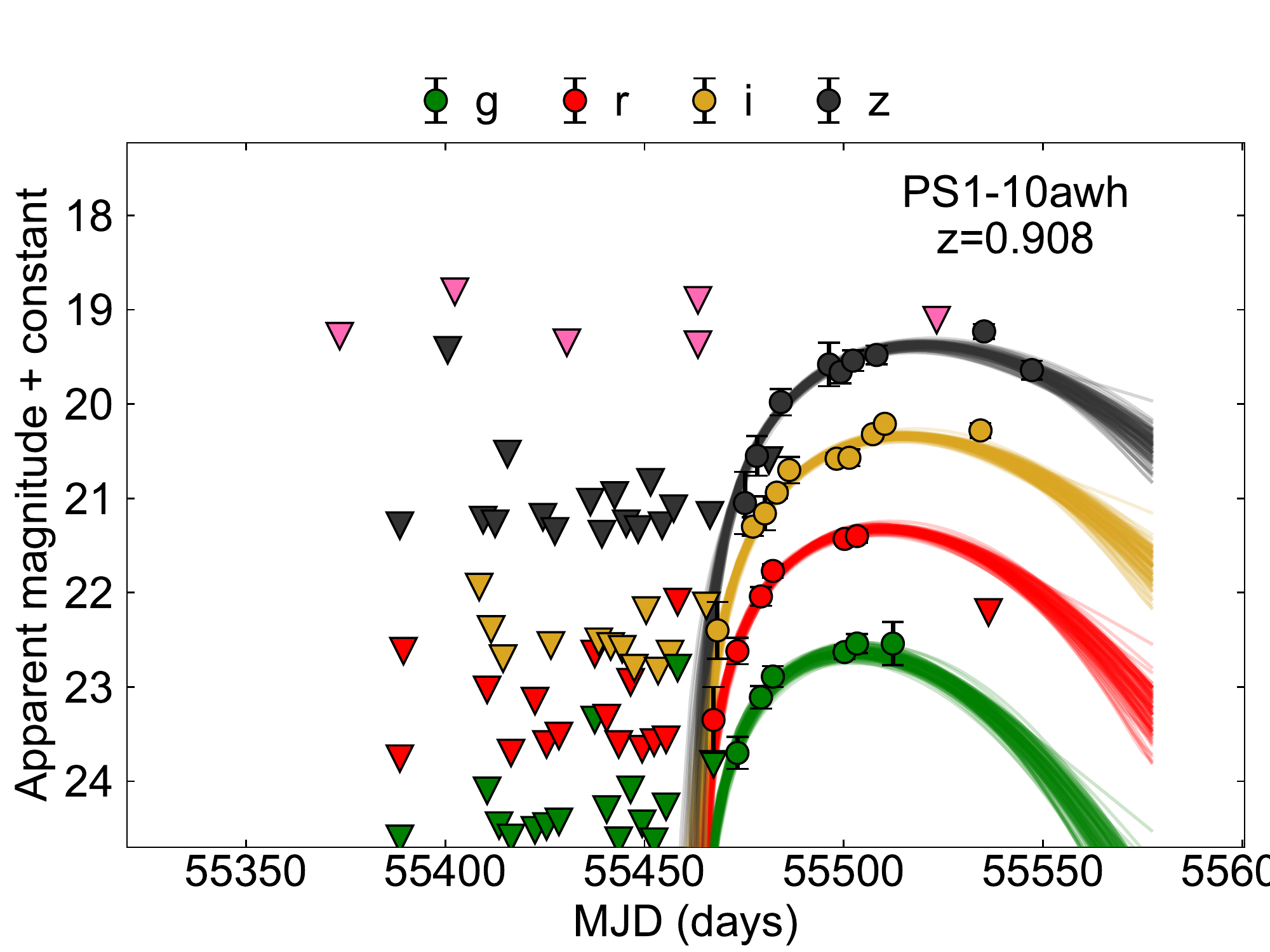}
\includegraphics[width=5.6cm]{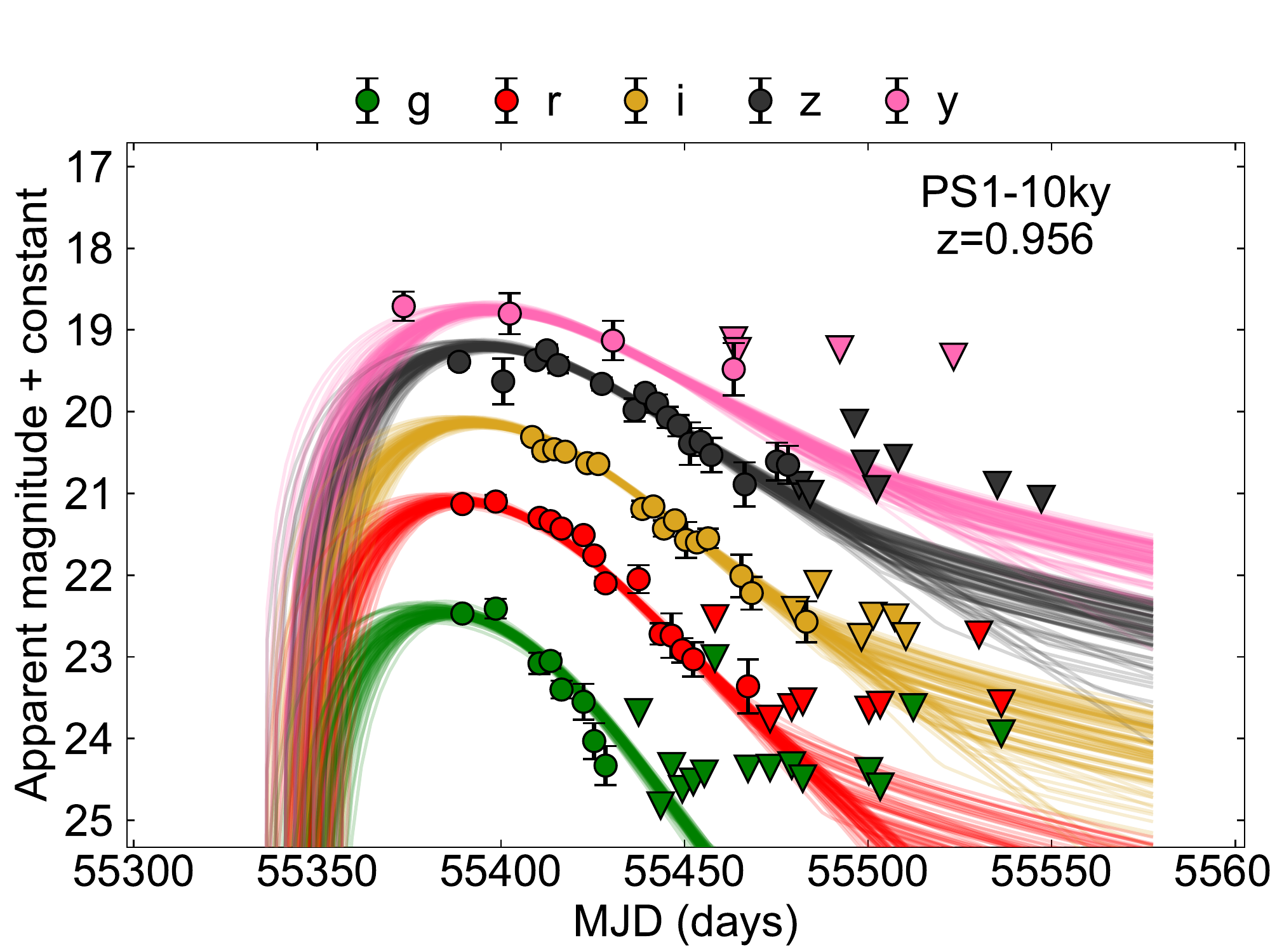}
\includegraphics[width=5.6cm]{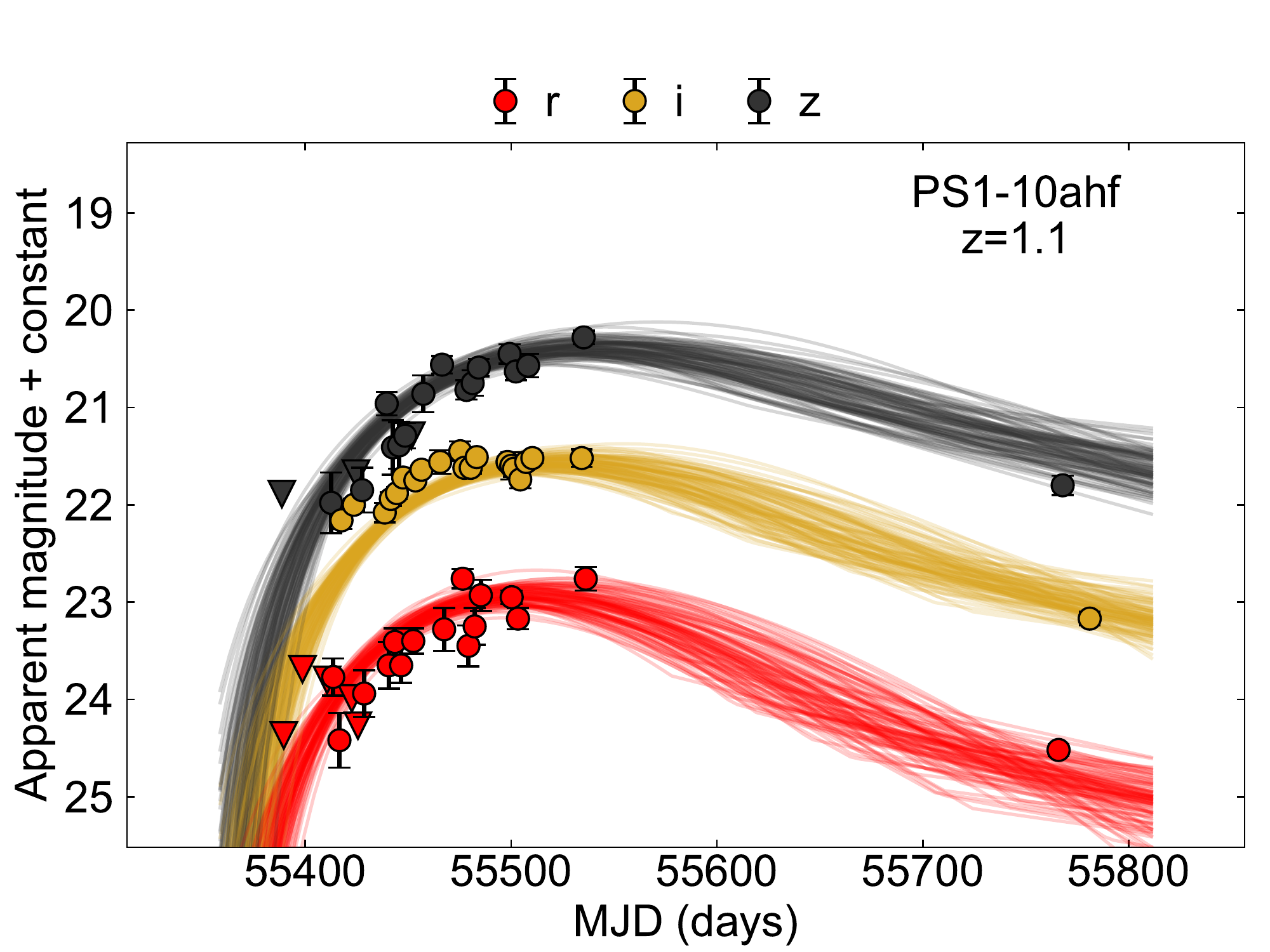}
\includegraphics[width=5.6cm]{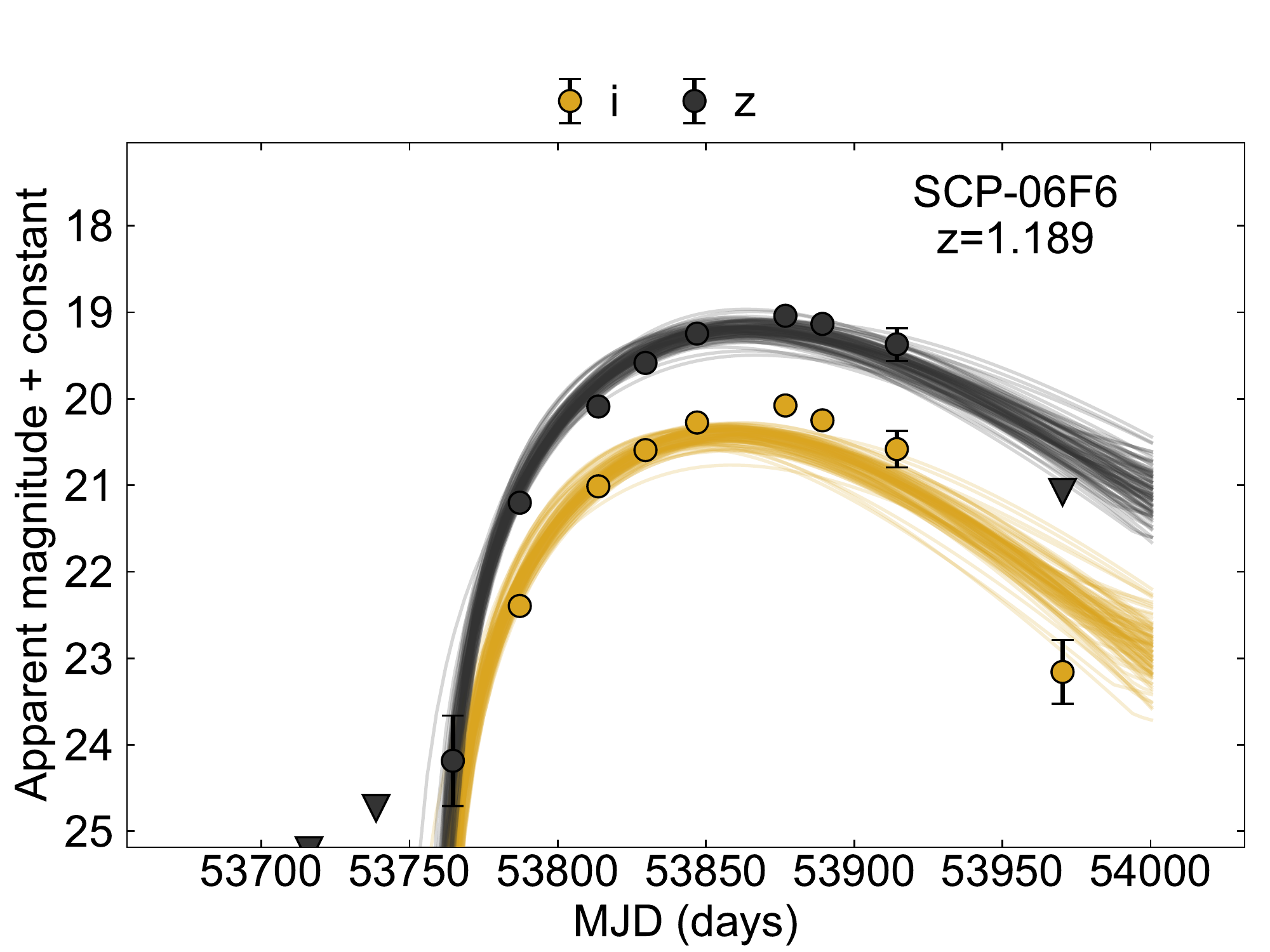}
\includegraphics[width=5.6cm]{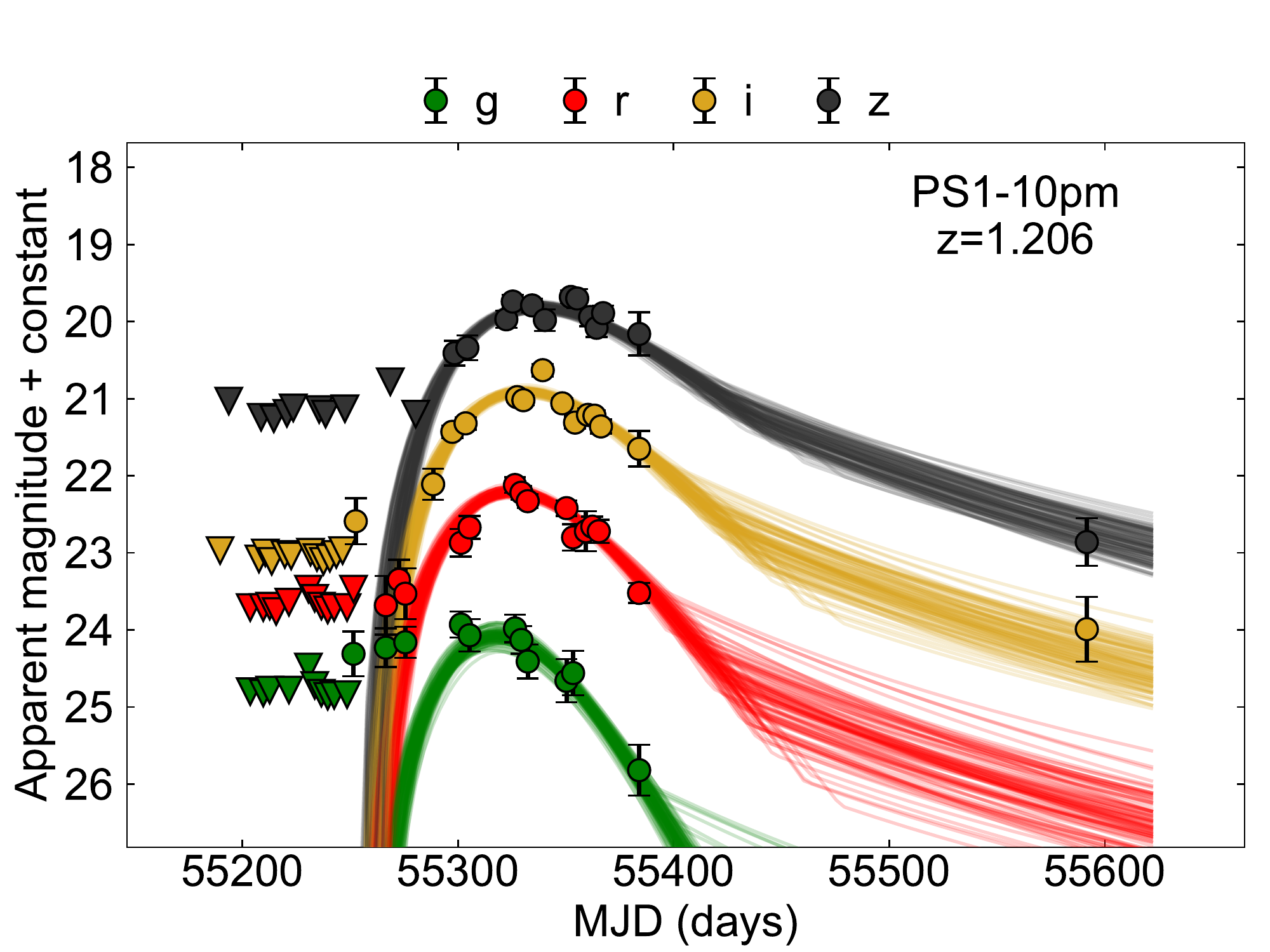}
\includegraphics[width=5.6cm]{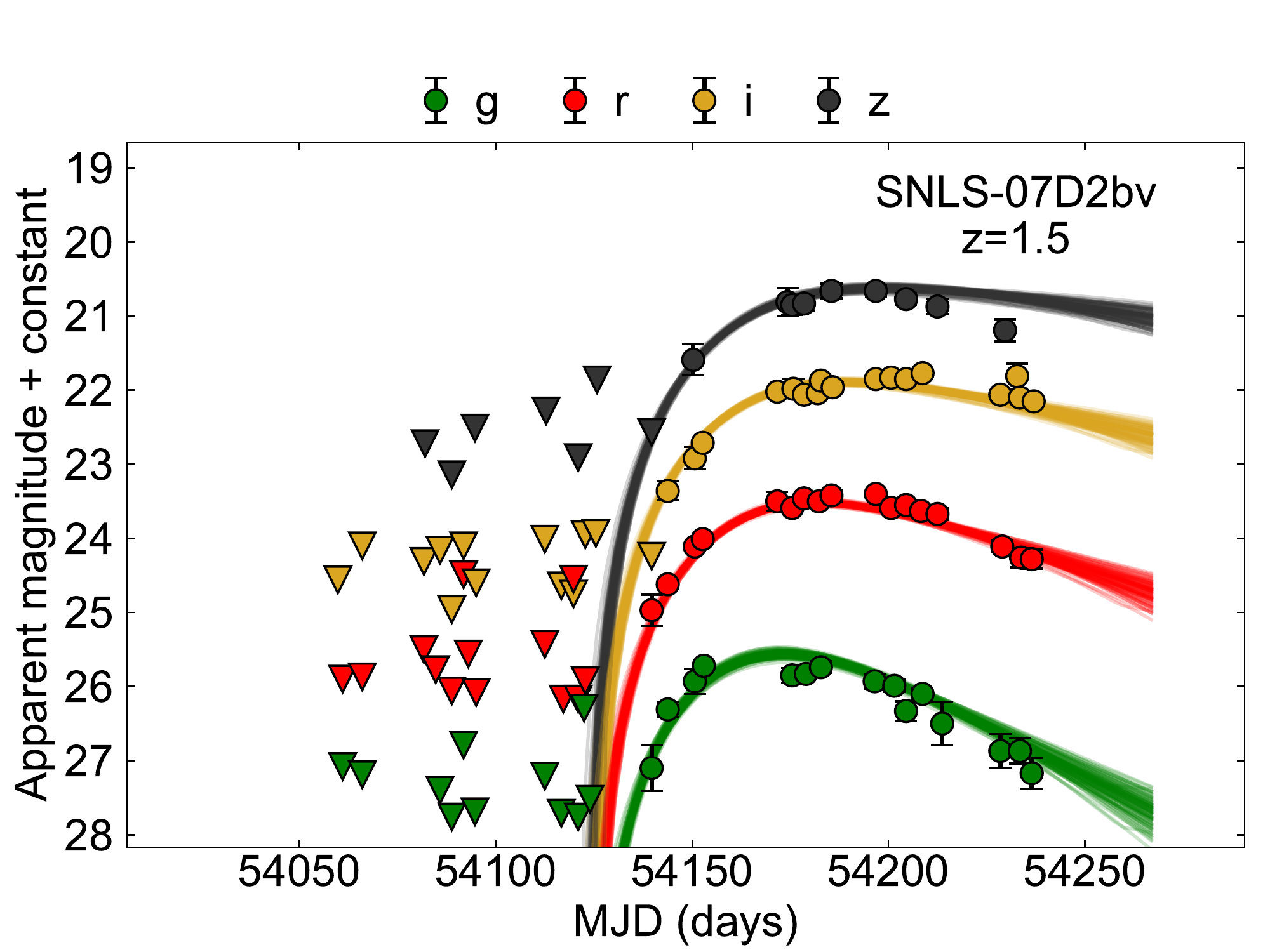}
\includegraphics[width=5.6cm]{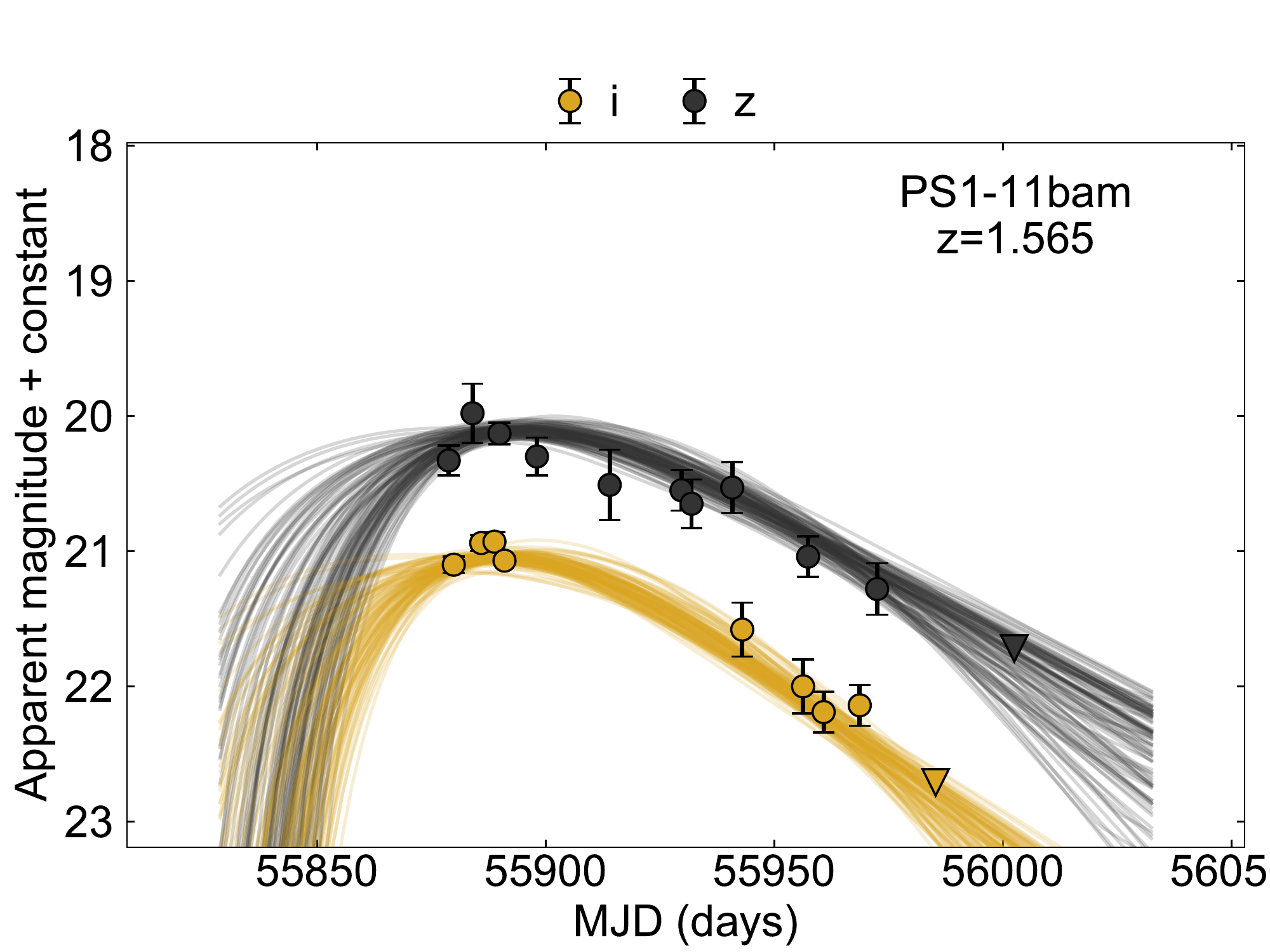}
\includegraphics[width=5.6cm]{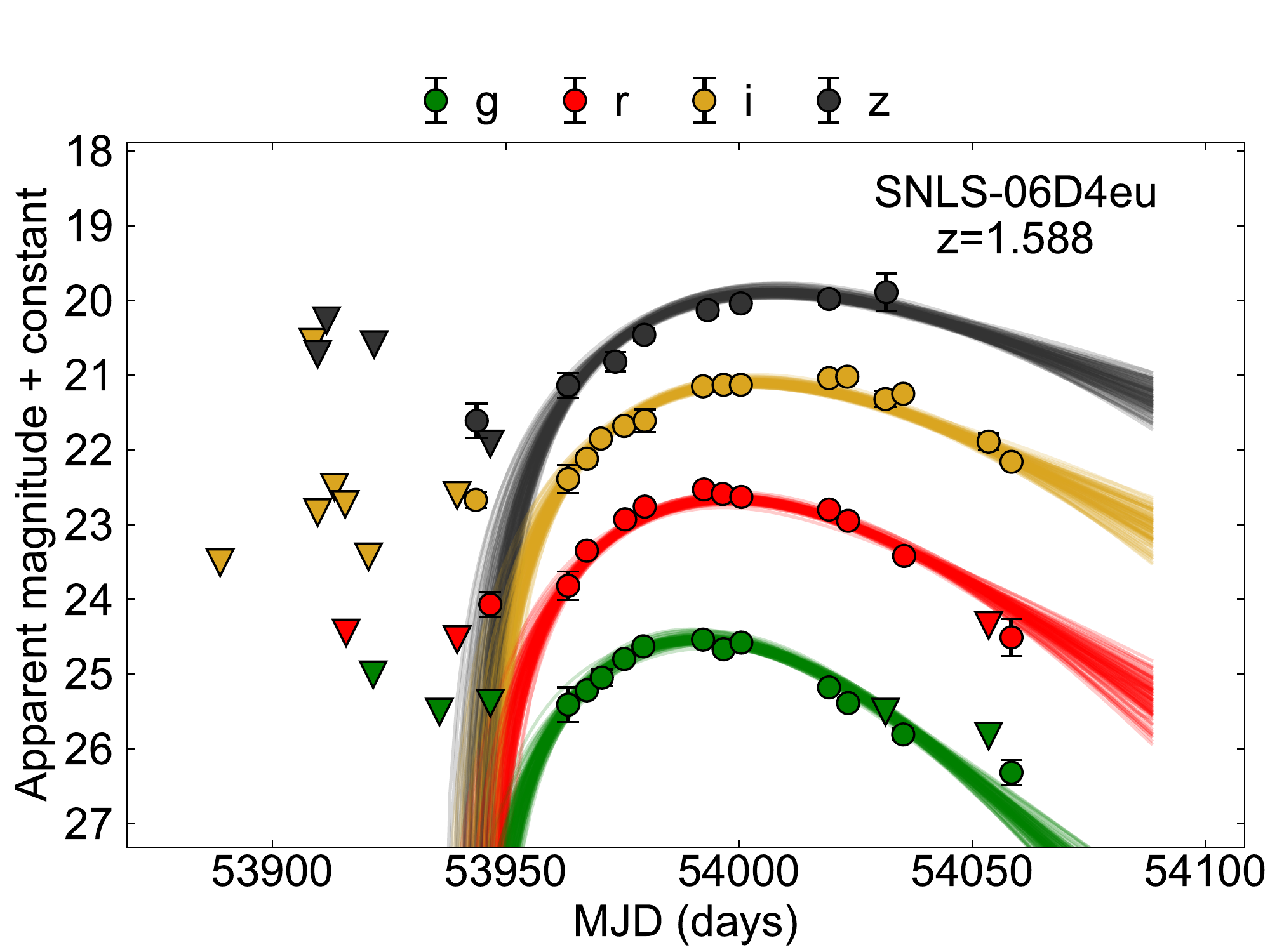}
\caption{Model fits---continued}
\end{figure*}

\section{Light curve fits}
\label{sec:fits}

We now present the results of our light curve fitting procedure. Figure \ref{fig:all_lcs} shows the model fits to all 38 SLSNe in the sample. Each panel shows the observed data from the \textit{Open Supernova Catalog} and the ensemble of light curves generated by the MCMC, with arbitrary offsets between the different bands for clarity. It can be seen from the spreads within each light curve ensemble that for most events, the data are sufficient to constrain the fits quite tightly.
The models are less tightly constrained for events with particularly noisy data (e.g.~SN\,2005ap). For events without constrained explosion epochs, we see that a range of model light curves with quite different rise times can equally well fit the peak and decline phases (SN\,2011kf). This is one of the key advantages of the MCMC approach compared to $\chi^2$ minimization---here we are not limited to picking one solution from an array of equally probable alternatives.
In addition, the colours are well-matched, with no large systematic band offsets, justifying our use of a simple absorbed blackbody SED. This is particularly true around maximum light.

We quantify the fit quality using a variance term, $\sigma$, which represents the additional uncertainty that, if added uniformly to all data points, would give a reduced $\chi^2$ equal to 1. The median value for all SLSNe is 0.12\,mag. This is similar to typical photometric errors in the data, confirming the good fit quality to the majority of the sample that can be seen by eye in Figure \ref{fig:all_lcs}. Only PTF11rks has $\sigma > 0.2$ at high significance. As we will show below, our posteriors are much narrower than our priors (Table \ref{tab:priors}), meaning that we have significantly narrowed the parameter space relevant to SLSNe. In the appendix, we provide another measure of fit quality for Bayesian models, the Watanabe-Akaike Information Criterion (WAIC) \citep{watanabe2010,gelman2014}; however, this is primarily intended for comparison between different models. This paper focuses only on magnetar models, but the WAIC score is important if one wishes to test these fits against other physical models in future. We will discuss this further in \citet{gui2017}.

\begin{figure*}
\centering
\includegraphics[width=\textwidth]{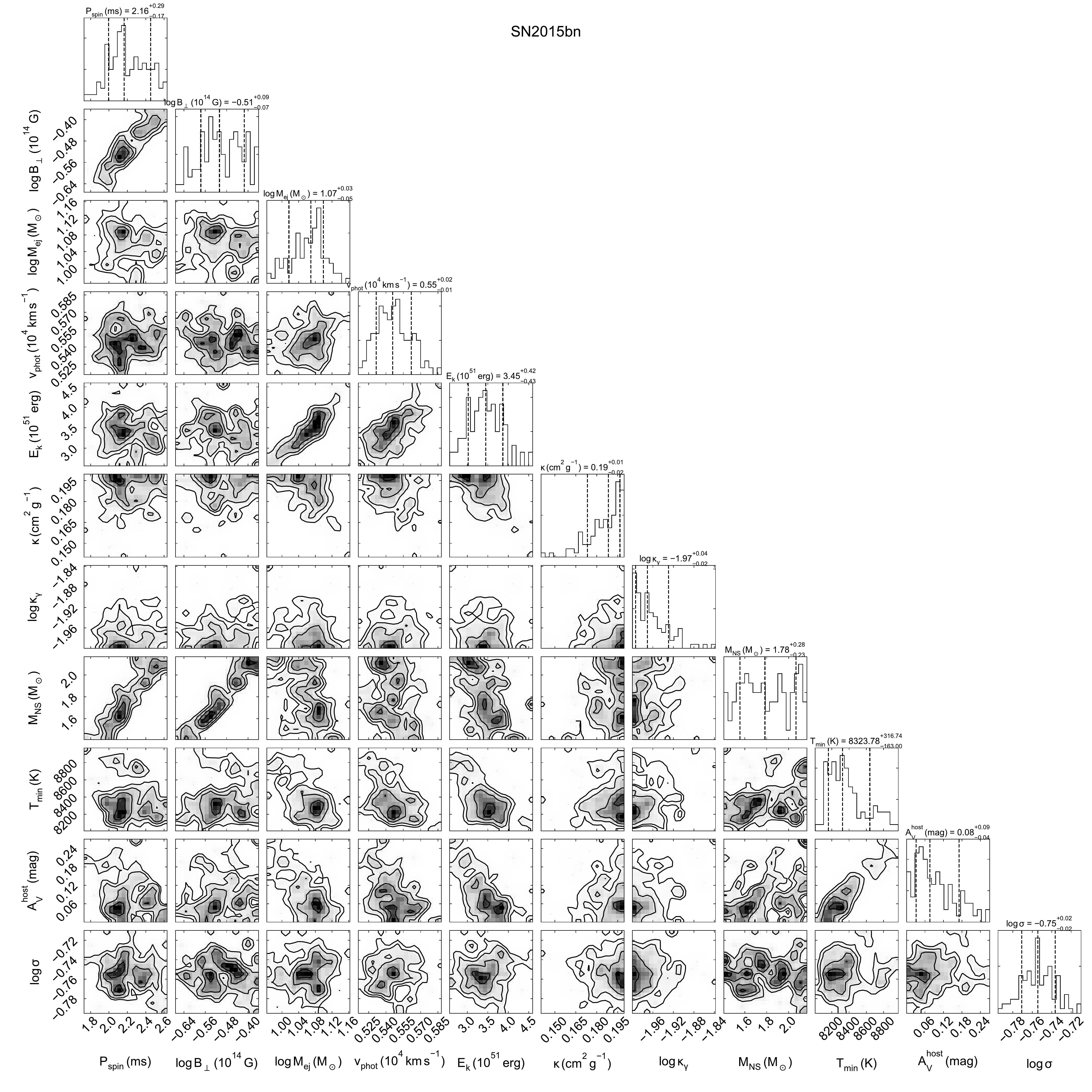}
\caption{Posteriors for magnetar model fit to SN\,2015bn. Medians and 1$\sigma$ ranges are labeled.}
\label{fig:15bn}
\end{figure*}

With the most data in the sample (by a factor $\gtrsim 3$), the event which provides the most stringent test of our model is SN\,2015bn. While the simple model cannot reproduce the `wiggles' most prominent in the bluer bands, it satisfactorily captures the rise, peak and decline over almost 500 days, and provides an excellent match to the colours across 16 filters. We present the posteriors for this fit in Figure \ref{fig:15bn}. Note that we also show the `posterior' for \Ek, derived from those for \Mej\ and \vp. The triangle plot shows degeneracies between some of the parameters. In particular, the values of \P, \B\ and \Mns\ exhibit strong correlations. This is not surprising, as these parameters are related through equations \ref{eq:emag} and \ref{eq:tmag}. Similarly, \k\ and \Mej\ are mostly constrained by equation \ref{eq:tdiff} and hence are also degenerate. The variance parameter in this case has a median $\sigma=0.18$\,mag, which is at the high end for our sample. This reflects the 'wiggles' in the light curve analysed by \citet{nic2016b}.

The most interesting physical parameters are \Mej, \P\ and \B, as these determine what conditions actually lead to SLSNe. Staying with the example of SN\,2015bn, we find \P\,$=2.2$\,ms, \B\,$=3.3\times 10^{13}$\,G, and \Mej\,$=11.7$\,\M. The uncertainties on these quantities are $\approx 10-20\%$ (Table \ref{tab:pars}). The parameter values are similar to the those inferred from previous modeling of the bolometric light curve \citep{nic2016b,nic2016c}\footnote{previous work assumed a spin axis misalignment of 45$^\circ$, so the directly comparable magnetic field is $B_{45} = B_\perp / \sin(45^\circ) = 4.7\times10^{13}$\,G}---however, only a small handful of SLSNe have sufficient UV-optical-NIR data to construct a reliable bolometric light curve for modeling comparable to SN\,2015bn. Furthermore, the ejected mass and kinetic energy (\Ek\,$=3.4 \times 10^{51}$\,erg) are consistent with estimates based on the nebular-phase spectrum \citep{nic2016c,jer2016b}.

The other `nuisance' parameters are typically not well constrained by data, and instead we marginalize over them to determine realistic distributions of the main parameters. However in the case of SN\,2015bn, some of these nuisance parameters are actually quite tightly constrained. The most probable fits have \k\,$\approx 0.2$\,\cmsqperg\ and \kg\,$\approx 0.01$\,\cmsqperg---i.e.~the ejecta are close to fully ionised, but the gamma-ray trapping at late times is low. This suggests substantial leakage of hard radiation from the magnetar out of the ejecta. Only a few events have sufficiently late observations to constrain \kg, but those that do favour similarly low values. The NS mass is not constrained by the model, but is degenerate with both \P\ and \B. The other nuisance parameters, \Av\ and \Tf, show a strong degeneracy with each other, but low values of \Av, as expected in dwarf galaxies, are preferred.

\begin{figure*}
\centering
\includegraphics[width=\textwidth]{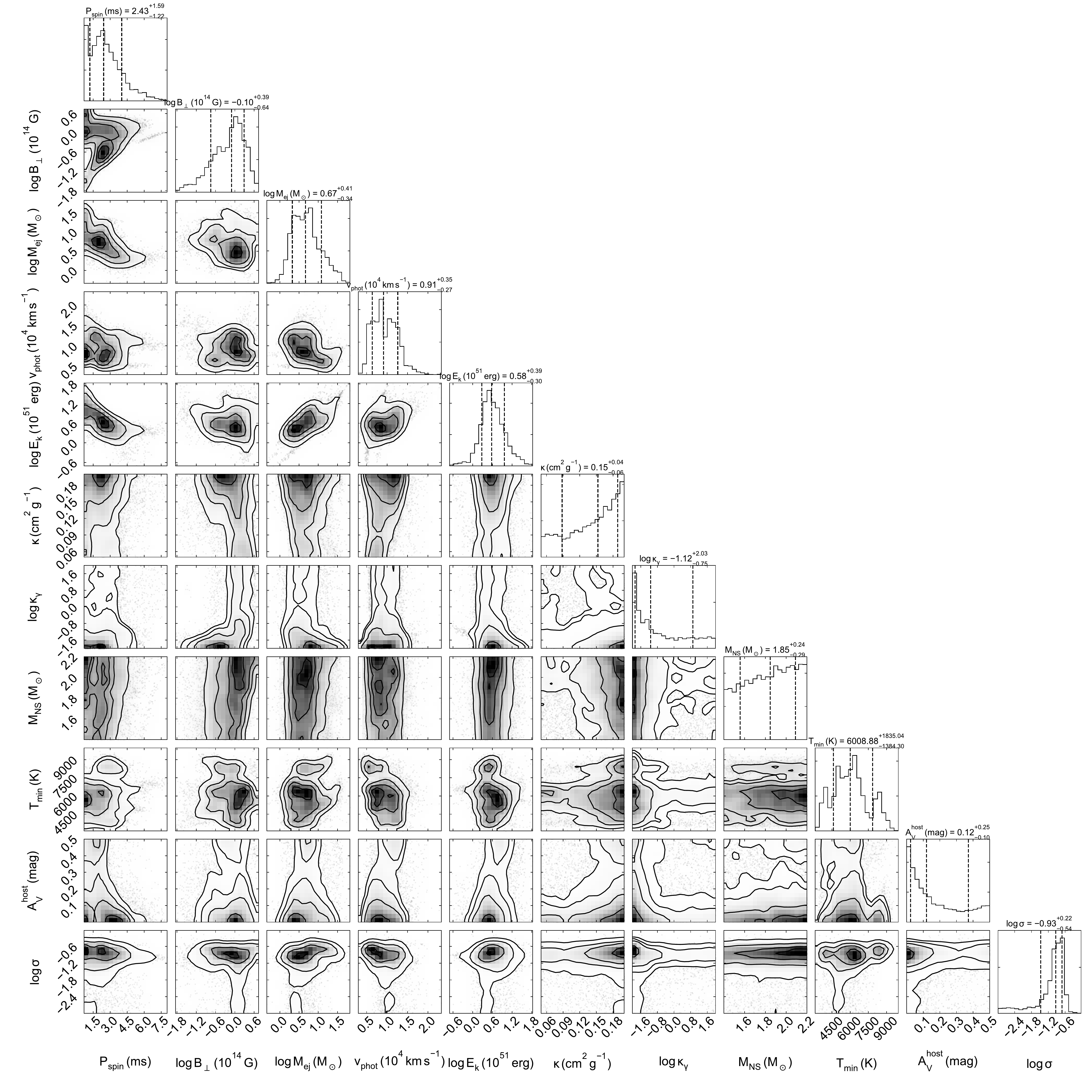}
\caption{Joint posteriors of all model parameters for the full SLSN sample. Medians and 1$\sigma$ ranges are labeled.}
\label{fig:joint}
\end{figure*}

We show the summed posteriors for all fits in Figure \ref{fig:joint}. We find the following median values for key parameters:
\begin{itemize}
\item \P\,$=2.4^{+1.6}_{-1.2}$\,ms
\item \B\,=$0.8^{+1.1}_{-0.6} \times 10^{14}$\,G
\item \Mej\,$=4.8^{+8.1}_{-2.6}$\,\M
\item \Ek\,=$3.9^{+5.9}_{-2.0} \times 10^{51}$\,erg.
\end{itemize}

The most uncertain parameter is \Ek, as in general the kinetic energy for a given mass depends sensitively on the density profile. This introduces a systematic uncertainty that is not included in our quoted statistical errors. To derive \Ek\ we have used the density profile given in equation 12 of \citet{mar2017}, which they showed to give a good match to 1D \texttt{KEPLER} simulations of exploding compact stars. For this density profile, \Ek\,=\,1/2\,\Mej\,\vej$^2$. 

Our median velocity is 50\% lower than the median velocity from \citet{Liu&Modjaz16} at 15 days after maximum light. As mentioned in section \ref{sec:code} this is primarily due to our simplifying assumption of a constant velocity at early times, whereas in reality the velocity at the photosphere decreases as the photosphere recedes in mass coordinate. Our time-averaged ejecta velocities are consistent with the measurements from \citeauthor{Liu&Modjaz16} later than 20-30 days after maximum light.

By assuming that \vej\,=\,\vp, we may be underestimating the kinetic energy, since the photospheric velocity is not necessarily representative of the fastest ejecta. If we instead use the maximum-light velocities measured by \citet{Liu&Modjaz16} to derive \Ek, we find a median value of $8.2 \times 10^{51}$\,erg.

\citet{nic2015b} previously estimated the median ejecta mass in SLSNe as 6.0\,\M---somewhat larger than our new estimate of 4.8\,\M\ (but well within our 1\,$\sigma$ range of 2.2--12.9\,\M). \citeauthor{nic2015b} took the opacity to be 0.1\,\cmsqperg, whereas here we let the opacity vary from 0.05--0.2\,\cmsqperg\ and find a median value of 0.15\,\cmsqperg. Given the difference in opacity (the derived mass scales as \Mej\,$\propto \kappa^{-1}$) these estimates are remarkably consistent, considering the methods used are very different.


In this section we have presented the probability distributions of the key magnetar model parameters for all published SLSNe observed around maximum light. While these posteriors do encompass the typical spin period and magnetic field values derived from previous studies, the realistic error bars and increased sample size allow us to examine exactly where SLSNe occur in this multidimensional parameter space. In contrast to the idea that magnetar models are overly flexible in fitting light curves, we find that the regions of interest are narrow relative to our uninformed priors. In the following section, we will examine what our derived posteriors tell us in terms of the SLSN physics and connection to observables.

\begin{figure*}
\centering
\includegraphics[width=15cm]{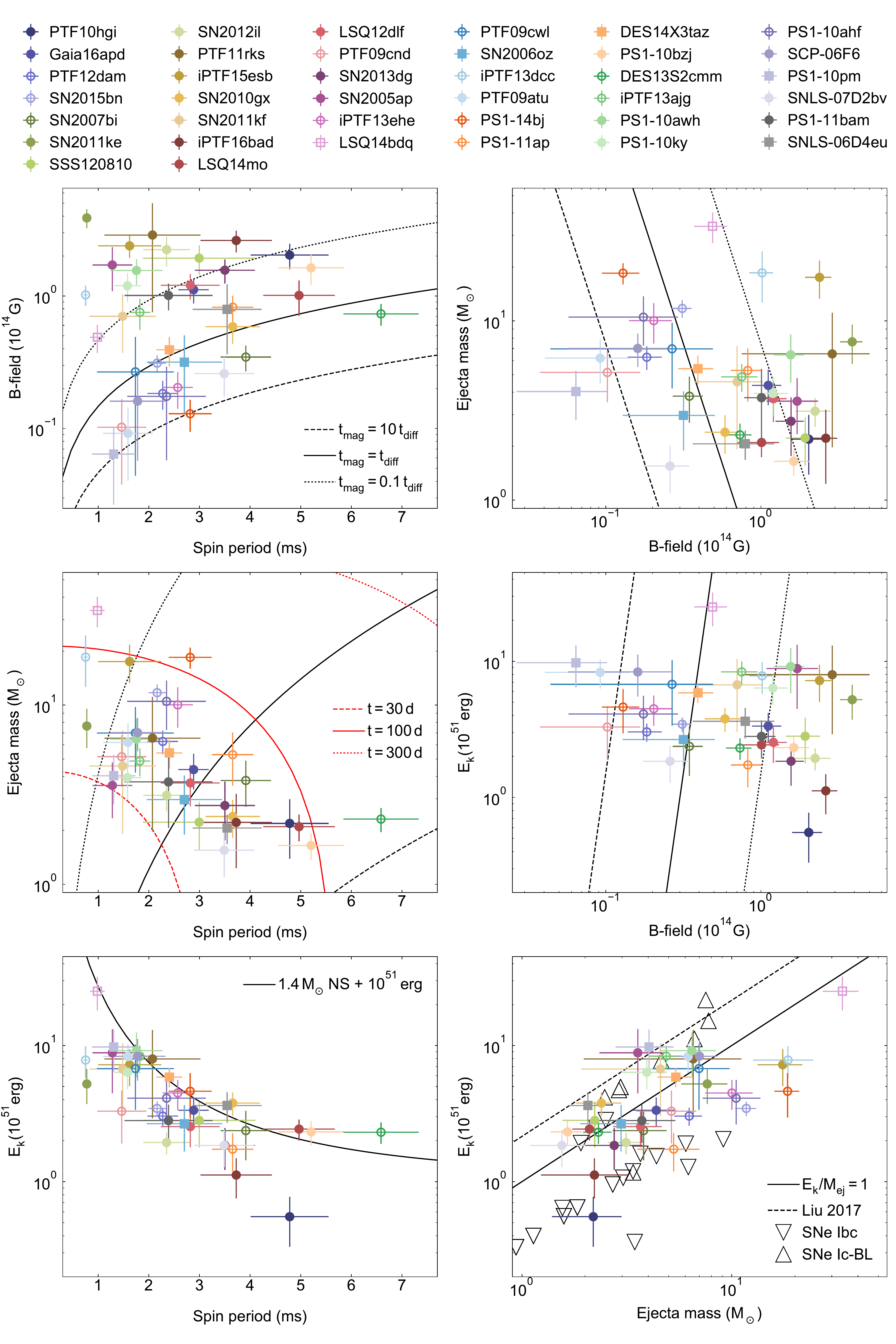}
\caption{Median values and 1-$\sigma$ errors of key parameters (\P, \B, \Mej, \Ek) for all SLSNe. Empty symbols correspond to slowly evolving SLSNe, while squares indicate an observed double-peak in the light curve. Data for other SN types comes from \citet{dro2011} and \citet{tad2014}. The various contours are described in the text.}
\label{fig:bpme}
\end{figure*}

\section{Analysis}
\label{sec:analysis}

\subsection{Fundamental properties}

Having derived a set of physical parameters for each SLSN, we now compare the properties of these events. In Figure \ref{fig:bpme} we plot each pair of variables from our best-fitting \P, \B, \Mej\ and \Ek\ values. Investigating the top 4 panels, there are no obvious strong correlations between parameters, nor any clear signs of separate clusters of events. Those with an early bump observed in their light curve are plotted with different symbols, but show no separation from the general distribution. Rather, the SLSNe appear to populate certain regions relatively uniformly, while avoiding others.

To show why this is the case, on each panel we plot lines of constant $t_{\rm mag}/t_{\rm diff}$; i.e.~we set equation \ref{eq:tmag} equal to equation \ref{eq:tdiff} up to some constant, $A$. Lines of $t_{\rm mag}/t_{\rm diff}=A$ follow the relation
\begin{equation}
\left(\frac{P}{\rm ms}\right)^2\,\left(\frac{B_\perp}{10^{14}\,{\rm G}}\right)^{-2} = 20.7\,A\,\left(\frac{M_{\rm ej}}{{\rm M}_\odot}\right)^{3/4}\,\left(\frac{E_{\rm K}}{10^{51}\,{\rm erg}}\right)^{-1/4}.
\end{equation}
On each panel, we assume median values for the variables that are not plotted, therefore these lines should be considered somewhat fuzzy in reality. However, it is striking that the vast majority of SLSNe fall between $0.1\lesssim t_{\rm mag}/t_{\rm diff}\lesssim 10$.

This supports previous work suggesting that SLSNe result when the engine timescale matches the ejecta diffusion timescale \citep{kas2010,met2015,nic2015b}. We find that this condition can be relaxed in the case of the shortest spin periods: in this regime, although the magnetar loses energy relatively quickly, the energy reservoir is sufficiently deep that a significant amount of rotational power remains after a diffusion timescale (i.e.~around light curve maximum). In the opposite extreme, it appears that it is very difficult to form SLSNe with $t_{\rm mag} \gg t_{\rm diff}$. In this case, the rotational energy is lost too gradually to have a large impact on the peak luminosity.

\begin{figure}
\centering
\includegraphics[width=\columnwidth]{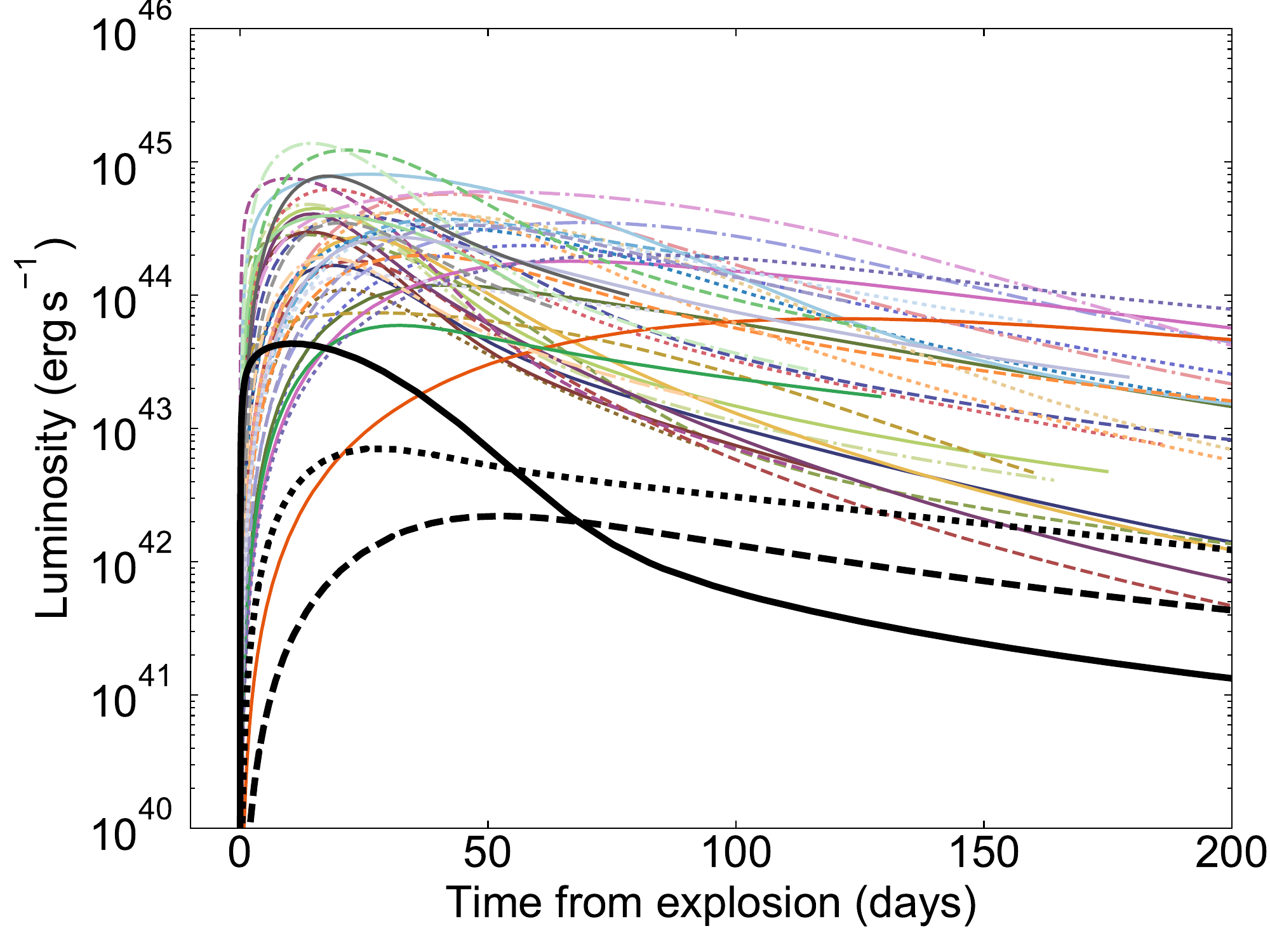}
\includegraphics[width=\columnwidth]{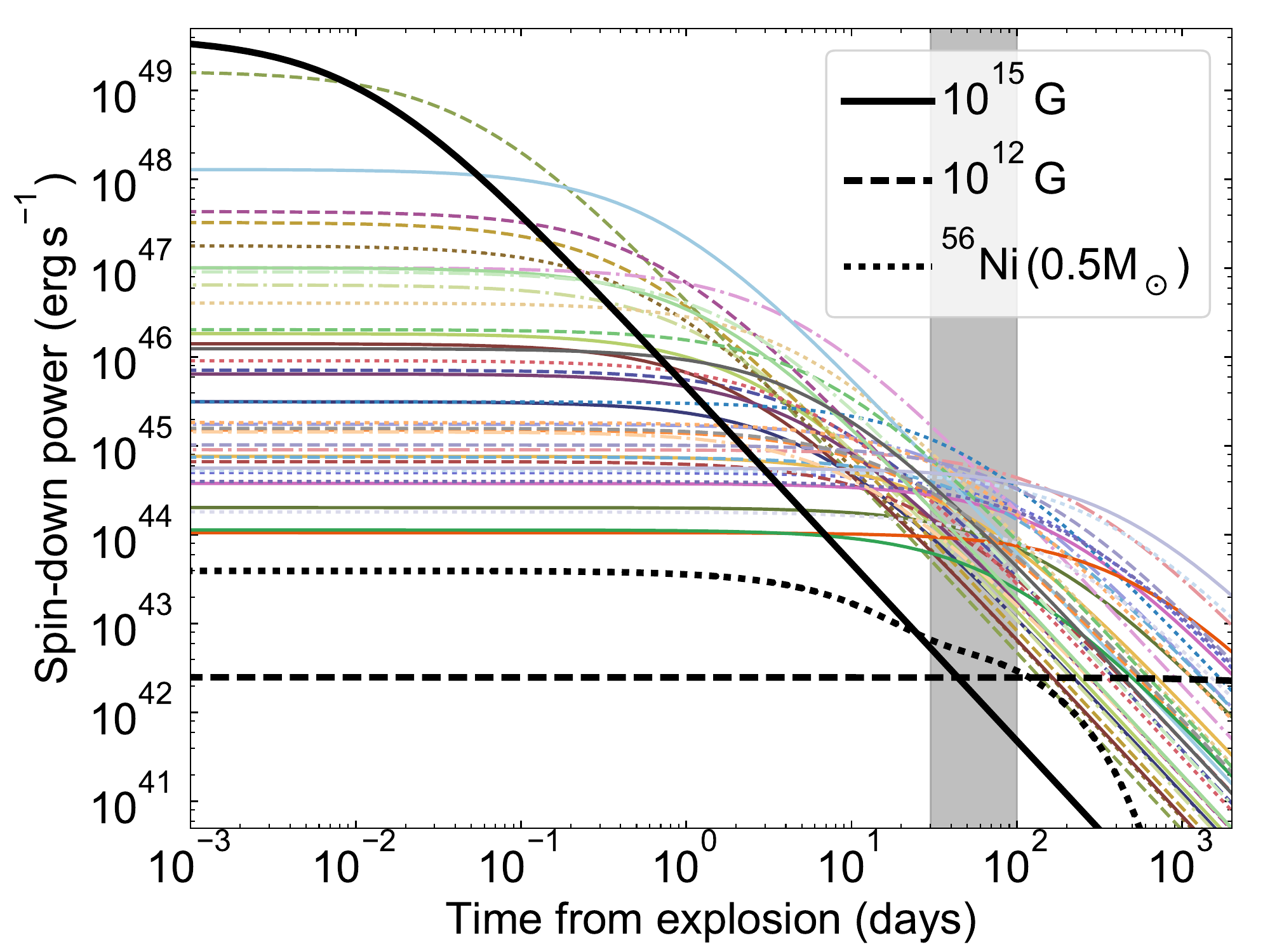}
\caption{Top: bolometric light curves from our model fits. Bottom: the spin-down power of the magnetar in each case. Despite a range of initial values, the spin-down power at maximum light converges to $\sim 10^{44}$\,\ergpers. Magnetars with \B\,$\ll 10^{13}$\,G or \B\,$\gg 10^{14}$\,G do not provide enough energy at 30-100\,d after explosion to power SLSNe, but in the latter case could generate a luminous transient with a shorter timescale (and could strongly affect the kinetic energy).}
\label{fig:eng}
\end{figure}

We show this explicitly in Figure \ref{fig:eng}, where we plot the bolometric light curves derived from each model fit, and the corresponding spin-down luminosities for the best-fit \P\ and \B. We find that the light curves peak at $\sim30-100$\,d after explosion, which we highlight on the lower panel showing the spin-down power. What is remarkable is that many different initial conditions (combinations of \P\ and \B), spanning almost 5 orders of magnitude in power input at the time of magnetar birth, converge to a typical power of a few $\times 10^{44}$\,\ergpers\ by the time energy can escape the ejecta. This gives the characteristic observed maxima according to the well-known `Arnett Law': at the light curve peak, luminosity is equal to the rate of power input.

We also plot in Figure \ref{fig:bpme} the engine power for hypothetical events with significantly longer and shorter spin-down times. In the case of a very powerful ($10^{15}$\,G) magnetic field, giving a very fast spin-down as would be required for example in a GRB-SN, the initial energy power is very large, but is lost too quickly to dominate the luminosity a month after explosion. For a weaker field ($10^{12}$\,G), the resultant input power is deposited over such a long time that there is never enough instantaneous power to produce an unusually bright light curve. In both cases, the luminosity at 30--100\,d is subdominant to \Ni\ decay, assuming a typical nickel mass of $\sim 0.5$\,\M\ for GRB-SNe. Therefore magnetars with much stronger or weaker magnetic fields would not produce SNe that look like the SLSNe in our sample---and in many cases would be difficult to distinguish from normal SNe.

The spin-down functions in Figure \ref{fig:eng} are also suggestive in that (following the Arnett Law) a SLSN peaking on a shorter timescale, say $\lesssim 10$\,d, could in principle be even more luminous than those we have observed. This raises the question of how fundamental is the typical light curve timescale of 30--100\,d? \citet{nic2015b} showed that there is a simple but surprisingly tight relationship between the spin-down timescale, the diffusion timescale, and the light curve width: $t_e \simeq t_{\rm mag} + t_{\rm diff}$, where $t_e$ is the time for which the SN is within a factor $e$ of its peak luminosity. Returning to Figure \ref{fig:bpme}, we plot lines of constant $t_e$ on the middle-left panel (again using equations \ref{eq:tmag} and \ref{eq:tdiff}). The data are bracketed by 30\,d\,$<t_e<100$\,d, as expected. More revealingly, these lines are steep functions of \P\ and \Mej. Getting a transient with a significantly shorter timescale requires both \P\,$\lesssim 2$\,ms and \Mej\,$\lesssim 3$\,\M. While not ruled out on physical grounds, such systems should be rare, given this small corner of parameter space, and possibly difficult to find and classify due to their short light curve timescales. However, this may be a promising target for future surveys, particularly at high redshift where increased luminosity increases the search volume, and time-dilation makes the light curve width more amenable to typical survey cadences. In the opposite case ($t_e>100$\,d), these high-mass, long-spin-down transients should be faint, slow evolving, and difficult to detect. Thus, their absence from the data could be indicative of a selection bias rather than intrinsic rarity.

In the lower panels of Figure \ref{fig:bpme}, we investigate the impact of \Ek.
There is a clear correlation with \P\; however, this is trivial as the magnetar rotational energy provides most of the available kinetic energy (section \ref{sec:const}). This is demonstrated explicitly by the curve showing the available rotational energy from a 1.4\,\M\ neutron star, which provides an upper envelope to the data.

In the final panel of Figure \ref{fig:bpme}, we also plot \Ek\ against \Mej. Again with the caveat that our kinetic energies are technically lower limits due to the simplified velocity profile we adopt, we see that SLSNe lie close to a specific kinetic energy $E_{\rm K}/M_{\rm ej}\approx 1$. We compare these to literature values for normal and broad-lined Type Ic SNe from \citet{dro2011} and \citet{tad2014}, which we note have been derived following similarly simplistic assumptions about the ejecta structure. The $E_{\rm K}/M_{\rm ej}$ ratios for SLSNe appear to span the range between normal and broad-lined Type Ic SNe.

\begin{figure}
\centering
\includegraphics[width=\columnwidth]{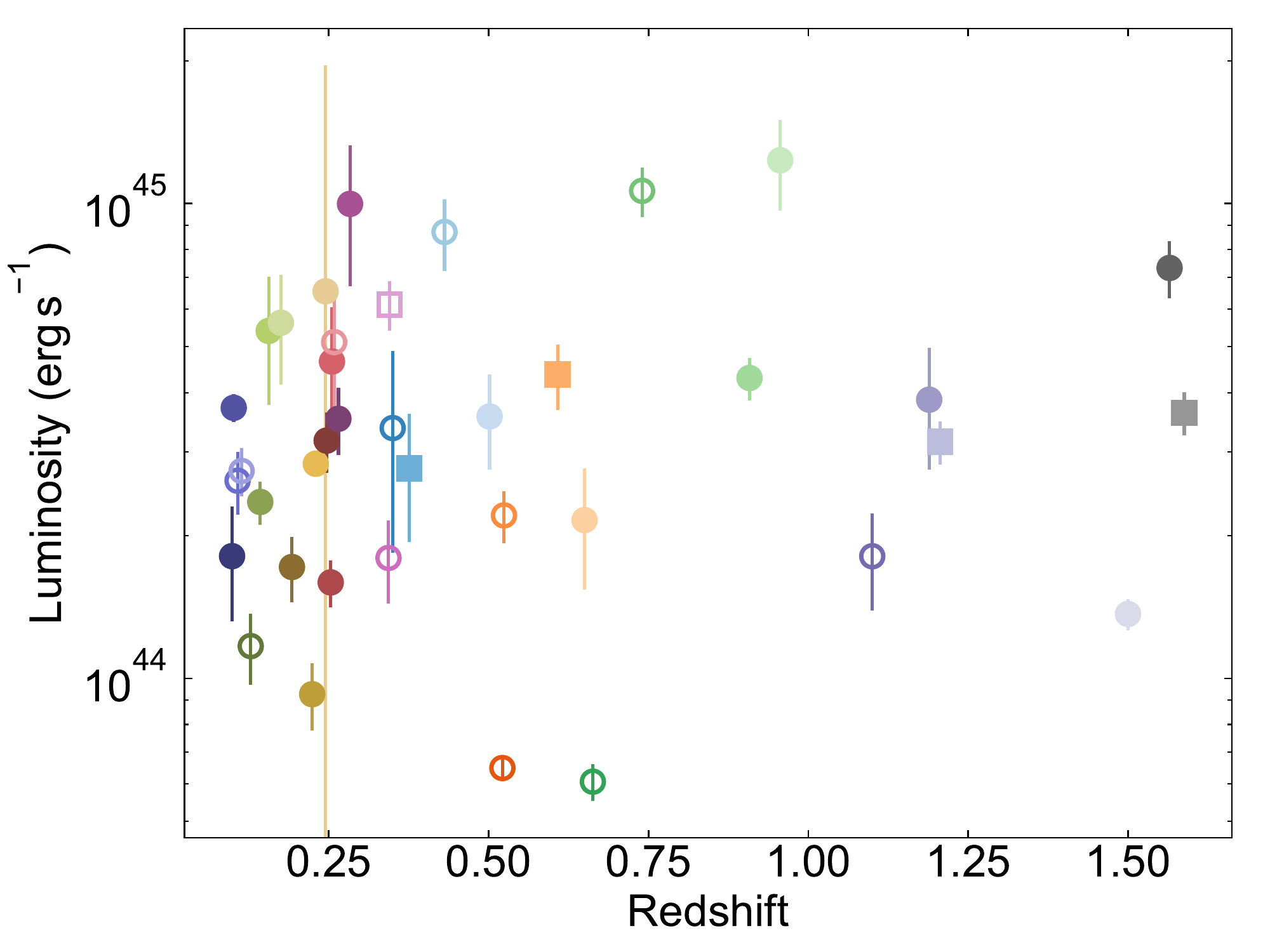}
\includegraphics[width=\columnwidth]{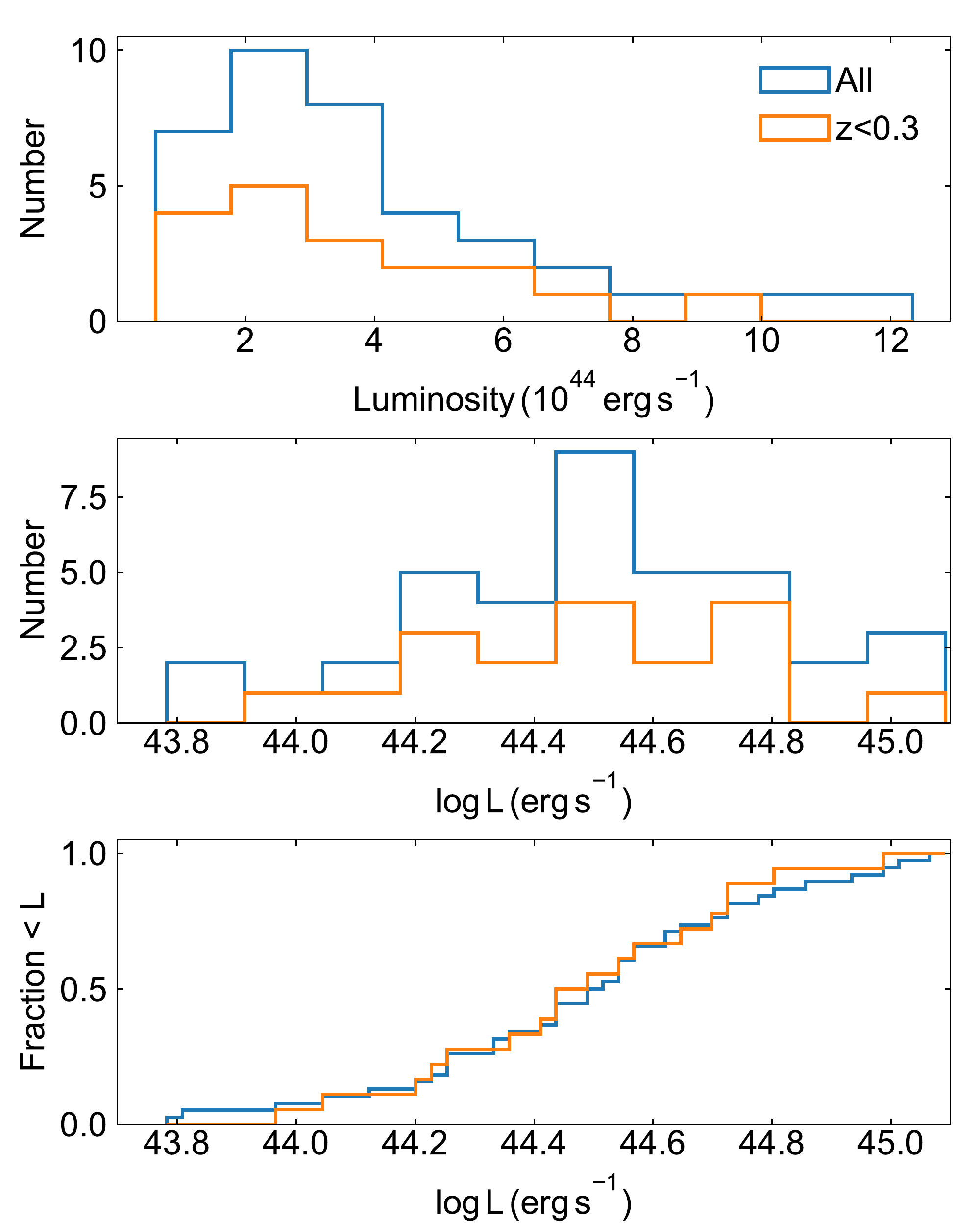}
\caption{An approximate luminosity function for SLSNe. The median luminosity of our sample is $3.2 \times 10^{44}$\,\ergpers. Despite the different selection effects present in the many surveys that discovered these events, the luminosity function appears to show little evolution with redshift.}
\label{fig:lum}
\end{figure}

Perhaps this is not surprising, as SLSNe show spectroscopic similarity to both classes of H-poor SNe \citep{pas2010}. What is particularly compelling, however, is that this supports an increasingly popular picture wherein SNe Ic, SLSNe, and broad-lined SNe Ic/GRB-SNe form a hierarchy of increasing engine power and/or decreasing engine timescale. In SLSNe, the neutron star remnant, due to a combination of large rotational energy and optimal spin-down time, deposits thermalized energy around the light curve peak as well as kinetic energy from the magnetar wind. In broad-lined and GRB SNe, the spin-down is much faster (or the engine is a black hole rather than a neutron star) such that most energy goes into expansion, and in the latter case a relativistic jet. Thus their kinetic energies may be even greater than those in SLSNe. We note, however, that if we instead take the SLSN velocities from \citet{Liu&Modjaz16} in our calculation of \Ek, we derive $E_{\rm K}/M_{\rm ej}\approx 2$ for many SLSNe (shown approximately as a dashed line in Figure \ref{fig:bpme}), comparable to GRB SNe. As recently noted by \citet{pre2017}, a true derivation of \Ek\ requires detailed spectroscopic models to probe the density structure.

Finally, we note that the typical masses of SLSN ejecta are significantly larger than for most SNe Ic, but appear to be similar to broad-lined SNe Ic (though the sample size for the latter is small), in agreement with the findings of \citet{nic2015b}. The range we find is $2\lesssim M_{\rm ej} \lesssim 20$\,\M. The exception is LSQ14bdq, for which our best-fit model favours a more massive ejecta with $\sim 30$\,\M. This is similar to the mass estimated by \citet{nic2015a}, however we note that this event became Sun-constrained just before reaching maximum light, and so the shape of the peak is not well constrained.

\begin{figure*}
\centering
\includegraphics[width=15cm,trim={0 53cm 0 0},clip]{BPME.pdf}
\includegraphics[width=15cm]{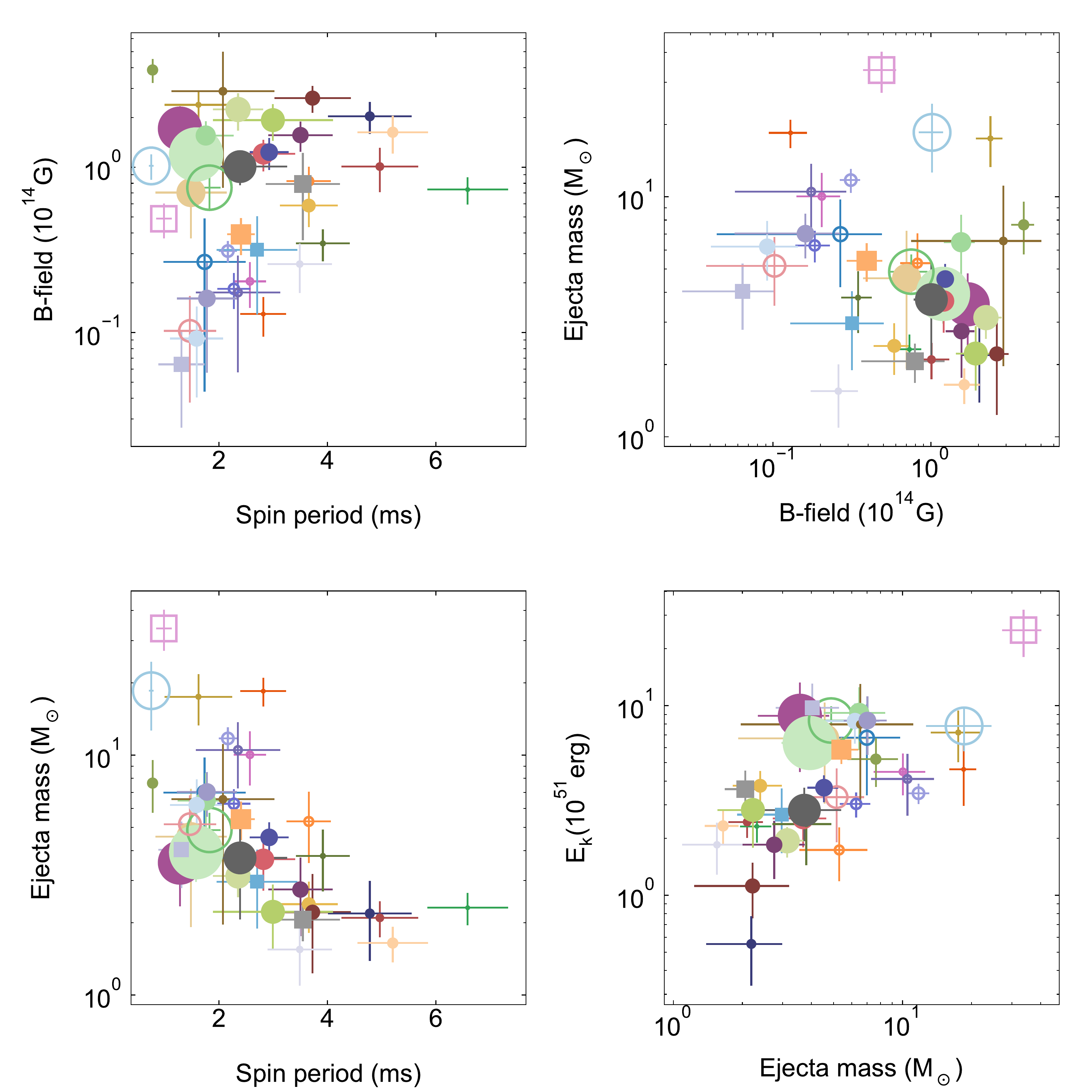}
\caption{A subset of the comparisons from Figure \ref{fig:bpme}, but here symbols have been scaled in proportion to the peak luminosity of each SLSN. The brightest events tend to have $P<2$\,ms and \B$\approx 10^{14}$\,G.}
\label{fig:Lprops}
\end{figure*}

\subsection{Connecting observables to physical properties}

We show the distribution of peak luminosities in our fits in Figure \ref{fig:lum}. This can be used to construct a rough luminosity function, though it is hard to account for observational bias given that these SLSNe come from a wide range of surveys. In particular, physically-related events may extend to lower luminosities but could be missed due to a classification bias against events that are not formally `super-luminous' (i.e.~$M<-21$). Such selection effects are difficult to acount for. Interestingly, the peak luminosities in our sample show negligible evolution with redshift. In the top panel we plot luminosity against redshift. The luminosity function for events at $z<0.3$, which is roughly the volume within which all major surveys are sensitive to SLSNe, is indistinguishable from that for the whole sample. We find a median peak luminosity of $3.2\times 10^{44}$\,\ergpers\ for the entire sample, with a 1$\sigma$ range $1.7-6.2\times 10^{44}$\,\ergpers.

\begin{figure*}
\centering
\includegraphics[width=15cm]{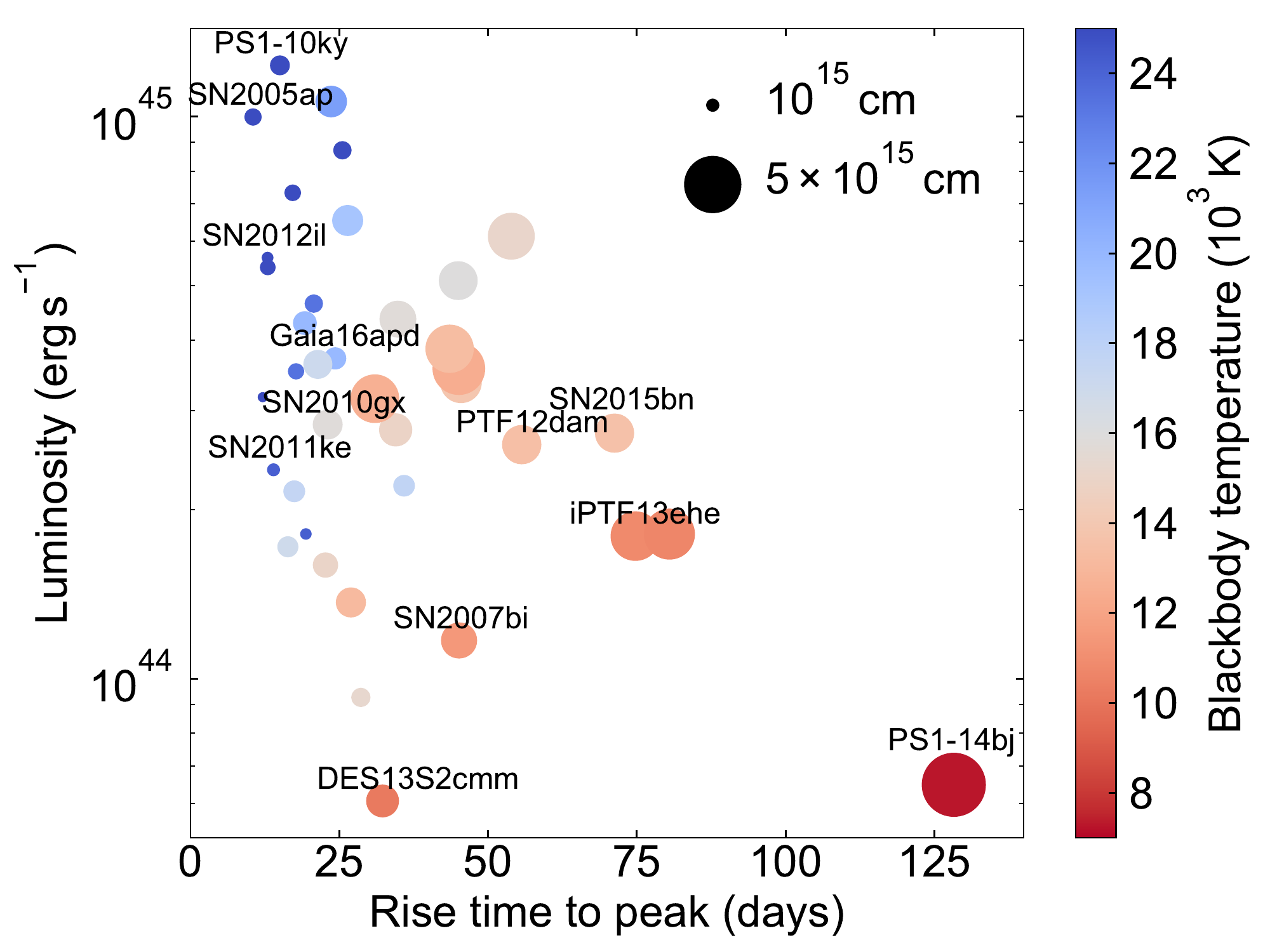}
\caption{Luminosity vs rise time for the SLSN light curve fits. Symbols are scaled according to the area of the photosphere and coloured according to temperature. Note that the bolometric light curves generally peak earlier than the optical light curves, thus the temperatures at optical maximum will be lower than those shown here. Bluer events result from high luminosities and short rise times (hence more compact photospheres).}
\label{fig:col}
\end{figure*}

It is interesting to ask which physical parameters are most important in determining the peak luminosity. In Figure \ref{fig:Lprops}, we repeat some of the panels from Figure \ref{fig:bpme} but with each datapoint scaled in proportion to the maximum luminosity of that SLSN. It is immediately apparent that no one parameter is a perfect predictor of luminosity. However, the brightest events fall in the region with short spin period and relatively high magnetic field. Ejecta mass seems to have relatively little effect, while the slight preference for higher kinetic energy is likely a reflection of the faster spin in such events.

One of the defining characteristics of SLSNe is their blue spectra, particularly at early times. However, recent observations of the UV spectrum of Gaia16apd from \citet{yan2016} have demonstrated an unexpected degree of diversity in the UV properties of SLSNe that imply a range of photospheric temperatures \citep{nic2017}. At the other extreme, PS1-14bj displayed a much redder spectrum at maximum than the rest of the population \citep{lun2016}. Despite this diversity, it is clear from our light curve fits that our simple SED model shown in Figure \ref{fig:sed} can reproduce the UV-optical colours at peak for all of the events in our sample.

In Figure \ref{fig:col}, we connect the observed colour diversity to the luminosities, rise times and radii of the SLSNe. The time taken to reach maximum light is sensitive to both the spin-down time and the diffusion time. The peak luminosity is additionally sensitive to the initial spin period, while the radius reached by this phase also depends on the velocity. Combining these factors, we show that the bluest SLSNe are those that reach a bright peak luminosity after a short rise, such that the photosphere is still relatively compact ($\sim10^{15}$\,cm) and the engine luminosity is higher up the spin-down curve (Figure \ref{fig:eng}).

This group includes the UV-bright Gaia16apd, but in fact our fits imply that a number of SLSNe should have had even higher temperatures at maximum light. However, most do not have the well-sampled UV data of Gaia16apd. One interesting exception in PS1-10ky \citep{chom2011} -- this is the most luminous model fit in the sample, and is predicted to have a high photospheric temperature at peak, but at $z=0.956$ the rest-frame UV is well sampled, and the colours are redder than for Gaia16apd. The reason for the discrepancy here is that our fit prefers a fairly large host galaxy extinction, \Av\,$=0.39$\,mag, which strongly reddens the UV colours.

\begin{figure*}
\centering
\includegraphics[width=15cm]{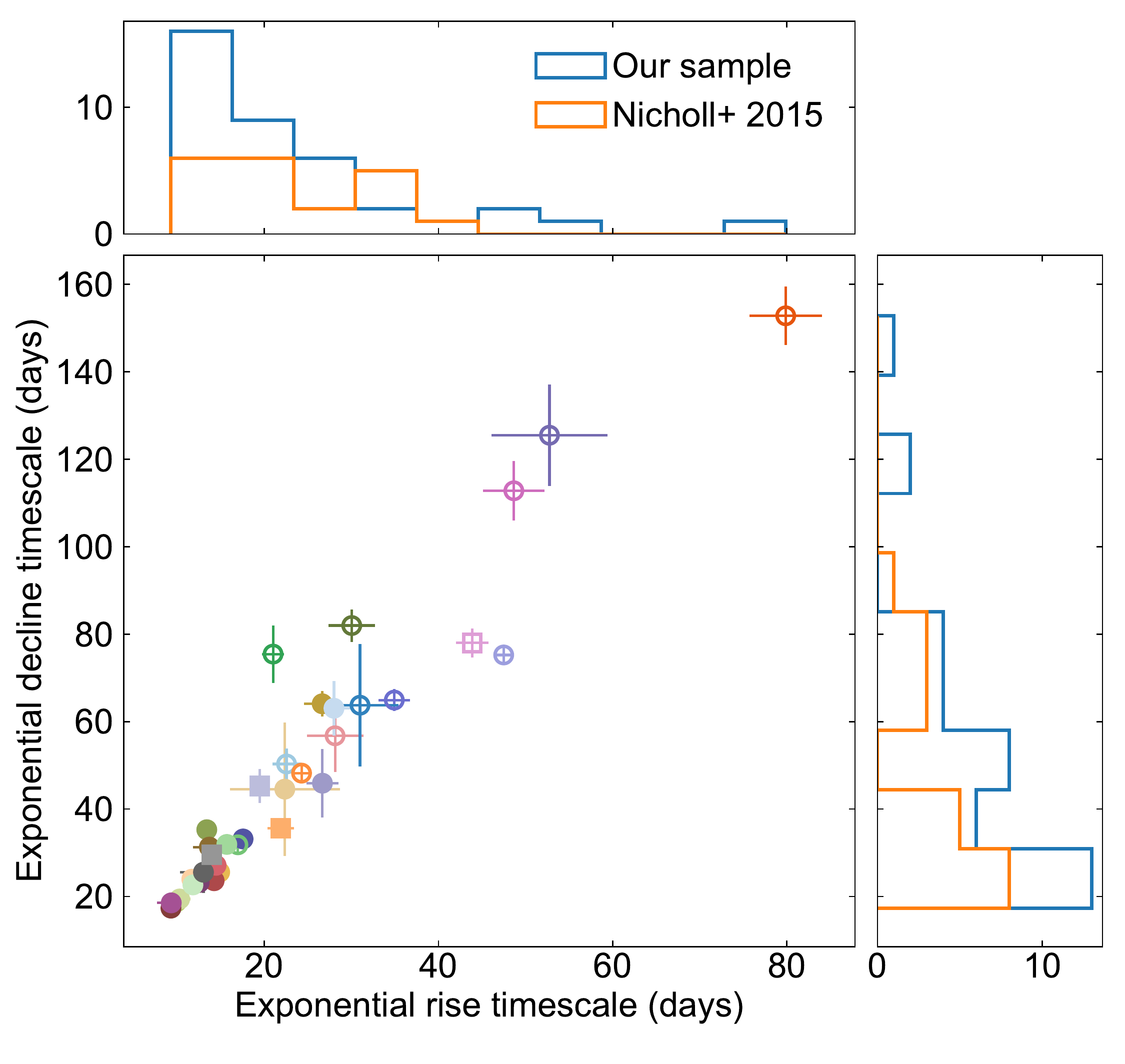}
\caption{Bolometric rise-decline relation \citep{nic2015b} for our light curve fits. The models lie close to the locus $t_{\rm rise}=t_{\rm dec}$. Top and right panels show the projected histograms. With our large sample, the distribution of timescales shows no sign of bimodality.}
\label{fig:time}
\end{figure*}

On the other hand, SLSNe with longer rise times are redder (though 'red' in this context generally means a maximum-light temperature $\gtrsim 12000$\,K) because the ejecta have had more time to expand, and their engines provide energy more gradually rather than rapidly powering an early maximum. The most extreme example by far is PS1-14bj, which has a temperature of only 7000\,K due to a lower luminosity and significantly more time to expand (though with a lower velocity than most of the other SLSNe, the radius---$5.6\times 10^{15}$\,cm---is only a factor of $\sim2$ greater than the mean).

This has an important implication for the spectroscopic evolution. It has been pointed out by \citet{yan2015} and \citet{lun2016} that iPTF13ehe and PS1-14bj had spectra before maximum light that resembled typical slow-declining SLSNe such as PTF12dam and SN\,2007bi at 50 days after maximum. We assert that this is a simple and intuitive consequence of these objects having already cooled significantly before they reach maximum light, due to their longer rise times. The point is that using the time of maximum light to define the phase of the spectroscopic evolution can be misleading if one is not careful: the most important physical parameter affecting the spectrum is the \emph{temperature}, which is a function of the rise time and velocity in addition to luminosity.

Our modeling makes a somewhat weaker prediction for the spectroscopic evolution at late times. Using equation \ref{eq:tneb}, we can estimate the time at which SLSN ejecta become optically thin, i.e.~evolve into the nebular phase. We find that the optical depth typically falls below 1 between 130-375\,d after explosion (1 sigma range) with median of 220\,d. The upper end of this range corresponds to the most massive SLSNe, and indeed is reasonably well matched to the very slow late-time evolution of SN\,2015bn \citep{nic2016c}. Nebular spectroscopy of the faster events is of course more challenging, but should soon be available to test the lower end of our suggested range.

\begin{figure*}
\centering
\includegraphics[width=15cm]{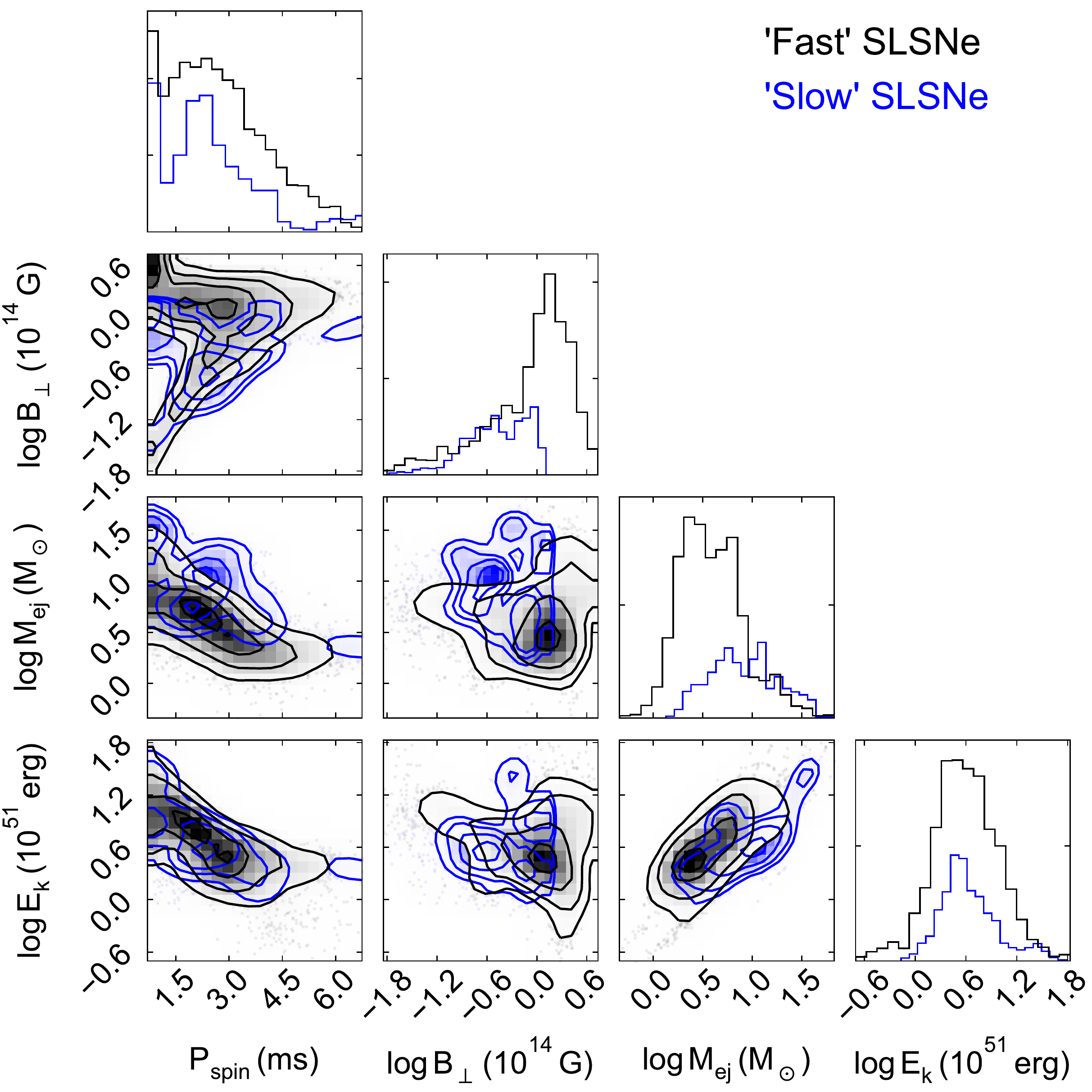}
\caption{Comparison of posteriors for key physical parameters, between SLSNe with fast and slow light curve evolution. The kinetic energy posterior is derived from ejecta mass and velocity. Slower evolving SLSNe favour lower $B$-field, higher \Mej. Their magnetar spin periods and kinetic energies are indistinguishable on average (and therefore the slow events have a lower specific energy), as are all other model parameters (not shown here).}
\label{fig:fastslow}
\end{figure*}

One caveat is that in reality, recombination may hasten the transition to the nebular phase. For example, although they still retain a clear continuum in the spectrum, well-observed (and mostly slow-declining) SLSNe such as SN\,2015bn, SN\,2007bi, PTF12dam, LSQ14an and Gaia16apd have shown [\ion{Ca}{2}] emission between $\sim 50-100$\,d after maximum light \citep{gal2009,nic2013,nic2015b,kan2016,nic2017,ins2017}. Polarimetry of SN\,2015bn from \citet{lel2017} indicates that this originates in an outer part of the ejecta, perhaps where the density is lower or the ejecta have already recombined.

\subsection{Fast vs slow SLSNe}

We now return to the question of whether SLSNe comprise separate sub-populations of fast and slowly declining events, as originally suggested by \citet{gal2012} based on the observations available at the time. \citet{nic2013} showed that the spectrum of slowly evolving SLSNe at early times closely resemble the more common faster events, and proposed that they were all related. The statistical study of \citet{nic2015b} found that there was not a significant gap in timescales between two sub-populations, though there did seem to be a possible lack of SLSNe with decline timescales of $\sim 50$\,d. \citet{kan2016b} suggested that Gaia16apd falls in this gap. However, \citet{ins2017} suggested that some spectroscopic differences between fast and slow events may be significant (though there is a potential bias in that slow declining events generally have much better data to probe their subtleties).

In Figure \ref{fig:time} we show the distributions of rise and decline times for our model fits. There is no evidence for separate populations of fast and slow declining light curves. We compare to the distributions measured (using a model-independent method) by \citet{nic2015b}. Our distributions are largely consistent with the previous results, but the larger sample size washes out the hints of bimodality that were visible in the smaller sample.

We also investigate any differences in the derived fit parameters for events that have been described in the literature as `slow'. As was apparent from Figure \ref{fig:bpme}, there is generally a lot of overlap between the locations in parameter space of fast and slow SLSNe, and no clear offsets in any one parameter. We make this more explicit in Figure \ref{fig:fastslow}, where we show the full posterior distributions for the main physical parameters separately for fast and slow events. The slower events favour a combination of low \B\ and larger \M, but in neither case is there a clear offset from the fast events; slow events simply favour the tails of continuous distributions. Moreover, the distributions in \P\ and \Ek\ are virtually identical for the two subsamples. In particular, \Ek\ should trace the explosion mechanism, and seems to be consistent with a common formation channel for all SLSNe.

Overall our results support a picture where fast and slow SLSNe form a continuum in timescales, determined by the range in engine and diffusion timescales that can result from relatively modest differences in the ejecta mass and magnetic field. Some differences in spectroscopic properties (particularly colour) between the fastest and slowest events is expected given the wide range of rise times, as this leads to a diversity in photospheric temperatures and radii at maximum light. We showed this explicitly in Figure \ref{fig:col}.

\subsection{On the possible connection to fast radio bursts}

Another transient phenomenon that has recently generated a lot of interest is the new class of fast radio bursts \citep[FRBs;][]{Lorimer+07}. After the discovery that at least one FRB repeats \citep{Spitler+16} and inhabits a dwarf galaxy that is indistinguishable from typical SLSN hosts \citep{Chatterjee+17,Tendulkar+16}, \citet{met2017} proposed that FRBs originate from the magnetar remnants of young SLSNe. \citet{nic2017b} then demonstrated that the FRB rate can match the SLSN rate if such magnetars produce FRBs for a few decades--centuries after birth (matching physical arguments from \citeauthor{met2017} based on opacity and energetics). \citet{nic2017b} were agnostic on how magnetars would produce these FRBs, but from our model fitting here we can infer the spin power at arbitrary times. We find that at 10 years after explosion, the spin-down luminosities range between $10^{39}-10^{43}$\,\ergpers. Given that FRBs can emit $\sim 10^{38}$\,erg in 1\,ms, the typical spin power is insufficient to power FRBs. This suggests that an alternative energy source, such as the magnetic powering argued by \citeauthor{met2017}, may be more likely.

\subsection{Relation to host galaxy properties}

\begin{figure*}
\centering
\includegraphics[width=15cm]{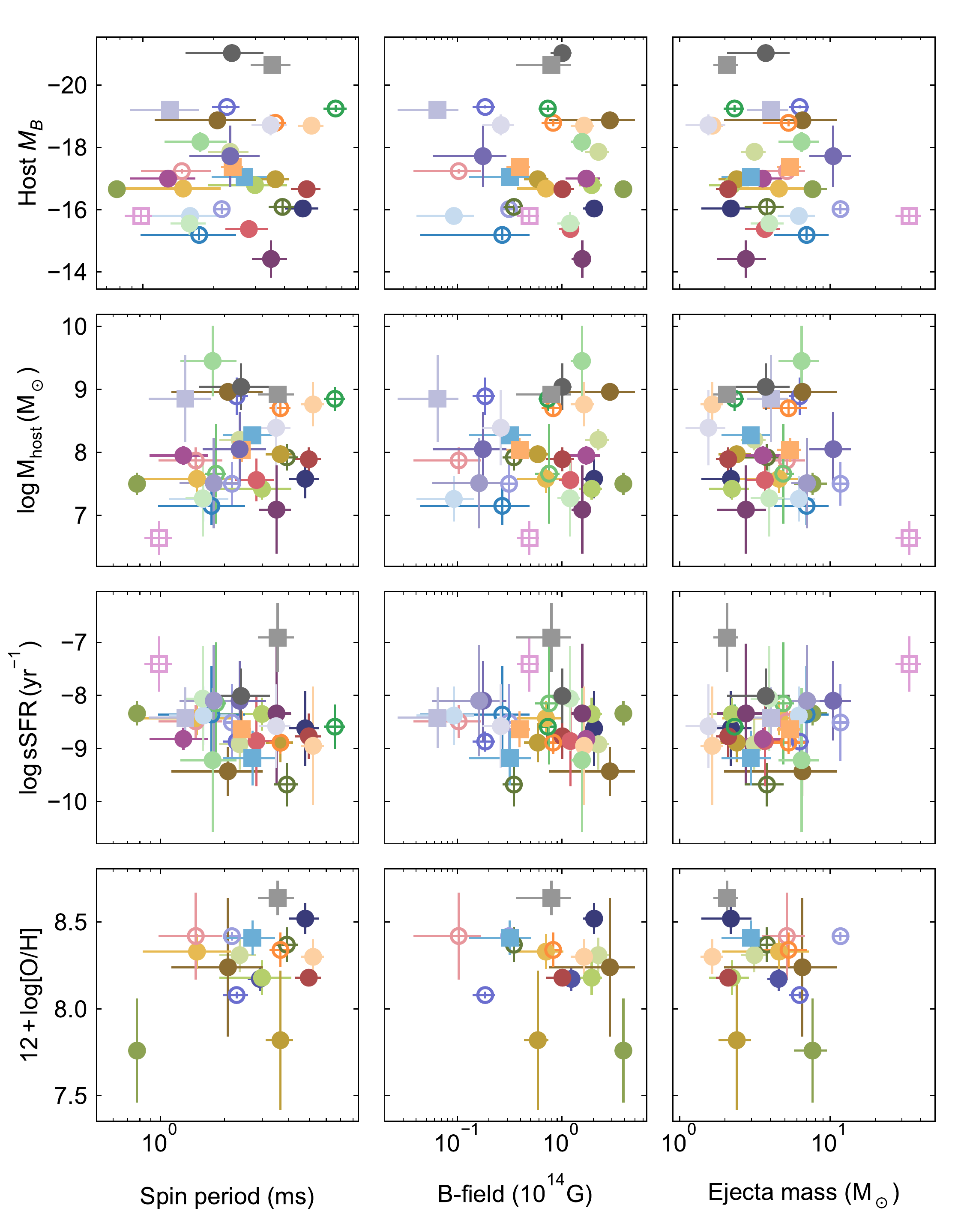}
\caption{Comparison between SLSN magnetar parameters and host galaxy properties. We do not find any significant correlations between SLSN and host properties.}
\label{fig:host}
\end{figure*}

\citet{chen2016} used the magnetar spin periods published in the literature for a sample of nearby SLSNe to compare against the metallicities of their host galaxies. For a sample of 10 objects, they found a possible correlation in the sense that faster rotation may occur in lower metallicity environments. The important implication would be that SLSN progenitors rotate faster at lower metallicity, perhaps due to reduced mass (and hence angular momentum) loss. However, the authors acknowledged that their sample was too small to draw strong conclusions.

We test this relation here using our larger and uniform sample. Figure \ref{fig:host} shows the oxygen abundance compared to \P, \B\ and \M\ for 17 SLSNe in our sample that have measured metallicities in the same scale (T.-W.~Chen, private communication). These abundances are reported in the $R_{23}$ calibration of \citet{kob2004}. We do not find a statistically significant relation between metallicity and any of our magnetar parameters.

While measuring metallicity requires host galaxy spectra (which can be challenging for such faint and often distant galaxies), large photometric samples of SLSN hosts are now available. \citet{schu2016} reported properties for the host galaxies of 53 Type I SLSNe, derived from SED modelling. We compare our magnetar parameters with the host galaxy absolute magnitude, stellar mass and specific star-formation rate for the 33 host galaxies from \citeauthor{schu2016} corresponding to SLSNe in our sample. However, as with metallicity, we do not find any correlations with host galaxy properties.  This indicates that within the range of environments that support SLSN production, there is no strong effect of the metallicity or any other parameter on the details of the engine and/or ejecta properties. 

\citet{per2016}, \citet{chen2016} and \citet{schu2016} recently proposed a metallicity `cutoff' for SLSN host galaxies. Below a threshold metallicity $Z \lesssim 0.5$\,\Z, the SLSN rate per galaxy appears to show no correlation with metallicity, but is sharply suppressed at higher metallicity. Our findings here, that the physical parameters of SLSNe are not correlated with those of their host galaxies, is consistent with this picture, and seems to indicate that as long as the environment is sufficiently metal-poor to allow SLSN production, the full magnetar parameter space is available to these events.

\section{Towards an understanding of the progenitors}
\label{sec:progenitors}

The distributions of explosion parameters reflect the properties of the progenitor stars. Here we use our posteriors in \Mej, $P$ and $B$ to consider the question of what kinds of stars lead to SLSNe. We do this using an order-of-magnitude comparison to the core-collapse SN rate, of which SLSNe comprise a fraction $\sim$\,few\,$\times 10^{-4}$.

Almost all SLSNe occur in galaxies with metallicity $Z \lesssim 0.5$\,\Z\ \citep{per2016,chen2016,schu2016}. Star formation at such metallicities accounts for approximately 20\% of the total star formation at low redshift \citep{chen2016}, and so a similar factor should be included in calculating the fraction of stars that can lead to SLSNe.

We next assume that the total masses in our models---\Mej\,$+$\,\Mns---is representative of the total carbon-oxygen core mass formed by the progenitor by the time of explosion. The lowest masses in our fits are $\gtrsim 2$\,\M, with a typical NS mass of 1.8\,\M. Therefore we estimate that SLSNe result from stars that form CO cores with $M_{\rm CO} \gtrsim 4$\,\M. Comparing to the simulations of massive, rotating stars (at comparable metallicity) from \citet{yoon2006}, these core masses imply zero-age main sequence masses $M_{\rm ZAMS} \gtrsim 20$\,\M. The most massive SLSNe, with \Mej\,$\sim 20$\,\M, likely require progenitors of $\sim 60$\,\M. Assuming a Salpeter initial mass function, the fraction of SN progenitors ($M_{\rm ZAMS} \gtrsim 8$\,\M) with $M_{\rm ZAMS} \gtrsim 20$\,\M is roughly 25\%.

This lower bound on the $M_{\rm ZAMS}$ could be relaxed somewhat if the progenitors evolve chemically homogeneously. In this scenario, rapid rotation enables efficient mixing of nuclear burning products from the core into the envelope, and unburned material into the core, preventing the formation of strong chemical gradients \citep{mae1987}. This leads to compact, hot stars and massive CO cores. While difficult to confirm observationally, several lines of evidence suggest that chemically homogeneous evolution may occur in at least some massive stars \citep[e.g.~see recent work by][and references therein]{man2016}. For progenitors with initial rotation rates $\gtrsim 70$\% of the Keplerian velocity at the stellar surface, \citeauthor{yoon2006} found that even their 12\,\M\ models could form CO cores with mass compatible to our fits.

Interestingly, many theoretical studies find that chemically homogeneous evolution should be more prevalent at low metallicity. Simulations by \citet{bro2011} suggest that it can occur at LMC/SMC metallicity for stars with $M_{\rm ZAMS} \gtrsim 15-20$\,\M for surface rotation $\gtrsim 400$\,\kms, but they did not find such evolution at Galactic metallicity. Given that the observed metallicity threshold for SLSN production is comparable to LMC metallicity, chemically homogeneous evolution could be an important factor in explaining the deficit of high-metallicity SLSNe.

In addition to its effect on the CO core mass, rapid rotation is of course a key ingredient in millisecond magnetar models. For our posterior range in the spin period ($\approx 1-5$\,ms), the specific angular momentum of the magnetar is $\sim 10^{15}$\,cm$^2$\,s$^{-1}$. \citet{yoon2006} found that this is typical of the final CO cores in their simulations for stars with initial rotational velocities $\gtrsim 200-300$\,\kms\ (depending on metallicity). \citet{demink2013} compiled from various surveys the observed rotational velocities of OB stars. In the LMC, roughly 10-15\% of these stars had $v \sin i > 200$\,\kms, and a few percent had $v \sin i > 300$\,\kms. Therefore massive stars can form cores with sufficient angular momentum to produce SLSNe if they are in roughly the fastest 10\% of rotators.

\citet{demink2013} further modeled the observed distribution in $v \sin i$ for OB stars. They found that virtually all stars with $v \sin i > 200$\,\kms\ acquired their rapid rotation through binary interaction---either as the result of a merger or, more commonly, as the mass-gaining secondary following Roche lobe overflow. This suggests that binarity may be essential in supplying the angular momentum necessary to make SLSNe. \citet{sana2012} estimate that up to 70\% of O stars may be in binaries close enough to exchange mass. \citet{demink2013} also found that the fraction of massive stars with rapid rotation increased at lower metallicity, due to reduced stellar winds and more compact stellar structure for a given mass.

The stellar models of \citet{yoon2006}, which we used to estimate the progenitor masses from our ejecta masses, were for single stars, which may be problematic if most SLSNe are from interacting binaries. \citet{yoon2010} presented similar calculations for binary models. Their grid spacing is coarser than \citet{yoon2006}, but they also find that stars with $M_{\rm ZAMS} \gtrsim 20$\,\M\ can produce the required CO core mass for SLSNe.

Finally, our sample of SLSNe requires magnetars with fields $B>3\times10^{13}$\,G. \citet{fer2006} modeled the radio luminosities of a galactic population of neutron stars from the Parkes Multi Beam Survey and found $\lesssim 10$\% have magnetic fields in this range. If the magnetar field is primarily due to flux conservation during the collapse of the star, this suggests progenitors with magnetic fields of $\sim 1000$\,G. Such fields have been measured for a small number of Galactic O stars \citep{don2002,don2006}. However, it is also possible that the magnetar acquires its field through a dynamo mechanism \citep[e.g.][]{dunc1992,mos2015}

The posteriors in our model fits showed no correlations between any of the properties discussed above ($Z$, \Mej, \P, $B$). We can therefore make a simple rate calculation by assuming that each of these variables is independent. If SLSNe come from massive stars at $Z<0.5$\,\Z, and are in the top 25\%, 10\% and 10\% for progenitor mass, rotational velocity and magnetization, respectively, their rate compared to other core collapse SNe is $0.2 \times 0.25 \times 0.1 \times 0.1 = 5 \times 10^{-4}$. This is comparable to the observed SLSN rate\footnote{However, if the strong magnetic field is generated by a dynamo mechanism (i.e.~is a consequence of rapid rotation), our rate estimate could be biased somewhat low, and an additional ingredient may be required} This gives a volumetric rate of $\sim 100$\,Gpc$^{-3}$\,yr$^{-1}$, and is close to the observed rate of SLSNe \citep[found to be $\approx 30-100$\,Gpc$^{-3}$\,yr$^{-1}$;][]{qui2011,mcc2015,pra2016}.

The key ingredient for making SLSN progenitors appears to be rapid rotation. Other factors such as large (but not extreme) progenitor mass and low metallicity are most likely a consequence of this requirement: massive stars are more likely to be fast rotators and are frequently born in close binaries, and low metallicity reduces stellar winds and therefore angular momentum loss. Additionally, the chemically homogeneous evolutionary channel becomes accessible at larger mass and lower metallicity, and may therefore play an important role in forming the massive, rapidly rotating stellar cores that lead to SLSNe.

\section{Conclusions}
\label{sec:conc}

We have presented a set of MCMC model fits to the multicolour light curves of 38 SLSNe (the entire published spectroscopic sample with observations at maximum light), using our new open source light curve fitting code \mosfit \citep{gui2017}.

Examining the posteriors, we find that SLSNe have spin periods $\approx 1-6$\,ms, magnetic fields $\sim~10^{14}$\,G, and relatively large ejecta mass and kinetic energy, with typical values of $\sim~5$\,\M\ and $\sim~4\times10^{51}$\,erg but extending to $\gtrsim 20$\,\M and $\sim~10^{52}$\,erg. The ratio \Ek/\Mej\,$\approx1-2$, depending on assumptions about the ejecta velocity structure, putting SLSNe intermediate between normal SNe Ic and GRB-SNe.

While some of these values are similar to previous studies, we have shown for a large sample of SLSNe that the range of likely parameters for the class are quite modest, and well-constrained by the existing data. Our reasoning is as follows: for ejecta masses from reasonable progenitors, light curve widths are typically $\sim 30-100$\,days around peak. To input the required power of $\sim 10^{44}$\,\ergpers\ at this time, the engine must have either a spin-down time comparable to the diffusion time (within an order of magnitude), or a very short spin period (and thus a large energy reserve). So although the properties of the ejecta and magnetar are technically decoupled in this model, most combinations of \P\ and $B$ do not result in a particularly luminous light curve for realistic ejecta. Fainter events likely exist, but they may be difficult to distinguish from normal \Ni-powered SNe.

We also used our fits to estimate a luminosity function for SLSNe, with a median peak luminosity of $3.2 \times 10^{44}$\,\ergpers. The decrease in number at higher luminosity is relatively smooth, while the low-luminosity end is likely truncated by selection and classification biases. The most luminous events tend to have shorter spin periods and stronger magnetic fields.

In contrast to other models that fit the bolometric light curves of SLSNe, our multicolour fits provide strong constraints on the temperature and radius of SLSNe photospheres. While their optical light curves peak on a timescale of $\approx 25-100$\,d, bolometric light curves may peak as early as 15\,d. At this time, the events can be extremely UV-bright, as was recently observed for Gaia16apd \citep{yan2016,nic2017,kan2016b}. Events with cool spectra at maximum, such as iPTF13ehe \citep{yan2015} and especially PS1-14bj \citep{lun2016}, have long rise times that result in larger photospheric radii and lower spin-down power at this phase. We therefore stress that the temperature evolution, rather than the time of maximum light, is the important parameter when comparing SLSN spectra.

These slow-rising (and fading) SLSNe have been the subject of much debate, in particular as to whether they form a separate subclass distinct from other SLSNe. Using the bolometric output of our model fits, we find a continuous distribution in timescales, and argue for a single SLSN population. The diversity in their light curves simply come from the range in diffusion and spin-down times.

Magnetars have also become a popular model to explain FRBs. We calculated the spin-down power for each magnetar model in our sample at 10\,yr after explosion. The range of luminosity, $10^{39}-10^{43}$\,\ergpers, seems to be insufficient to power the most luminous FRBs. Magnetic (rather than rotational) powering of FRBs is still possible.

We compared the derived parameters of SLSNe with those of their host galaxies to determine whether any particular environmental variable correlates with a key property of the explosion, and found no significant correlations. This is in contrast to the recent study by \citet{chen2016}, who found a possible correlation between metallicity and spin period but for a much smaller SLSN sample.

Finally, we examined the implications of our magnetar parameters for constraining the progenitor stars of SLSNe. The ejecta masses imply progenitors with initial masses $M_{\rm ZAMS}\gtrsim20$\,\M, while the core angular momentum requirements suggest SLSNe come from the fastest rotating massive stars, which would likely require close binaries. The role of low metallicity is most likely to reduce angular momentum loss via stellar winds, and possibly even to enable chemically homogeneous evolution in some cases.

This study suggests several possibilities to make further progress in understanding SLSNe. On the theoretical side, our proposed progenitor scenario should be further tested using stellar models evolved to core collapse and simulations of magnetar formation such as those carried out by \citet{mos2015}. Our models neglected the early bumps observed in some SLSN light curves (but showed that on average their main peaks were indistinguishable from other SLSNe), so future work should aim to connect the properties of the two peaks. Progress in this direction has been made recently by \citet{mar2017}.

We now have a model code and a set of physical parameters that enables users to generate simulated model light curves that can reproduce the full SLSN sample. With these simulations readily available, future surveys could greatly boost their efficiency in photometrically classifying these events. Given the vast number of transients expected in the fast-approaching era of LSST, template light curves for these and other transient classes are essential.

\acknowledgments

We thank V.~Ashley Villar and Peter K.~G.~Williams for testing and development of \mosfit. This work benefited from many helpful conversations at the MIAPP workshop `SLSNe in the next decade'. M.N.~thanks Ting-Wan Chen and Yuqian Liu for providing metallicity and velocity measurements, respectively. The computations presented in this work were performed on Harvard University's Odyssey computer cluster, which is maintained by the Research Computing Group within the Faculty of Arts and Sciences. The Berger Time Domain group is supported in part by NSF grant AST-1411763 and NASA ADA grant NNX15AE50G.

\bibliographystyle{yahapj}
\bibliography{mybib}

\appendix
\section{appendix section}

\setcounter{figure}{0} 
\setcounter{table}{0} 

\begin{figure*}
\centering
\includegraphics[width=\textwidth]{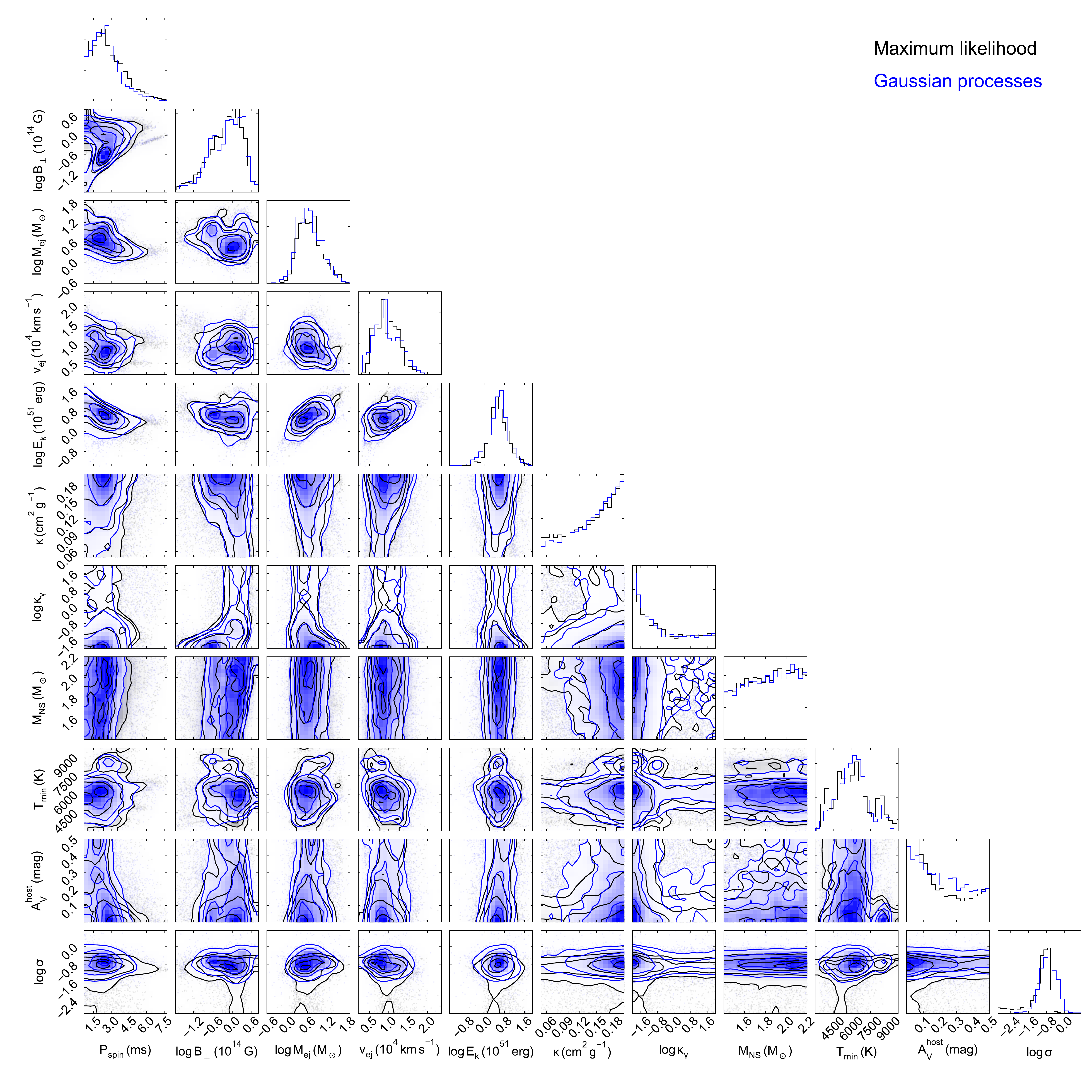}
\caption{Comparison of summed posteriors for fits using Maximum Likelihood Analysis (no explicit model for covariance in the data) vs Gaussian processes (including covariance; see \citealt{gui2017}). The overall sample properties are seen to be unchanged by the choice of fitting method.}
\label{fig:joint}
\end{figure*}

\begin{figure*}
\centering
\includegraphics[width=15cm]{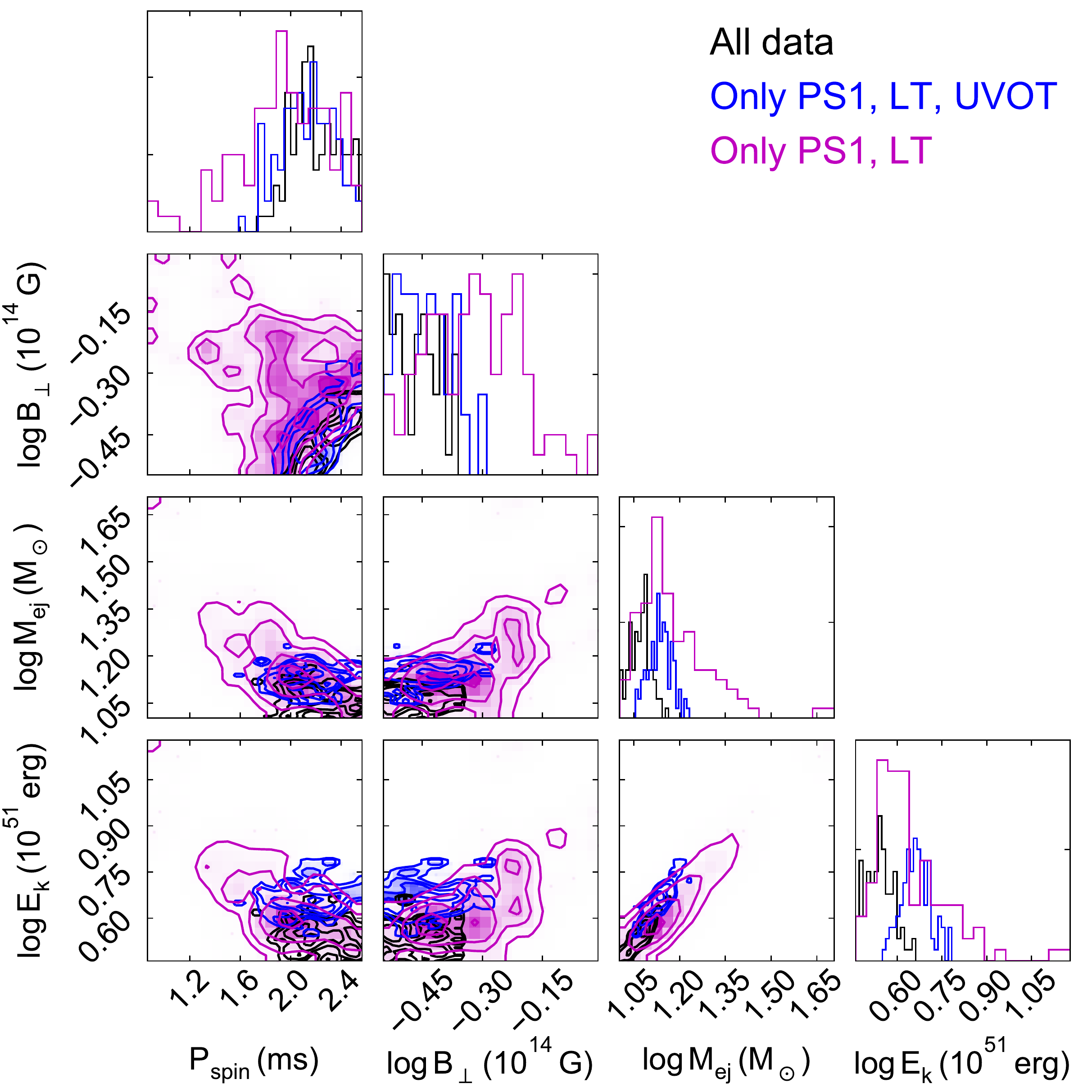}
\caption{Comparison of posteriors for SN\,2015bn with various cuts on the data. Only a subset of parameters are shown for clarity. Black: same as Figure \ref{fig:15bn}. Blue: Only includes PS1 public data, sparsely sampled Liverpool Telescope optical photometry ($ugriz$) and UV from \textit{Swift}/UVOT (hence no NIR data, no densely sampled optical, no deep late-time follow-up -- data used in fit makes up 40\% of the total available). Magenta: Same as blue, but also excluding UVOT data (thus using only 18\% of data). Posteriors overlap to $1\sigma$ or better in all cases, though the medians can differ by up to $\sim 0.1$\,dex. Excluding UV data primarily serves to broaden the posteriors, as the temperature is less tightly constrained. The inferred engine properties are very similar in all cases; mass and energy are more sensitive to the data cuts due to degeneracy in the diffusion time. Importantly, we find that the solution is not dependent on having extremely well-sampled light curves, giving us confidence for the rest of the sample. While variations in the data quality may broaden or introduce small shifts in the posteriors, this will not change the overall parameter space spanned by the sample or significantly affect our conclusions.}
\label{fig:15bn_compare}
\end{figure*}

\setlength\tabcolsep{3pt}
\begin{table*}
\caption{Medians and 1$\sigma$ bounds for all parameters and all SLSNe.}
\label{tab:pars}
\begin{center}
\begin{tabular}{ccccccccccccc}
	&	\P	&	\B	&	\Mej	&	$\langle$\vp$\rangle$*	&	$E_{\rm min}\dagger$	&	$\kappa$	&	$\kappa_\gamma$	&	\Mns	&	\Tf	&	\Av	&	$\sigma$	& WAIC$\ddagger$\\
	&	(ms)	&	($10^{14}$\,G)	&	(\M)	&	($10^3$\,\kms)	&	($10^{51}$\,erg)	&	(\cmsqperg)	&	(\cmsqperg)	&	(\M)	&	($10^3$\,K)	&	(mag)	& (mag)\\
\hline 
PTF10hgi	&	$4.78_{0.77}^{0.89}$	&	$2.03_{0.45}^{0.45}$	&	$2.19_{0.80}^{1.80}$	&	$5.12_{0.31}^{0.36}$	&	$0.55_{0.22}^{0.49}$	&	$0.10_{0.04}^{0.06}$	&	$0.06_{0.03}^{0.03}$	&	$1.85_{0.30}^{0.22}$	&	$6.58_{0.20}^{0.23}$	&	$0.11_{0.07}^{0.11}$	&	$0.12_{0.01}^{0.01}$	&96.83\\
Gaia16apd	&	$2.93_{0.36}^{0.34}$	&	$1.23_{0.27}^{0.26}$	&	$4.54_{0.71}^{1.59}$	&	$9.02_{0.18}^{0.24}$	&	$3.69_{0.59}^{1.38}$	&	$0.16_{0.04}^{0.02}$	&	$0.57_{0.39}^{13.11}$	&	$1.83_{0.27}^{0.29}$	&	$8.00_{0.16}^{0.12}$	&	$0.02_{0.02}^{0.03}$	&	$0.12_{0.00}^{0.01}$	&567.56\\
PTF12dam	&	$2.28_{0.30}^{0.32}$	&	$0.18_{0.05}^{0.04}$	&	$6.27_{0.95}^{1.23}$	&	$7.01_{0.25}^{0.31}$	&	$3.03_{0.46}^{0.69}$	&	$0.16_{0.04}^{0.02}$	&	$0.01_{0.00}^{0.00}$	&	$1.83_{0.27}^{0.26}$	&	$6.48_{0.21}^{0.28}$	&	$0.16_{0.10}^{0.09}$	&	$0.22_{0.02}^{0.01}$	&109.39\\
SN2015bn	&	$2.16_{0.17}^{0.29}$	&	$0.31_{0.05}^{0.07}$	&	$11.73_{1.34}^{0.83}$	&	$5.46_{0.14}^{0.16}$	&	$3.45_{0.43}^{0.42}$	&	$0.19_{0.02}^{0.01}$	&	$0.01_{0.00}^{0.00}$	&	$1.78_{0.23}^{0.28}$	&	$8.32_{0.16}^{0.32}$	&	$0.08_{0.04}^{0.09}$	&	$0.18_{0.01}^{0.01}$	&587.65\\
SN2007bi	&	$3.92_{0.50}^{0.53}$	&	$0.35_{0.08}^{0.13}$	&	$3.80_{1.09}^{1.52}$	&	$7.90_{1.41}^{0.95}$	&	$2.37_{0.94}^{0.98}$	&	$0.16_{0.05}^{0.03}$	&	$0.06_{0.02}^{0.03}$	&	$1.81_{0.26}^{0.24}$	&	$8.38_{0.36}^{0.42}$	&	$0.07_{0.06}^{0.11}$	&	$0.13_{0.01}^{0.01}$	&169.22\\
SN2011ke	&	$0.78_{0.06}^{0.09}$	&	$3.88_{0.64}^{0.32}$	&	$7.64_{1.89}^{6.96}$	&	$8.15_{0.32}^{0.23}$	&	$5.22_{1.50}^{4.60}$	&	$0.13_{0.06}^{0.04}$	&	$4.75_{4.24}^{47.75}$	&	$2.05_{0.19}^{0.10}$	&	$5.52_{0.17}^{0.20}$	&	$0.06_{0.05}^{0.08}$	&	$0.19_{0.02}^{0.02}$	&129.84\\
SSS120810	&	$3.00_{1.11}^{0.90}$	&	$1.93_{0.48}^{0.45}$	&	$2.22_{0.66}^{1.25}$	&	$11.13_{0.88}^{0.93}$	&	$2.82_{1.04}^{1.40}$	&	$0.14_{0.06}^{0.03}$	&	$0.22_{0.12}^{3.00}$	&	$1.88_{0.35}^{0.22}$	&	$3.80_{0.18}^{0.23}$	&	$0.33_{0.19}^{0.12}$	&	$0.20_{0.03}^{0.03}$	&31.00\\
SN2012il	&	$2.35_{0.46}^{0.51}$	&	$2.24_{0.57}^{0.33}$	&	$3.14_{0.58}^{0.97}$	&	$7.93_{0.75}^{0.57}$	&	$1.94_{0.35}^{0.65}$	&	$0.08_{0.01}^{0.03}$	&	$3.18_{2.89}^{31.64}$	&	$1.90_{0.34}^{0.19}$	&	$6.27_{0.15}^{0.24}$	&	$0.12_{0.08}^{0.10}$	&	$0.11_{0.01}^{0.02}$	&56.56\\
PTF11rks	&	$2.07_{0.95}^{2.77}$	&	$2.88_{2.13}^{1.22}$	&	$6.54_{4.57}^{4.08}$	&	$12.11_{1.89}^{1.81}$	&	$7.97_{5.06}^{9.13}$	&	$0.16_{0.05}^{0.03}$	&	$0.25_{0.23}^{20.23}$	&	$1.87_{0.30}^{0.22}$	&	$8.28_{0.59}^{0.45}$	&	$0.42_{0.07}^{0.06}$	&	$0.29_{0.03}^{0.02}$	&44.94\\
iPTF15esb	&	$1.62_{0.62}^{0.65}$	&	$2.39_{0.46}^{0.56}$	&	$17.50_{4.22}^{11.92}$	&	$6.28_{0.39}^{0.44}$	&	$7.22_{2.20}^{4.93}$	&	$0.14_{0.05}^{0.03}$	&	$1.21_{1.17}^{13.79}$	&	$1.80_{0.21}^{0.29}$	&	$5.37_{0.32}^{0.38}$	&	$0.20_{0.13}^{0.15}$	&	$0.17_{0.01}^{0.01}$	&196.86\\
SN2010gx	&	$3.66_{0.54}^{0.60}$	&	$0.59_{0.15}^{0.28}$	&	$2.39_{0.58}^{0.61}$	&	$12.65_{0.63}^{0.55}$	&	$3.78_{0.74}^{0.84}$	&	$0.18_{0.03}^{0.02}$	&	$0.02_{0.01}^{0.02}$	&	$1.79_{0.24}^{0.28}$	&	$3.99_{0.12}^{0.11}$	&	$0.02_{0.01}^{0.02}$	&	$0.14_{0.01}^{0.01}$	&248.46\\
SN2011kf	&	$1.48_{0.66}^{1.16}$	&	$0.70_{0.33}^{0.56}$	&	$4.57_{2.66}^{17.85}$	&	$11.46_{1.45}^{1.15}$	&	$6.72_{3.69}^{20.25}$	&	$0.16_{0.05}^{0.03}$	&	$0.04_{0.02}^{0.04}$	&	$1.85_{0.30}^{0.26}$	&	$5.86_{0.25}^{0.26}$	&	$0.23_{0.18}^{0.20}$	&	$0.06_{0.03}^{0.02}$	&66.53\\
iPTF16bad	&	$3.73_{0.70}^{0.65}$	&	$2.62_{0.49}^{0.55}$	&	$2.22_{0.98}^{1.05}$	&	$7.11_{0.59}^{0.71}$	&	$1.12_{0.36}^{0.29}$	&	$0.07_{0.02}^{0.04}$	&	$1.96_{1.76}^{18.69}$	&	$1.79_{0.24}^{0.23}$	&	$6.28_{0.65}^{0.35}$	&	$0.05_{0.04}^{0.08}$	&	$0.08_{0.01}^{0.01}$	&91.29\\
LSQ14mo	&	$4.97_{0.71}^{0.65}$	&	$1.01_{0.30}^{0.27}$	&	$2.10_{0.36}^{0.42}$	&	$10.74_{0.41}^{0.52}$	&	$2.43_{0.43}^{0.50}$	&	$0.17_{0.02}^{0.03}$	&	$0.02_{0.00}^{0.01}$	&	$1.85_{0.27}^{0.22}$	&	$4.97_{0.16}^{0.17}$	&	$0.08_{0.06}^{0.10}$	&	$0.00_{0.00}^{0.01}$	&128.35\\
LSQ12dlf	&	$2.82_{0.58}^{0.55}$	&	$1.20_{0.26}^{0.31}$	&	$3.68_{0.96}^{2.28}$	&	$8.28_{0.24}^{0.25}$	&	$2.54_{0.75}^{1.49}$	&	$0.11_{0.04}^{0.04}$	&	$2.36_{2.00}^{18.09}$	&	$1.77_{0.25}^{0.31}$	&	$3.77_{0.14}^{0.14}$	&	$0.29_{0.11}^{0.14}$	&	$0.08_{0.01}^{0.01}$	&141.34\\
PTF09cnd	&	$1.46_{0.48}^{0.38}$	&	$0.10_{0.06}^{0.09}$	&	$5.16_{1.64}^{2.41}$	&	$8.56_{1.41}^{1.53}$	&	$3.29_{1.38}^{3.82}$	&	$0.16_{0.05}^{0.03}$	&	$0.01_{0.00}^{0.01}$	&	$1.82_{0.23}^{0.22}$	&	$4.44_{0.41}^{0.36}$	&	$0.17_{0.12}^{0.18}$	&	$0.13_{0.02}^{0.01}$	&58.03\\
SN2013dg	&	$3.50_{0.59}^{0.60}$	&	$1.56_{0.32}^{0.41}$	&	$2.75_{0.99}^{1.63}$	&	$8.38_{0.51}^{0.44}$	&	$1.85_{0.63}^{1.10}$	&	$0.12_{0.04}^{0.06}$	&	$0.04_{0.02}^{0.02}$	&	$1.80_{0.21}^{0.22}$	&	$5.07_{0.31}^{0.23}$	&	$0.07_{0.06}^{0.10}$	&	$0.01_{0.01}^{0.02}$	&125.80\\
SN2005ap	&	$1.28_{0.39}^{0.57}$	&	$1.71_{0.63}^{0.75}$	&	$3.57_{1.23}^{3.04}$	&	$15.22_{2.20}^{2.51}$	&	$8.85_{4.39}^{9.50}$	&	$0.15_{0.06}^{0.03}$	&	$0.09_{0.08}^{5.05}$	&	$1.89_{0.35}^{0.16}$	&	$5.77_{0.95}^{0.89}$	&	$0.25_{0.16}^{0.15}$	&	$0.01_{0.00}^{0.01}$	&57.05\\
iPTF13ehe	&	$2.57_{0.30}^{0.38}$	&	$0.20_{0.06}^{0.09}$	&	$10.03_{2.55}^{2.28}$	&	$6.66_{0.31}^{0.27}$	&	$4.48_{1.12}^{0.97}$	&	$0.16_{0.04}^{0.03}$	&	$0.04_{0.02}^{0.02}$	&	$1.87_{0.25}^{0.22}$	&	$5.01_{0.15}^{0.19}$	&	$0.15_{0.11}^{0.16}$	&	$0.05_{0.01}^{0.01}$	&121.38\\
LSQ14bdq	&	$0.98_{0.15}^{0.20}$	&	$0.49_{0.12}^{0.13}$	&	$33.71_{6.56}^{6.16}$	&	$8.71_{0.66}^{0.61}$	&	$25.06_{6.99}^{8.59}$	&	$0.19_{0.02}^{0.01}$	&	$0.01_{0.00}^{0.00}$	&	$1.80_{0.20}^{0.27}$	&	$6.78_{0.29}^{0.49}$	&	$0.37_{0.14}^{0.09}$	&	$0.12_{0.02}^{0.02}$	&53.51\\
PTF09cwl	&	$1.74_{0.76}^{0.66}$	&	$0.27_{0.22}^{0.59}$	&	$6.98_{2.78}^{9.32}$	&	$9.11_{2.08}^{4.14}$	&	$6.78_{3.41}^{11.54}$	&	$0.17_{0.06}^{0.02}$	&	$0.03_{0.02}^{2.05}$	&	$1.86_{0.32}^{0.27}$	&	$3.91_{0.26}^{0.32}$	&	$0.32_{0.19}^{0.11}$	&	$0.06_{0.06}^{0.22}$	&-8.21\\
SN2006oz	&	$2.70_{0.75}^{0.74}$	&	$0.32_{0.19}^{0.24}$	&	$2.97_{1.08}^{2.58}$	&	$9.46_{0.75}^{0.69}$	&	$2.66_{1.00}^{2.31}$	&	$0.13_{0.04}^{0.06}$	&	$0.39_{0.35}^{13.20}$	&	$1.80_{0.23}^{0.28}$	&	$5.93_{1.05}^{0.72}$	&	$0.11_{0.08}^{0.15}$	&	$0.01_{0.01}^{0.02}$	&76.58\\
iPTF13dcc	&	$0.81_{0.08}^{0.15}$	&	$0.98_{0.15}^{0.21}$	&	$23.96_{7.99}^{12.51}$	&	$5.23_{0.17}^{0.29}$	&	$6.61_{2.10}^{3.75}$	&	$0.10_{0.03}^{0.05}$	&	$0.68_{0.64}^{19.91}$	&	$1.97_{0.23}^{0.15}$	&	$6.27_{0.52}^{0.52}$	&	$0.05_{0.03}^{0.09}$	&	$0.21_{0.02}^{0.02}$	&76.56\\
PTF09atu	&	$1.59_{0.49}^{0.43}$	&	$0.09_{0.05}^{0.08}$	&	$6.20_{1.72}^{2.01}$	&	$11.76_{0.78}^{1.10}$	&	$8.30_{2.00}^{2.69}$	&	$0.16_{0.05}^{0.03}$	&	$0.02_{0.01}^{0.02}$	&	$1.88_{0.30}^{0.18}$	&	$4.98_{0.45}^{0.66}$	&	$0.21_{0.14}^{0.17}$	&	$0.06_{0.02}^{0.02}$	&64.84\\
PS1-14bj	&	$2.82_{0.42}^{0.54}$	&	$0.13_{0.03}^{0.09}$	&	$18.46_{2.50}^{2.72}$	&	$5.07_{0.89}^{0.58}$	&	$4.61_{1.65}^{1.66}$	&	$0.18_{0.02}^{0.02}$	&	$0.01_{0.00}^{0.00}$	&	$1.85_{0.36}^{0.25}$	&	$8.99_{0.30}^{0.27}$	&	$0.05_{0.03}^{0.07}$	&	$0.20_{0.02}^{0.02}$	&60.03\\
PS1-11ap	&	$3.66_{0.41}^{0.45}$	&	$0.82_{0.18}^{0.14}$	&	$5.29_{1.74}^{2.62}$	&	$5.73_{0.23}^{0.20}$	&	$1.73_{0.55}^{0.90}$	&	$0.10_{0.04}^{0.07}$	&	$7.22_{6.24}^{35.67}$	&	$1.87_{0.30}^{0.22}$	&	$8.11_{0.19}^{0.33}$	&	$0.06_{0.04}^{0.07}$	&	$0.19_{0.01}^{0.01}$	&246.12\\
DES14X3taz	&	$2.41_{0.26}^{0.30}$	&	$0.39_{0.10}^{0.13}$	&	$5.41_{0.99}^{0.79}$	&	$10.46_{0.63}^{0.94}$	&	$5.87_{0.93}^{0.78}$	&	$0.18_{0.02}^{0.02}$	&	$0.01_{0.00}^{0.00}$	&	$1.87_{0.28}^{0.21}$	&	$6.17_{0.28}^{0.34}$	&	$0.44_{0.10}^{0.05}$	&	$0.15_{0.03}^{0.03}$	&49.51\\
PS1-10bzj	&	$5.21_{0.65}^{0.76}$	&	$1.63_{0.42}^{0.24}$	&	$1.65_{0.28}^{0.51}$	&	$11.80_{0.63}^{0.73}$	&	$2.32_{0.48}^{0.80}$	&	$0.16_{0.03}^{0.03}$	&	$6.15_{5.72}^{27.05}$	&	$1.86_{0.33}^{0.22}$	&	$6.18_{0.93}^{0.55}$	&	$0.11_{0.09}^{0.16}$	&	$0.10_{0.02}^{0.02}$	&69.21\\
DES13S2cmm	&	$6.59_{0.75}^{0.70}$	&	$0.73_{0.14}^{0.20}$	&	$2.31_{0.36}^{0.47}$	&	$9.99_{0.75}^{0.58}$	&	$2.31_{0.41}^{0.55}$	&	$0.16_{0.03}^{0.02}$	&	$4.48_{3.75}^{36.82}$	&	$1.76_{0.26}^{0.31}$	&	$7.05_{0.18}^{0.26}$	&	$0.06_{0.05}^{0.08}$	&	$0.10_{0.02}^{0.02}$	&123.44\\
iPTF13ajg	&	$1.82_{0.21}^{0.25}$	&	$0.75_{0.20}^{0.15}$	&	$4.87_{0.87}^{0.89}$	&	$13.08_{0.58}^{0.57}$	&	$8.35_{1.61}^{1.54}$	&	$0.17_{0.02}^{0.02}$	&	$1.58_{1.46}^{28.31}$	&	$1.89_{0.32}^{0.21}$	&	$4.75_{0.55}^{0.31}$	&	$0.45_{0.06}^{0.03}$	&	$0.13_{0.01}^{0.01}$	&142.84\\
PS1-10awh	&	$1.76_{0.52}^{0.65}$	&	$1.56_{0.34}^{0.44}$	&	$6.47_{1.94}^{3.78}$	&	$11.83_{0.92}^{1.22}$	&	$9.16_{3.33}^{6.01}$	&	$0.16_{0.04}^{0.03}$	&	$2.02_{2.00}^{31.19}$	&	$1.80_{0.28}^{0.27}$	&	$5.90_{0.87}^{1.10}$	&	$0.08_{0.05}^{0.08}$	&	$0.02_{0.02}^{0.02}$	&187.13\\
PS1-10ky	&	$1.58_{0.27}^{0.38}$	&	$1.19_{0.29}^{0.26}$	&	$3.94_{0.98}^{2.85}$	&	$12.41_{0.70}^{0.99}$	&	$6.36_{1.60}^{3.55}$	&	$0.14_{0.05}^{0.04}$	&	$4.44_{4.15}^{39.88}$	&	$1.85_{0.31}^{0.23}$	&	$5.90_{0.73}^{1.10}$	&	$0.39_{0.12}^{0.08}$	&	$0.04_{0.01}^{0.02}$	&188.56\\
PS1-10ahf	&	$2.35_{0.78}^{0.57}$	&	$0.17_{0.12}^{0.11}$	&	$10.50_{3.23}^{3.57}$	&	$6.37_{1.02}^{0.94}$	&	$4.10_{1.48}^{1.70}$	&	$0.15_{0.05}^{0.04}$	&	$0.07_{0.05}^{12.80}$	&	$1.85_{0.25}^{0.25}$	&	$6.85_{0.89}^{0.92}$	&	$0.26_{0.17}^{0.13}$	&	$0.22_{0.02}^{0.02}$	&45.95\\
SCP-06F6	&	$1.78_{0.55}^{0.53}$	&	$0.16_{0.10}^{0.19}$	&	$7.02_{1.49}^{1.43}$	&	$11.13_{1.93}^{2.40}$	&	$8.35_{2.85}^{6.24}$	&	$0.18_{0.04}^{0.02}$	&	$0.01_{0.00}^{0.01}$	&	$1.75_{0.22}^{0.27}$	&	$5.70_{0.77}^{0.73}$	&	$0.18_{0.12}^{0.23}$	&	$0.25_{0.05}^{0.07}$	&3.79\\
PS1-10pm	&	$1.31_{0.43}^{0.53}$	&	$0.06_{0.04}^{0.06}$	&	$4.03_{1.24}^{1.85}$	&	$15.75_{1.01}^{1.24}$	&	$9.76_{3.34}^{4.76}$	&	$0.15_{0.06}^{0.03}$	&	$0.02_{0.01}^{0.02}$	&	$1.85_{0.29}^{0.26}$	&	$6.47_{0.48}^{0.70}$	&	$0.10_{0.06}^{0.11}$	&	$0.08_{0.03}^{0.03}$	&55.10\\
SNLS-07D2bv	&	$3.49_{0.60}^{0.57}$	&	$0.26_{0.09}^{0.11}$	&	$1.55_{0.46}^{1.00}$	&	$10.89_{0.47}^{0.39}$	&	$1.85_{0.57}^{1.06}$	&	$0.12_{0.05}^{0.05}$	&	$2.19_{2.08}^{21.20}$	&	$1.80_{0.31}^{0.26}$	&	$5.85_{0.99}^{1.14}$	&	$0.03_{0.02}^{0.03}$	&	$0.09_{0.01}^{0.01}$	&152.86\\
PS1-11bam	&	$2.39_{0.87}^{0.74}$	&	$1.01_{0.23}^{0.39}$	&	$3.73_{1.66}^{3.58}$	&	$8.94_{1.56}^{1.38}$	&	$2.81_{0.88}^{1.24}$	&	$0.09_{0.03}^{0.06}$	&	$1.18_{1.07}^{22.89}$	&	$1.83_{0.27}^{0.26}$	&	$6.19_{1.22}^{0.96}$	&	$0.06_{0.04}^{0.07}$	&	$0.02_{0.02}^{0.05}$	&22.93\\
SNLS-06D4eu	&	$3.55_{0.68}^{0.58}$	&	$0.79_{0.43}^{0.20}$	&	$2.06_{0.39}^{0.50}$	&	$13.19_{0.80}^{0.76}$	&	$3.63_{0.92}^{1.08}$	&	$0.17_{0.03}^{0.02}$	&	$0.17_{0.14}^{15.93}$	&	$1.88_{0.28}^{0.23}$	&	$6.05_{1.16}^{1.00}$	&	$0.04_{0.03}^{0.07}$	&	$0.13_{0.02}^{0.02}$	&57.26\\

\hline
\end{tabular}
\end{center}

* This is the time-averaged photospheric velocity, rather than the velocity of the fastest material.\\
$\dagger$ For our assumed density profile \citep{che1989,mar2017}, \Ek\,$=1/2$\,\Mej\,\vej$^2$, where \vej\ is the characteristic velocity of the fastest ejecta. The minimum kinetic energy of each SLSN is calculated by assuming \vej\,=\,\vp, i.e.~the photosphere encloses most of the ejecta. If instead we assume the velocity of the fastest material is represented by the maximum-light velocities measured by \citet{Liu&Modjaz16}, the typical \Ek\ is larger by a factor $\approx2$.\\
$\ddagger$ The Watanabe-Akaike information criteria \citep[or ``widely applicable Bayesian criteria''][]{watanabe2010,gelman2014}.

\end{table*}

\end{document}